\RequirePackage{rotating}
\documentclass[fleqn,usenatbib]{mnras}

\usepackage[T1]{fontenc}
\usepackage{units}

\DeclareRobustCommand{\VAN}[3]{#2}
\let\VANthebibliography\thebibliography
\def\thebibliography{\DeclareRobustCommand{\VAN}[3]{##3}\VANthebibliography}

\usepackage{graphicx}	
\usepackage{amsmath}	
\usepackage{amssymb}	
\usepackage{newtxtext,newtxmath}
\usepackage{multirow}
\usepackage{lscape}
\usepackage{makecell}
\usepackage{subfigure}
\usepackage{float}
\usepackage{newtxtext,newtxmath}
\usepackage{lineno}
\setcitestyle{notesep={ }}
\makeatletter
\newcommand{\labtext}[2]{%
  \@bsphack
  \csname phantomsection\endcsname 
  \def\@currentlabel{#1}{\label{#2}}%
  \@esphack
}

\newcommand{\redcheck}{{\color{black}\checkmark}}
\newcommand{\myhash}{\#}

\setcitestyle{notesep={}}

\makeatother

\title[Extensive Periodicity Analysis in Jetted AGN]{Extensive Analysis of $\gamma$-Ray Periodicity in Jetted AGN from the 4FGL Catalog Using \textit{Fermi}-LAT Observations}
\author[P. Peñil et al.]{
P. Peñil,$^{1}$\thanks{E-mail: ppenil@clemson.edu}
A. Dom\'inguez,$^{2}$\thanks{E-mail: alberto.d@ucm.es}
S. Buson,$^{3,4}$\thanks{E-mail: sara.buson@uni-wuerzburg.de}
M. Ajello,$^{1}$
S. Adhikari,$^{1}$
A. Rico,$^{1}$
\\
$^{1}$Department of Physics and Astronomy, Clemson University, Kinard Lab of Physics, Clemson, SC 29634-0978, USA\\
$^{2}$IPARCOS and Department of EMFTEL, Universidad Complutense de Madrid, E-28040 Madrid, Spain\\
$^{3}$Julius-Maximilians-Universität Würzburg, Fakultät für Physik und Astronomie, Emil-Fischer-Str. 31, D-97074 Würzburg, Germany\\
$^{4}$Deutsches Elektronen-Synchrotron DESY, Platanenallee 6, 15738 Zeuthen, Germany\\
}

\date{Accepted 2025 September 15. Received 2025 September 15; in original form 2025 April 16}

\pubyear{2025}

\begin{document}
\label{firstpage}
\pagerange{\pageref{firstpage}--\pageref{lastpage}}
\maketitle
\begin{abstract}
The quest to uncover periodic patterns within the $\gamma$-ray emissions of jetted active galactic nuclei (AGN) has recently emerged as a focal point in astrophysics. One of the primary challenges has been the necessity for prolonged exposures in the $\gamma$-ray energy band. In our investigation, we leverage 12 years' worth of observations from the \textit{Fermi}-LAT to systematically explore periodicity across 1492 jetted AGN cataloged in 4FGL, representing the largest sample analyzed to date. Our analysis involves a robust pipeline employing nine distinct techniques designed to detect potential periodic emissions within their $\gamma$ rays. We note that 24 objects with previous hints of periodicity are deliberately excluded in the present work since they were reanalyzed in a dedicated paper using a similar methodology. Using this thorough approach, we do not find any evidence for periodic signals in the 1492 jetted AGN $\gamma$-ray light curves analyzed here.
\end{abstract}


\section{Introduction} \label{sec:intro}
In jetted active galactic nuclei (AGN), the emission is predominantly from the jet and shows significant variability. This variability occurs across the entire electromagnetic spectrum over different timescales, ranging from minutes to years \citep[e.g.,][]{sillanpaa_oj287, urry_multiwavelengh, sarkar_cta1002_fast_period}. Evidence of periodic patterns has been observed in some objects \citep[e.g.,][]{ackermann_pg1553, jorge_2022}. This observed periodic variability in these jetted AGN is commonly associated with a stochastic process \citep[e.g.,][]{covino_negation}, often stemming from either the accretion disk and subsequently transmitted to the jet \citep[e.g.,][]{rieger_2019} or originating directly within the jet itself \citep[e.g.,][]{giannios_mini_jets}. However, recent discoveries of periodic behavior in certain jetted AGN have sparked debate regarding its underlying causes \citep[e.g.,][]{ren_s5_1044+71, sagar_pg1553}. The identification of periodic emissions in jetted AGN holds the potential to reveal crucial insights into various aspects. Specifically, the periodic behavior has been explained through various geometrical models. Some explanations include the orbital motion of blobs within the jet under the influence of magnetic fields \citep{mohan_qpo_blobs}, twisted and inhomogeneous jet structures \citep{raiteri2017}, the formation of helical jets or particles moving along helical trajectories \citep{rieger2004}, and the precession of the jet, often referred to as the lighthouse effect \citep{camenzind1992}. Additionally, a periodic emission might even hint at the existence of a supermassive black hole binary \citep[SMBHB, e.g.,][]{tavani_blazar, dey2018authenticating, oneill2022}. Nonetheless, a key question remains: is this periodicity a genuine recurring pattern, or is it due to noise?

The search for long-term $\gamma$-ray periodicity in jetted AGN, particularly periods extending beyond a year, has been challenging due to the absence of continuous, extended monitoring. However, the \textit{Fermi}-Large Area Telescope (LAT), operational for over 12 years, stands out as an ideal tool for this purpose. Its capacity to regularly scan the entire sky with high sensitivity makes it well-suited for prolonged monitoring. Previously, efforts to identify periodic behavior have mainly concentrated on a select few objects \citep[e.g.,][]{gong_pks_0405_385, zhang_pks0521_36}. More recent studies have expanded this analysis to encompass a larger sample of blazars, examining tens of these objects \citep[e.g.,][]{yang_carma, ren_s5_1044+71}. \citet{tarnopolski20} examined eleven blazars and found a 3$\sigma$ periodicity only in PKS 2155$-$304, confirming results by \citet{penil_2022} using a methodology analogous to that employed in this paper. In a similar vein, \citet{benkhali_power_spectrum} analyzed six blazars and reported $>$3$\sigma$ significance in five cases of the six blazars analyzed, most of which overlap with those studied by \citet{penil_2022}. However, the apparent periods in blazar light curves (LCs) can also arise from purely stochastic processes \citep[e.g.][]{vaughan_criticism, covino_negation}, leading to spurious detections of periodicity.

In a study by \citet{penil_2020} - hereafter referred to as P20 - over 351 jetted AGN, mostly blazars, were investigated using 9 years of $\gamma$-ray data from the \textit{Fermi}-LAT. A dedicated analysis pipeline was developed to systematically search for periodic patterns. This initial analysis unveiled 24 blazars exhibiting evidence of periodic emissions in their $\gamma$ rays at a local significance $\geq$2$\sigma$. Building upon this, in \citet{penil_2022} the same sample of blazars was re-examined using 12 years of \textit{Fermi}-LAT observations, using an extended energy coverage, and with an improved analysis pipeline similar to the one used in the present work.

In this paper, our focus revolves around a systematic exploration of periodic $\gamma$ rays within a sample consisting of 1492 jetted AGN. To carry out this investigation, we rely on data collected by the \textit{Fermi}-LAT over a span of 12 years, starting from August 2008 and extending through December 2020. Our approach builds upon the pipeline used in P20 as a foundation, with specific adjustments made to address certain limitations. We specifically adjusted the pipeline to resolve previous issues and introduced new analytical methods, thereby enhancing the robustness of the results obtained. These modifications are implemented to ensure a more dependable and comprehensive assessment of periodic $\gamma$ rays within this extensive dataset.

The paper is structured as follows. In $\S$\ref{sec:sample}, we introduce the blazar sample used and outline the methodology employed for the data reduction. Next, $\S$\ref{sec:methodology} presents the periodicity analysis methodology. This section details the specific modifications integrated into the pipeline for this study. In $\S$\ref{tab:power_indices}, we present the estimation of the power spectral density (PSD) and the statistical corrections derived from them. $\S$\ref{sec:correction} covers additional statistical corrections that were implemented to enhance the reliability and robustness of our results. $\S$\ref{sec:results} presents the results derived from our study. Lastly, we summarize the findings in $\S$\ref{sec:summary}.

\section{Blazar sample} \label{sec:sample}
The fourth Fermi-LAT source catalog (4FGL), derived from the initial eight years of LAT observations, encompasses over 5000 $\gamma$-ray sources. A development, the 4FGL-DR2 catalog, has been released, leveraging the first 10 years of LAT observations \citep[][]{4fgl_dr2}. Out of these $\gamma$-ray sources, over 3300 are categorized as jetted AGN, with $\approx$97\% of them being blazars. We identify the variable jetted AGN, utilizing the variability index, which serves as a numerical measure designed to evaluate and quantify the degree of variability displayed by blazars based on changes in their brightness over time \citep[][]{abdollahi_4fgl}. In the context of 4FGL-DR2, a jetted AGN is deemed variable when the variability index $\geq$18.48 \citep[as indicated by][]{abdollahi_4fgl}, resulting in a total of 1620 jetted AGN, accounting for 48.9\% of the initial sample. Following this, we refine the selection of jetted AGN by excluding those with 50\% or more upper limits in the LC (as we did in P20). Consequently, our final sample consists of 1492 jetted AGN, representing 45.1\% of the catalog sample. It is important to mention that, for the analysis conducted in this study, we intentionally excluded the 24 blazars that are analyzed with the same methodology in a different paper \citep[][]{penil_2022}. The utilization of the same pipeline in both papers ensures consistency between the studies. 

\subsection{{\it Fermi}-LAT Data Analysis}\label{sec:fermi_analysis}
The {\it Fermi}-LAT data reduction is performed by using the Python package \texttt{fermipy}\footnote{We use the version 0.19.0} \citep{Wood:2017yyb} and applied to each jetted AGN included in our sample. The specific procedure is explained below. We select the photons belonging to the \texttt{Pass 8 SOURCE} class \citep[][]{atwood_source_class, bruel_pass8}, in a region of interest (ROI) of 15$^\circ$ $\times$ 15$^\circ$ square, centered at the target. The ROI model includes all 4FGL-DR2 catalog sources \citep[][]{4fgl_dr2} located within 20$^{\circ}$ from the ROI center, as well as the Galactic and isotropic diffuse emission \footnote{\url{https://fermi.gsfc.nasa.gov/ssc/data/access/lat/BackgroundModels.html}} (\texttt{gll\_iem\_v07.fits} and \texttt{iso\_P8R3\_SOURCE\_V2.txt}). We apply a zenith angle cut of $\theta < 90^{\circ}$ to minimize the contamination from $\gamma$ rays produced in the Earth’s upper atmosphere. To avoid potential spurious effects in the periodicity analysis, we remove the time periods coinciding with solar flares and $\gamma$-ray bursts detected by the LAT. Finally, the standard data quality cuts ($ \rm DATA\_QUAL > 0) \&\& (LAT\_CONFIG == 1$) are applied. 

We use the \texttt{P8R3\_SOURCE\_V2} instrument response functions to carry out a binned analysis in the 0.1-800 GeV energy range using ten bins per decade in energy and 0.1$^{\circ}$ spatial bins. Considering a full-time range of 2008 Aug 04 15:43:36 UTC to 2020 Dec 10 00:01:26 UTC, we perform a maximum likelihood analysis. Each source is modeled using the spectral shapes and parameters reported in 4FGL as starting values for the fit. We first perform a fit of the ROI by means of the \textit{fermipy} method ``optimize'' to ensure that all spectral parameters are close to their global likelihood maxima. This is done by iteratively optimizing the components of the ROI model in sequential steps, starting from the largest components.\footnote{\url{https://fermipy.readthedocs.io/en/latest/fitting.html}}  
The initial results are evaluated for potential newly-detected sources with an iterative procedure by a test statistic (TS) map since our data span a different integration time to 4FGL. The TS is defined as $2\log(L/L_0)$, where \textit{$L_0$} is the likelihood without the source and \textit{$L$} is the likelihood of the model with a point source at a given position. TS=25 corresponds to a statistical significance of $\gtrsim4.0\sigma$~\citep[according to the prescription adopted in][]{mattox1996, abdollahi_4fgl}. We produce a TS map with a putative point source at each map pixel and evaluate its significance over the current best-fit model. The test source is modeled with a power-law spectrum where only the normalization can vary in the fit process, whereas the photon index is fixed at 2. We search for significant peaks (TS$>$25) in the TS map, with a minimum separation of 0.5$^{\circ}$ from existing sources in the model. We add a new point source to the model at the position of the most significant peak found. Then, the ROI is fitted again, and a new TS map is produced. This process is iterated until no more significant excesses are found, generally leading to adding two point-sources. 

\begin{figure*}
	\centering
 	\includegraphics[scale=0.2295]{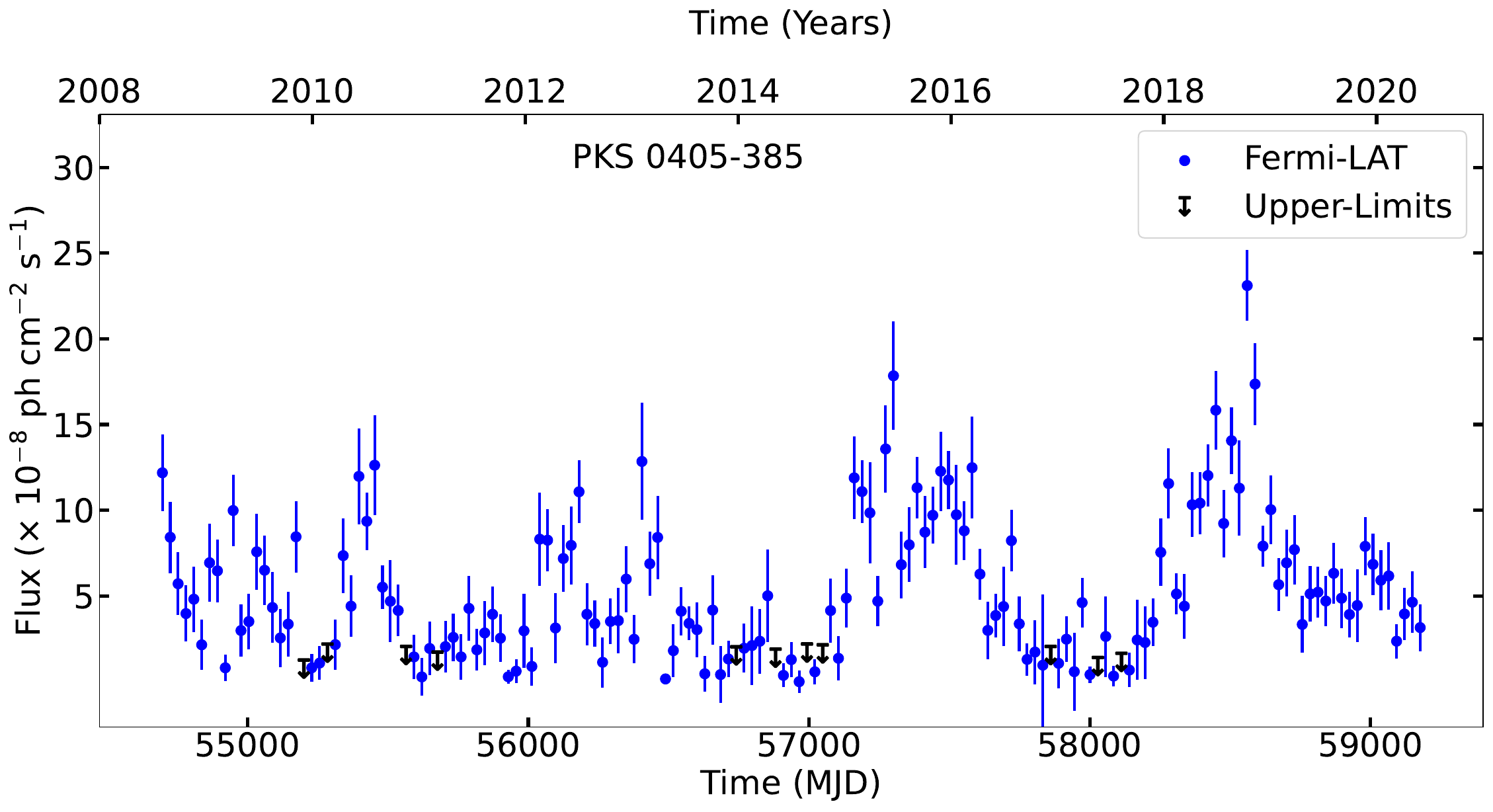}
 	\includegraphics[scale=0.2295]{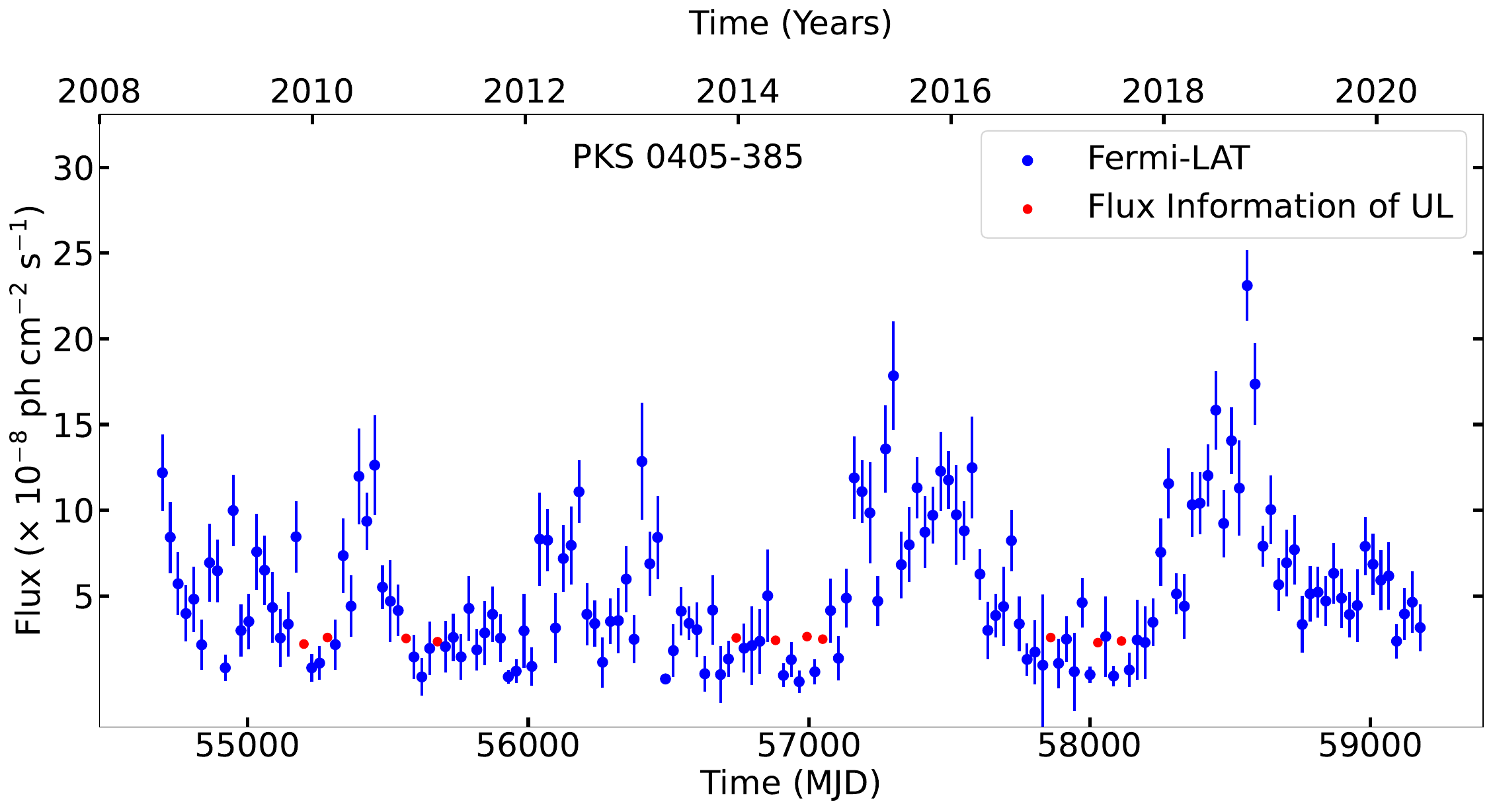}
	\caption{\textit{Left}: Light curve of the blazar PKS 0405$-$385. \textit{Right}: Light curve for the analysis with the information of the upper limits by using the likelihood profile of the flux distribution for extracting the 95\% upper limit, represented by the red points}.
	\label{fig:lc_ul}
\end{figure*}

Each LC is produced by splitting the data into 28-day bins and performing a full likelihood fit in each time bin. The best-fit ROI model obtained from the full-time interval analysis is employed to get the likelihood fit of each time bin. Initially, we try a fit the normalization of the target and of all sources in the inner $3^\circ$ of the ROI, along with the diffuse components. For a non-converging fit, the number of free parameters is progressively and iteratively restricted until the fit converges. This iterative process starts by fixing sources in the ROI that are weakly detected (i.e., with TS$<$4). After that, we fix sources with TS$<$9. Then, we fix sources up to $1^\circ$ from the ROI center and those with TS$<$25. Finally, all parameters except the target source’s normalization are fixed. We provide fluxes in each time bin when TS$>$1, yet when TS$<$1, we use the likelihood profile of the flux distribution for extracting the 95\% upper limit (see Figure \ref{fig:lc_ul}), as done by \citet{penil_2025_trends}.

To maintain consistency with P20, we adopt a strategy where the data are divided into 28-day bins. By utilizing 28-day binned $\gamma$-ray LCs, we can effectively manage the computational load while still retaining the ability to capture and analyze long-term variations in the observational data. 

This paper introduces three enhancements from the previous systematic periodicity analysis outlined in P20. Firstly, we expand our data analysis scope to include data above $\geqslant$0.1 GeV, contrasting the previous threshold of $\geqslant$1 GeV. This modification improves the signal-to-noise ratio and significantly reduces the number of upper limits in the LCs. This significant improvement dramatically increases the number of blazars that can be reliably analyzed. Specifically, the number of jetted AGN analyzed in this paper is substantially increased to 1492, in contrast to the 351 in the previous P20 sample. Finally, in P20, the removal of upper limits in LCs resulted in gaps within the data, causing an absence of observations in those intervals. This affects the analysis by removing information associated with upper limits, impacting the detection capacity of periodicity-search algorithms (see P20). Therefore, the use of the upper limits maximizes the available data. To address this, we employ the information for these LC points with low statistics by using the likelihood profile of the flux distribution for extracting the 95\% upper limit (see Figure \ref{fig:lc_ul}), as done by \citet{penil_2025_trends}. 

This strategy of using the information associated with the upper limits can introduce several potential issues. One concern is that this method might either overestimate or underestimate the true flux of the source, depending on the assumptions made about the flux distribution. Additionally, interpreting upper limits can be problematic. While upper limits indicate that a certain flux level has not been detected, they do not necessarily mean the flux is at or near that level. An LC of upper limits could be mistakenly interpreted as a period of inactivity or a decrease in brightness when, in fact, the source might still be active but below the detection threshold. As a result, the reconstructed LC may not accurately reflect the true variability of the object, particularly if many upper limits are present. This can lead to misleading interpretations of the source’s behavior and result in biased estimates of LC parameters, such as offset or amplitude. If upper limits are clustered around intervals of high variability, the reconstructed LC could falsely suggest epochs of heightened activity or even fake detections of periodic behavior.

However, removing the upper limits, thereby creating gaps in the LCs, has additional drawbacks that can impact the analysis. These gaps can affect the reliability of periodicity detection, potentially masking true periodic signals or introducing spurious ones \citep[][]{sagar_pg1553, sagar_segundo}. As demonstrated in P20, the impact of gaps becomes more significant when the gap probability in the LC is $\geq$50\%. Following a similar approach for handling upper limits, we set the threshold for the number of upper limits in the LC to the same fraction of missing data, aiming to minimize potential distortions in our analysis.

\section{Methodology} \label{sec:methodology}
\subsection{Initial Considerations } \label{sec:metho_introduction}
A central challenge in the search for periodic patterns in blazar LCs is distinguishing truly periodic signals, such as, for instance, those expected from SMBHB \citep[e.g.,][]{sagar_pg1553}, or jet precession \citep[][]{camenzind1992}, or kink events \citep[e.g.,][]{penil_kink_2025}, from behaviors arising from stochastic variability \citep[e.g.,][]{covino_negation}.

The uncertainty surrounding the detection of any period is often attributed to noise, a primary source of distortion in periodicity searches, leading to erratic fluctuations in brightness. This red noise, characterized as a stochastic process, is capable of producing false detections of periodic behavior \citep[e.g.,][]{vaughan_criticism}. In the context of LCs analysis in astronomy, such noise is linked to stochastic variations, exhibiting a more pronounced impact at higher periods. The PSD of this red noise is typically modeled by a power-law function \citep[$A*f^{-\beta}$ where $f$ is frequency,][]{rieger_2019}.

Hence, any statistical indication of a period demands evaluation with random noise as the null hypothesis. This evaluation involves reporting test statistics that can signify potential periodic behavior. Moreover, these test statistics must be assessed under an appropriate red-noise null hypothesis, employing a stochastic model calibrated to the observed LC (as detailed in $\S$\ref{sec:significance_correction}). Additionally, we explore the ``look-elsewhere'' effect's impact on our outcomes ($\S$\ref{sec:global_significance}). There are various factors contributing to uncertainty in the results of periodicity studies. For instance, previous research by our team involved conducting diverse tests to scrutinize the impact of periodicity searches. These tests include examining the influence of gaps in the LC \citep[see P20 and][]{sagar_pg1553} and flares \citep[][]{penil_flare_2025}. Such comprehensive testing aims to address potential limitations and ensure the robustness of our periodicity analysis.

The complexity of periodicity analysis has notably increased due to the availability of large datasets of LCs from telescopes such as \textit{Fermi}-LAT. Consequently, the demand for systematic methodologies to search for periodic patterns has grown. In response, we develop a pipeline capable of analyzing extensive datasets by employing various standard periodicity search algorithms (refer to $\S$\ref{sec:metho_pipeline}). This study enhances our pipeline by incorporating new periodicity search methods ($\S$ \ref{sec:new_methods}). Furthermore, we refine the estimation of test statistics associated with each identified period from these methods (as detailed in $\S$\ref{sec:boostrap} and $\S$\ref{sec:power_indices_significance}). These improvements aim to enhance the precision and reliability of our analysis in detecting potential periodic behaviors within the data.

As in our previous study, P20, our current research focuses on identifying long-term periodicity, spanning approximately years. This focus is particularly aimed at identifying jetted AGN that could potentially be candidates for SMBHB systems within the gas-driven regime. Therefore, in this study, we specifically search for periodicities within the range of 1 to 6 years.

\subsection{Systematic Periodicity Analysis} \label{sec:metho_pipeline}
We establish a systematic pipeline for periodicity analysis, as outlined in P20. This pipeline is structured across four distinct stages. In the first stage, we select a variable jetted AGN with $<$50\% of upper limits in its LC.

The subsequent stages in our pipeline conduct the periodicity search utilizing 10 established methods. These methods are reinforced with statistical techniques to derive the test statistics for the identified periods by each method. Finally, the most significant candidates demonstrating potential periodicity are selected. In this paper, we introduce modifications to the pipeline compared to the one presented in P20. 

\subsection{Replacing Bootstrap with Red-Noise LC Simulations} \label{sec:boostrap}
The pipeline used in P20 had certain limitations that we addressed in this current work. The pipeline's use of the bootstrap technique for inferring significance posed another challenge. The bootstrap \citep[and Fisher's method of randomization, which is based on bootstrap][]{linnel_pdm} generates pseudo-LCs resembling white noise. It was established in the literature \citep[e.g.][]{bhatta_s5_0716} that blazar LCs more accurately match a pink-noise LC (with $\beta\approx$1.0) rather than a white-noise distribution. Consequently, using the bootstrap method led to inconsistencies in inferring significance.

To overcome this, the pipeline in this work introduces modifications. The bootstrap technique is replaced by the artificial LCs method. We employ the approach based on Timmer\&Koenig technique \citep[denoted as TK, explained in $\S$\ref{sec:pipeline},][]{timmer_koenig_1995}. 

The TK method is a fast technique to generate synthetic LCs by first specifying a PSD that represents the desired variability characteristics of the source; in our case, this follows a power-law form. To account for variability, we generate random spectral index values in the range of [0.6–1.2] based on the results of \citet{bhatta_s5_0716}. The power spectrum provides the frequencies at which the brightness of the object varies and how strong these variations are at each frequency. Next, random phases are assigned to these frequencies, assuming a normal distribution, which ensures the variability remains unpredictable. The modified data are then transformed back into the time domain using an inverse Fourier transform, resulting in a synthetic LC that contains the statistical properties of the observed data. The resulting LCs from this technique are Gaussian distributed, while the $\gamma$-ray flux distribution of blazars tends to be log-normal distributed \citep[e.g.,][]{rieger_2019, bhatta_s5_0716}. Hence, the generated LCs are exponentiated to reflect a log-normal distribution, as suggested by studies such as \citet{shah_expontiate}. However, according to \citet{chakraborty_bending_power_law}, the exponential transformation applied to Gaussian simulations does not maintain the log-normal distribution. These artificial LCs are also adjusted according to the standard deviation and median of the original LC, maintaining the same sampling and time coverage. 

The TK method tends to overestimate the test statistics, as shown in \citet{jorge_2022}. Therefore, for the periodicity candidates identified by the pipeline, we apply a test statistics correction using the method of \citet[][, hereafter EM]{emma_lc}. The EM method generates synthetic LCs that match the PSD, probability density function, and sampling of real source LCs, although it requires more computational time than TK. This correction is detailed in $\S$\ref{sec:correction}. To maintain the independence of each test and analysis, we generate a new set of LCs for each, ensuring that the statistical results are not biased by reusing the same data.

\subsection{Considerations on the REDFIT method} \label{sec:redfit}
The REDFIT method, introduced by \citet{redfit_schulz}, is excluded from our analysis due to conceptual limitations in its treatment of stochastic variability in astronomical LCs. Specifically, REDFIT models the noise as an AR(1) process, an Ornstein-Uhlenbeck process \citep[][]{uhlenbeck_ornstein_1930}, which is a short-memory Gaussian-Markov process. This type of process emphasizes short-timescale variations and does not capture the long-memory behavior characteristic of true red noise, where power increases with timescale. As discussed by \citet{vaughan_criticism}, this makes the AR(1) null hypothesis inappropriate for modeling blazar variability. We also note that the damped random walk model, often equated with AR(1), shares these limitations and is similarly inadequate for capturing the complex temporal structure of AGN variability \citep[][]{kasliwal_dwr_2015}.

Furthermore, the test statistics estimated by REDFIT rely on a single best-fitting model, not adequately accounting for the uncertainty in model parameters. This oversight leads to an underestimation of the range of potential power-spectral values and, thus, an overestimation of the test statistics for the signal, as discussed in critiques by \cite{vaughan_bayesian, vaughan_criticism}.

Finally, our data is not well-suited for the range covered by the REDFIT method. This is due to the AR(1) model utilized in REDFIT, which normalizes the observed variance and assumes that the red noise acts as the background against which periodic signals are detected. However, in our specific case, this assumption does not hold true, as the hypothetical periodic signal could potentially dominate the variance of the LC itself.

\subsection{New Periodicity analysis methods}\label{sec:new_methods}
The pipeline has been further refined by adding a new analysis stage. This stage integrates novel analysis methods aimed at bolstering the robustness of our pipeline. Specifically, the periodicity candidates identified from the initial two stages are subject to analysis through two additional complementary methods: autocorrelation and autoregressive models. 

\subsubsection {Autocorrelation}
Physically, auto-correlation measures the likeness of a signal to itself and can effectively detect periodicity \citep[e.g.,][]{mcquillan2013}. In our pipeline, we use the \textit{z}-transformed discrete correlation function \citep[\textit{z}-DCF,~][]{zdfc_alexander}, as it enhances the conventional approach of the DCF \citep[see][ for more details]{zdfc_alexander}. This method calculates the correlation output, which is then fitted by a smooth curve\footnote{We use the function \texttt{savgol filter} of the Python package \texttt{Scipy}}, reducing low-frequency variability without altering the signal trend \citep[][]{savitzky_golay_filter}. This smooth curve aids in identifying the minima and maxima. The periods are then computed from the distance between consecutive maxima and minima in the smooth curve. The median of these periods is considered the signal's period, and its uncertainty is calculated using the method described in \citet{mcquillan2013}:
\begin{equation}\label{eq:mad}
\sigma\textsubscript{P} = \frac{1.483\times \text{MAD}}{\sqrt{N-1}},
\end{equation} 
Where $N$ is the number of peaks in the correlation, and $\text{MAD}$ is the median of the absolute deviations of the periods inferred from the different peaks. 

The test statistic is obtained by applying the method based on EM, simulating 50,000 LCs, and using 500 iterations to fit the original LC. We use the Python implementation of \citet[][]{connolly_code}.

\subsubsection {Autoregressive models}
Autoregressive models were proposed due to their efficiency in implementing periodicity analysis in astronomical LCs \citep{scargle_1981, caceres_arima}. These models, specifically, enable the modeling of LCs incorporating both stochastic and deterministic behaviors \citep[][]{feigelson_arima}. As a result, autoregressive models prove robust against stochastic noise, allowing for a more precise and accurate search for periodicity \citep[][]{caceres_arima}.

\paragraph*{AGN Behavior Modeling}
To characterize the stochastic properties of the AGN LCs, we consider autoregressive models that capture different forms of temporal dependence: Autoregressive Moving Average, the Autoregressive Integrated Moving Average, and the Autoregressive Fractionally Integrated Moving Average \citep[ARMA, ARIMA, and ARFIMA, respectively, ][]{chatfield_arima, feigelson_arima}. The simplest case is given by ARMA($p$,$q$) models, which describe short-memory processes with correlations that decay exponentially over time \citep[][]{feigelson_arima}. These models are defined by two parameters, $p$, which denotes the order (complexity) of changes in the signal level, and $q$, which represents the order of response to random shocks \citep[][]{feigelson_arima}. An extension is provided by ARIMA($p$,1,$q$) models, in which the additional unit-root component allows for integrated, random-walk–like behavior that produces long-term drifts in the LCs. The parameter $d$ in ARIMA is called the ``degree of differencing'' and signifies the stationarity of the LC, where $d$=0 indicates stationarity (indicating if the LC is characterized by a constant mean and variance). We also consider ARFIMA(0,$d$,0) models, where the fractional differencing parameter $d$ introduces long-memory behavior, leading to power-law autocorrelation decay and variability consistent with red-noise processes \citep[][]{feigelson_arima}. Finally, we test the more general ARFIMA ($p$,$d$,$q$) models, which combine the long-memory scaling with autoregressive and moving-average terms. This hybrid formulation is capable of representing both the scale-free correlations and additional short-memory components. In ARFIMA, the $d$ parameter characterizes the type of process. For instance, a value in the range of 0$<d<$1 corresponds to a long memory process, while -0.5$<d<$0 signifies a short memory process. A stationary LC typically features a $d$ parameter in the [0-0.5] range.

The AGN LCs characterization is performed by analyzing 100 AGNs of our sample, fitting their LCs with the autoregressive models described above. To compare the different fits, we use the Bayesian Information Criterion \citep[BIC, ][]{bic_schwarz}, which balances goodness of fit against model complexity. Lower BIC values indicate a more adequate description of the data, allowing us to identify whether the variability is best represented by short-memory dynamics, ARMA($p$,$q$), integrated short-memory behavior, ARIMA($p$,1,$q$), pure long-memory processes, ARFIMA(0,$d$,0), or a hybrid combination of both, ARFIMA($p$,$d$,$q$). To further assess the adequacy of the best-fit model for each LC, we apply the Ljung–Box test \citep{ljung_test}, which evaluates whether significant autocorrelation remains in the residuals. A model is regarded as statistically adequate when the test yields a p-value$\geq$0.05, indicating residuals consistent with white noise.

As a result of our analysis, we find that in 92.3\% of the blazar LCs, the ARFIMA($p$,$d$,$q$) model provides the best description of the variability. In the remaining 7.7\% of cases, the best fit is obtained with ARFIMA(0,$d$,0). All of these best-fit models pass the Ljung–Box test, confirming that they provide statistically adequate representations of the observed variability. This outcome highlights that the majority of sources require a hybrid model in which both long-memory scaling and short-memory components are present, while a smaller fraction can be explained by pure long-memory behavior alone. The absence of cases in the analyzed subsample favoring ARMA or ARIMA models indicates that blazar variability is poorly explained by processes limited to exponentially decaying correlations or purely integrated random walks. Instead, these results suggest that blazar LCs are predominantly governed by a red-noise continuum with persistent long-memory correlations, supplemented by localized structures that may arise from flaring episodes or quasi-periodic modulations.

\paragraph*{Periodicity Analysis}
We utilize two models, the ARFIMA and ARIMA. In this study, we use the ARFIMA method for application to stationary LCs. While ARFIMA can model non-stationary LCs, it is not recommended due to potential inconsistencies in its applications \citep[][]{arfima_baillie_1996, feigelson_arima}. ARIMA models are preferred for non-stationary LCs because they are specifically designed to address this characteristic through integer differencing \citep[see,~][]{caceres_arima}. To verify the stationarity, we perform the augmented Dickey-Fuller test on the LCs \citep[][]{dickey_fuller}. 

The process of searching for periodicity using the previous autoregressive models involves several steps. Initially, we choose the best-fit ARFIMA/ARIMA model using Akaike's Information Criterion \citep[AIC,~][]{akaike_criterio}, which is a penalized likelihood measure for model selection \citep[][]{Zhang_2018, Shen_2025}. Then, we derive the residuals from the original LC and the selected ARFIMA/ARIMA model. These residuals undergo analysis through the Autocorrelation Function \citep[e.g.,][]{zhang_pks0301, caceres_arima}. Any significant periodicity is identified based on a $\rm{\geqslant2\sigma}$ threshold \citep[a criterion found in related literature,~][]{zhang_periodicity_arima, yang_carma}. The quality of fit between the LC and its ARFIMA/ARIMA model is assessed by the Ljung-Box test. A well-fitted ARFIMA/ARIMA model is indicated by a p-value$\geq$0.05.

\begin{figure*}
\hspace*{2.75cm}\includegraphics[width=1.55\columnwidth]{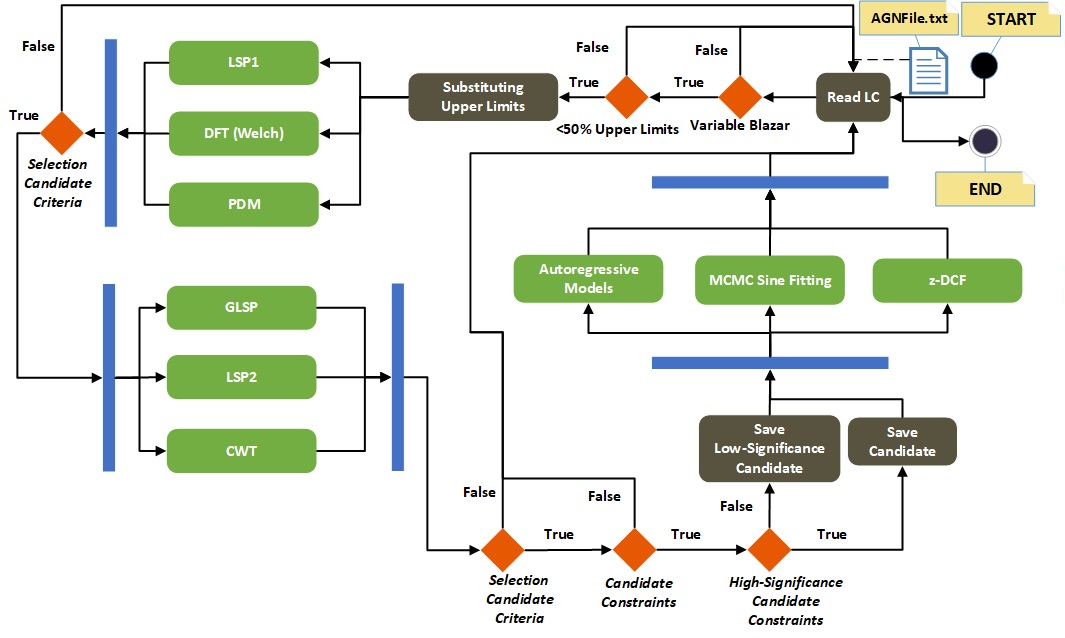}
\caption{The new Periodicity-search pipeline summarized activity diagram of Unified Modeling Language. It is organized in four stages. In this initial stage, we identify variable jetted AGN with $<$50\% of upper limits present in their LCs. The second stage involves conducting a rapid periodicity analysis of the selected sources. In the subsequent stage, selected sources undergo a more comprehensive analysis using methods that require more computational time. In the final stage, the most significant sources exhibiting evidence of periodicity are selected.}
\label{fig:study_flow}
\end{figure*}

The ARFIMA/ARIMA analysis is performed by using the R packages \texttt{stats} \citep[][]{manual_r}, and \texttt{arfima} \citep[][]{arima_hyndman_2008}. This R functionality is accessible from a Python environment via the package \texttt{rpy2} \citep[][]{JSSv035c02}.

\subsection{New Periodicity-Search pipeline} \label{sec:pipeline}
Figure \ref{fig:study_flow} illustrates the updated periodicity search pipeline. The listed methods include the technique for simulating artificial LCs (refer to $\S$\ref{sec:boostrap}) and specifying the quantity of these simulated LCs required to estimate the test statistic of the detected peaks:

\begin{enumerate}
    \item The Lomb-Scargle periodogram \citep[LSP,][]{Lomb_1976, Scargle_1982} is employed in two different scenarios. In the first scenario (LSP1), the test statistic is derived by overlaying a red noise spectrum on the LSP \citep{power_law}. In the second scenario (LSP2), 150,000 LCs are simulated using TK.    
    \item Generalized Lomb-Scargle periodogram \citep[GLSP,][]{lomb_gen}. We obtain the test statistic using the same procedure as LSP2.
    \item Phase Dispersion Minimization \citep[PDM,][]{pdm_stellingwerf}. The test statistic is estimated using the same procedure as LSP2.
    \item Enhanced Discrete Fourier Transform with Welch's method \citep[DFT-Welch,][]{welch_method}. We obtain the test statistic using the same procedure as LP2. 
    \item Continuous wavelet transform \citep[CWT,][]{wavelet_torrence}. The test statistic is obtained using the same procedure as LSP2. 
    \item Markov Chain Monte Carlo Sinusoidal Fitting \citep[MCMC Sine,][]{emcee}. To perform the parameter estimation, we use 100 walkers, 20,000 iterations, and 3000 ``burn-in'' steps to enable the stabilization of the MCMC.
\end{enumerate}

\subsubsection {Pre-analysis Selection}\label{sec:pre_analysis}
\begin{figure}
	\centering
	\includegraphics[scale=0.223]{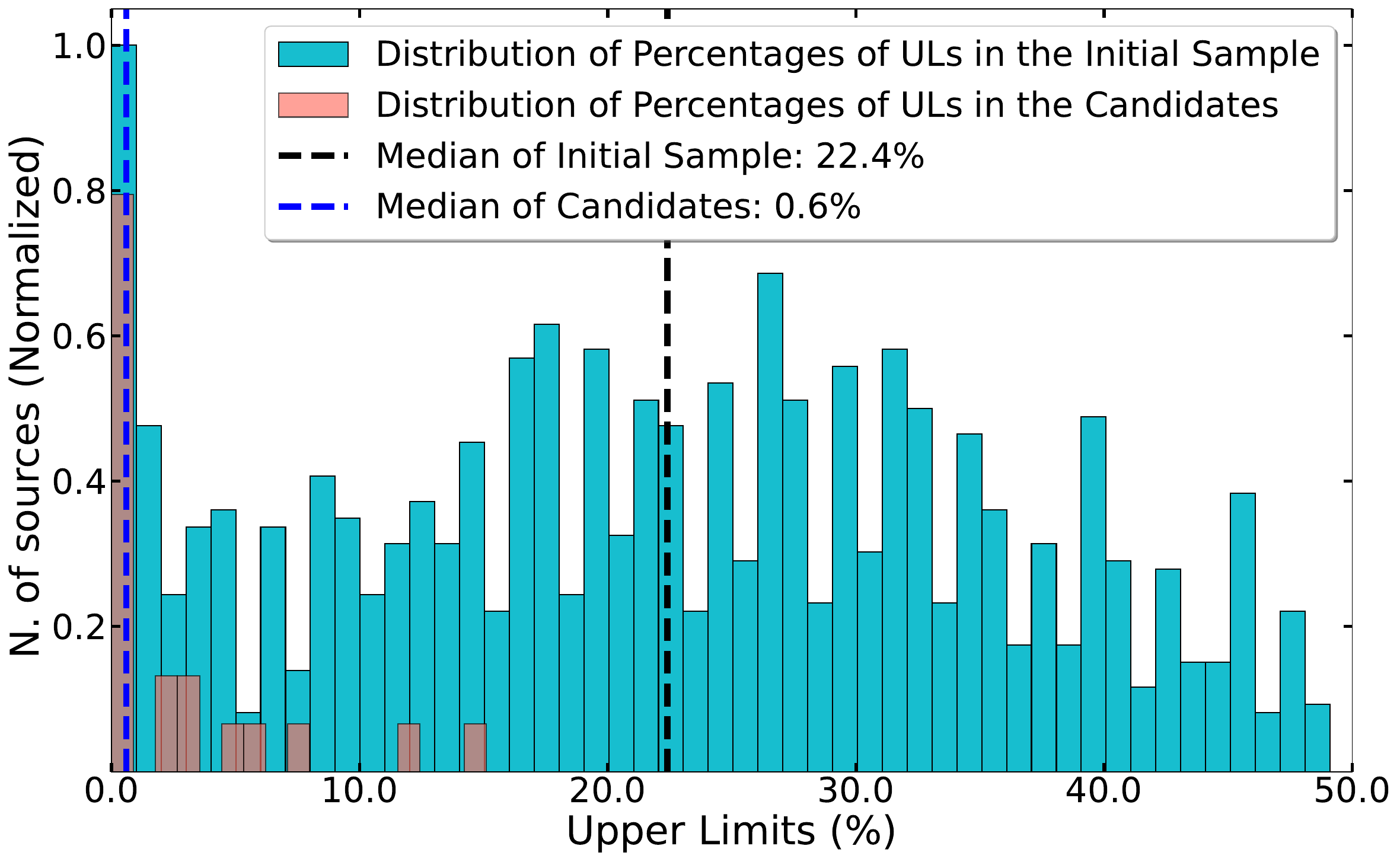}
	\caption{The distribution of upper limits across the analyzed LCs. The blue bars represent the normalized distribution of upper limits in the initial sample, with normalization based on the bin containing the highest number of sources. This distribution shows a peak concentration at 0\% (i.e., detections in all time bins) and a median of 22.4\%, indicating that most sources have moderate upper limits. In contrast, the orange bars illustrate the subset of sources exhibiting trends, also normalized based on the bin with the highest number of sources. This subset shows a lower overall percentage of upper limits, with a peak at 0\% (detections in all time bins) and a median of 0.6\%. }\label{fig:ul_distribution}
\end{figure}

From an initial set of over 3300 jetted blazars, we selected 1492 jetted AGN that are variable and have less than 50\% upper limits in their LCs for our analysis sample. Then, upper limits in the LC are replaced with the flux value that maximizes the likelihood function for that specific time bin. This process is illustrated in Figure \ref{fig:study_flow} as the ``Substituting Upper Limits'' stage. In Figure \ref{fig:ul_distribution}, we illustrate the distribution of upper limits within our sample. The bulk of sources exhibit upper limits falling between 15\% and 30\%, with a peak at 0\%. The median of this distribution, at 22.4\%, implies that most datasets contain relatively small proportions of upper limits.

\subsubsection {Coarser analysis}\label{sec:fast_analysis}
Similar to our approach in P20, our systematic periodicity analysis is structured in stages. The first stage involves methods that demand less computational time, allowing for a swift periodicity assessment: LSP1, DFT, and PDM. To identify potential periodicity candidates, we use criteria similar to those established in P20. These criteria rely on a two-step filter: (i) each analysis method's periodogram must exhibit a peak above 1$\sigma$, and (ii) at least two periodograms should display a peak above $\geq$2$\sigma$. These criteria provide flexibility in the selection process, preventing the oversight of potential periodicity candidates. You can find these criteria represented in Figure \ref{fig:study_flow} by the tag ``Selection Candidate Criteria.''

Following this initial and rapid search, the remaining sample comprises 453 jetted AGN, representing 30.6\% of the 1492 variable jetted AGN.

\subsubsection {Finer Analysis}\label{sec:fine_analysis}
The subset of 453 jetted AGN is now subjected to analysis in the subsequent stage, which involves methods demanding more computational power: GLSP, LSP2, and CWT. Utilizing the same selection criteria as previously described, we aim to identify potential periodicity candidates. This secondary assessment yields 115 jetted AGN, constituting 7.7\% of the sample. 

\subsubsection {Final Selection}\label{sec:final_selection}
The selection of jetted AGN displaying evidence of periodicity involves a new requirement: a minimum of three methods must identify a detection at $\geq$3$\sigma$ at the same period. However, if more than three methods yield significant results but at different periods, the candidate is excluded. Exceptions to this selection criterion are allowed, particularly if an algorithm fails to report a period that matches those identified by other methods. This condition is represented in Figure~\ref{fig:study_flow} by the tag ``Candidate Constraints.''

Subsequently, we further refine our selection by choosing highly significant periodicity candidates. For this, we mandate that at least four methods must indicate $\geq$4$\sigma$ significance at the same period. This criterion is depicted in Figure~\ref{fig:study_flow} under the tag ``High-Significance Candidate Constraints.'' With this criterion, the likelihood of false detection of spurious periodicity is maintained at $<$0.5\%, as discussed in P20. 

The sample of blazars obtained from the analysis comprises a total of 16 blazars (constituting 1.0\% of the 1492 blazars). Among these, 2 blazars meet the criteria for being considered as high-significance candidates (see Table \ref{tab:a_candiadtes_periods}), as determined by the previously established selection criteria.

\subsubsection {Complementary analysis}
Finally, the selected subsample undergoes further analysis using complementary methods. This final stage involves applying several techniques: autoregressive models, autocorrelation analysis, and MCMC Sine fitting. These approaches provide a more comprehensive validation of the periodicity candidates, ensuring that the conclusions drawn are supported by multiple independent lines of evidence.

\section{Power-Spectral Density} \label{tab:power_indices}
Estimating the power-spectral index serves as a means to get insights into the potential source of variability in the $\gamma$-ray flux. This is accomplished by fitting the PSD using a power-law model (as detailed in $\S$\ref{sec:boostrap}). The power-law fit characterizes the stochastic nature of $\gamma$-ray variability in jetted AGN. It also offers valuable information regarding the involvement of the accretion disk in jet emission, as noted in studies such as \citep[][]{ryan_variability, bhatta_s5_0716}.

Several issues arise when fitting power laws to the PSD of LCs. These include spectral leakage from finite sampling and windowing effects, which redistribute power between frequencies and can mimic or mask features of interest \citep[e.g., ][]{deeter_1982_leakage}. In addition, the statistical dependence of neighboring periodogram points violates the assumption of independence often implicit in classical regression approaches, leading to correlated errors that bias the inference of spectral slopes \citep[e.g.,][]{power_law}. The ambiguity introduced by smoothing choices further complicates interpretation, as different averaging schemes or binning strategies can yield divergent estimates of the underlying PSD continuum \citep[e.g.,][]{uttley_2002}. Finally, the least-squares regression tends to overweight low-frequency points with large variance, systematically biasing slope estimates and understating uncertainties in the presence of red noise \citep[e.g.,][]{vaughan_bayesian}. In blazar LCs, where red-noise–like variability dominates, apparent PSD excesses at specific frequencies can reflect stochastic fluctuations rather than genuine quasi-periodic oscillations. To mitigate such effects, significance is commonly assessed through Monte Carlo simulations of red-noise LCs \citep[][]{timmer_koenig_1995, emma_lc} or autoregressive model–based approaches \citep[e.g.,][]{tarnopolski20}.

In this study, the power spectral indices are estimated through Maximum Likelihood and Markov Chain Monte Carlo (ML-MCMC) analysis\footnote{Utilizing the Python package emcee}. Table \ref{tab:slopes} shows the estimations of these power-spectral indices, falling within the range of [0.6-1.2]. These values are consistent with those proposed in earlier studies such as \citet{bhatta_s5_0716}. The reported results align with findings detailed in \citet{yang_carma}. The mean of the power-spectral indices listed in Table \ref{tab:slopes} is $\sim$0.8, with a standard deviation of 0.1. These values are consistent with those presented in \citet{bhatta_s5_0716}.

In addition, the PSDs are also fitted by a bending power law (BPL) since this approach provides a more realistic model of blazars' variability on timescales from weeks to years \citep{chakraborty_bending_power_law}. The BPL fit is performed according to the expression of \citet{chakraborty_bending_power_law}:
\begin{equation} \label{eqn:bpl1} 
  P(\nu) = A \left( 1 + \left\{ \frac{\nu}{\nu_{b}} \right\}^{\alpha} \right)^{-1}, 
\end{equation} 

\begin{figure*}
	\includegraphics[width=\columnwidth]{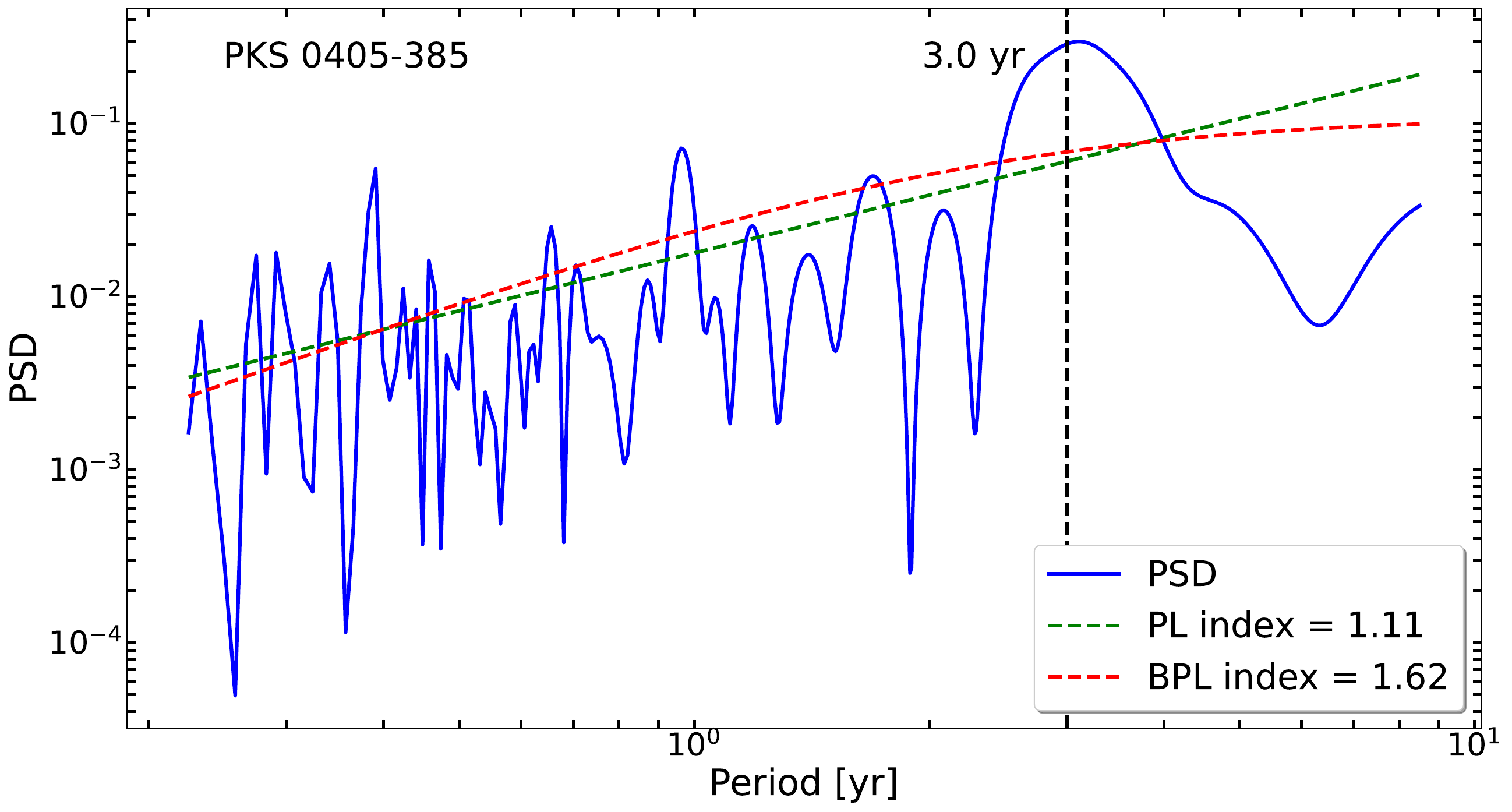}
        \includegraphics[scale=0.20]{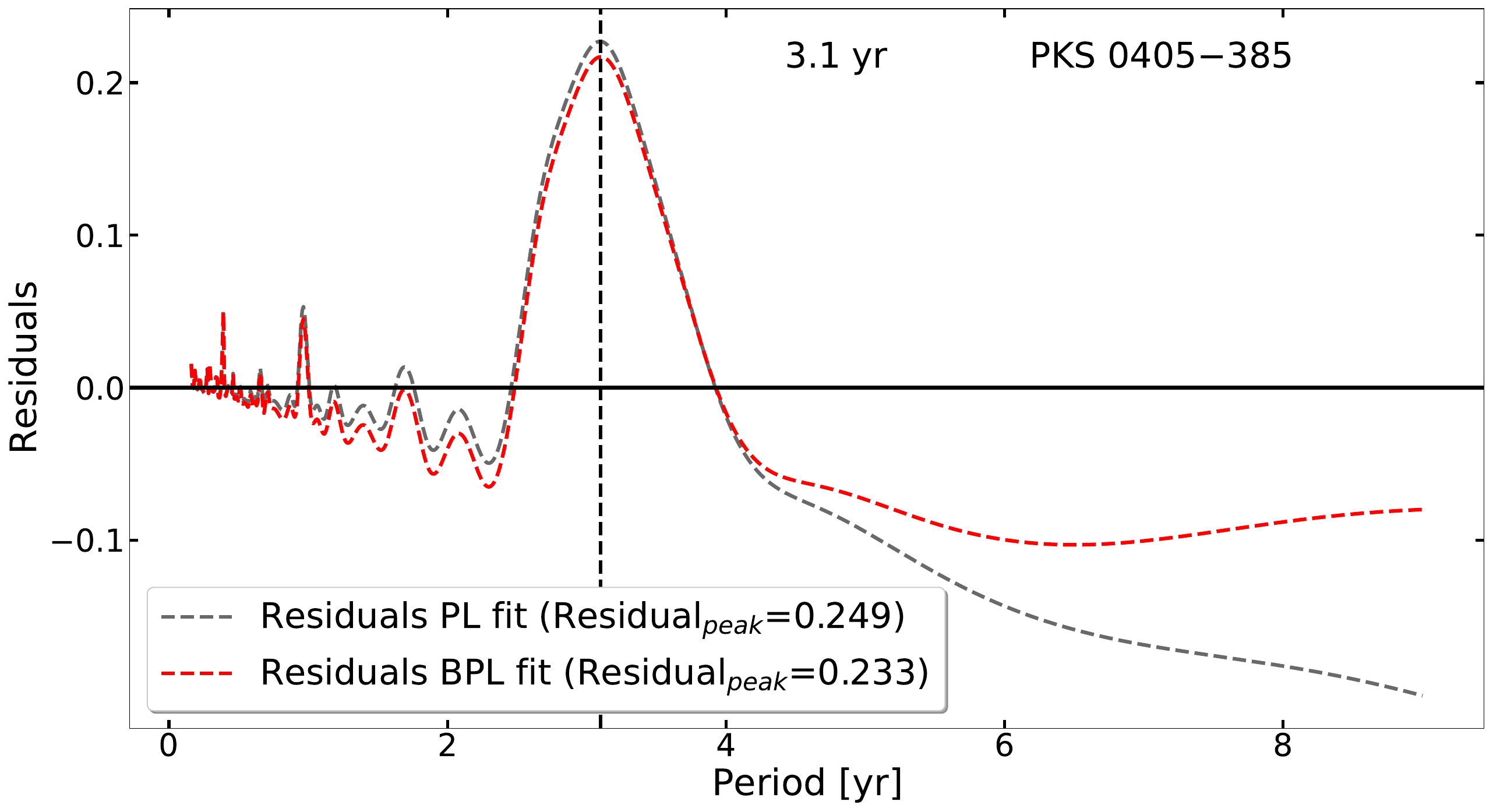}
	\caption{\textit{Left}: Examples of PSD fits for PKS 0405$-$385 using both the PL and BPL approaches, utilizing the power-spectral indices included in Table \ref{tab:slopes}. The vertical line denotes the peak associated with the period of 3.1 years. \textit{Right}: The residuals of the PSD fits for PKS 0405$-$385 are compared between the PL and BPL approaches. From the analysis, it is evident that the BPL approach yields smaller residuals compared to the PL approach. This suggests that the BPL model provides a better fit to the observed data for PKS 0405$-$385. The reduced residuals in the BPL approach indicate that this model captures the underlying features of the data more effectively.} \label{fig:reconstrucion_psd}
\end{figure*}

The BPL fit involves parameters such as $A$ for normalization, $\nu_{b}$ representing the break frequency, and $\alpha$ denoting the spectral index. These fits are obtained using the ML-MCMC analysis and are presented in Table \ref{tab:slopes}. In cases such as PKS 0215+015 and MH 2136$-$428 (refer to Figure \ref{fig:reconstrucion_psd} and Figure \ref{fig:psd_blazars}), both PSD fits show compatible slopes. This similarity suggests that the quantity of power at higher frequencies (beyond $\nu_{b}$, associated with white noise) is relatively disregarded compared to the power at lower frequencies (associated with pink-red noise).

\subsection{Evaluating the fits of the Power-Spectral Density}\label{sec:power_indices_significance}
We utilize the AIC for each PSD model to assess which one provides a more accurate fit. Subsequently, we employ the concept of relative likelihood of models (RLM) to compare the different PSD models. This comparison is conducted by considering a p-value$\rm{\leq0.05}$ and applying the following expression:
\begin{equation} \label{eqn:rml} 
\exp\left(\frac{\text{AIC}_{\text{min}} - \text{AIC}_{i}}{2}\right)
\end{equation}

Here, $\text{AIC}_{\text{min}}$ represents the minimum AIC value among the models, and $\text{AIC}_{i}$ represents the AIC value of the other model under consideration.

The BPL generates a better fit to the PSD in all cases, as shown in Table \ref{tab:correction_psds_results}. This better fit shows that the test statistic obtained by using the BPL is smaller than the one produced by the PL (with the exception of PMN J0533$-$5549), with a difference in the range of 10-20\% in 70\% of blazars and $\approx$5\% in 15\% of blazars. Therefore, the PL tends to result in a higher estimate of the test statistics. 

\subsection{Test Statistics based on the PSDs}
We derive new test statistics utilizing the previously estimated PSDs. Specifically, employing the method EM, we generate 150,000 artificial LCs to calculate test statistics for the identified periods and estimate the likelihood of false positives.

We compare the test statistic obtained with TK (see Table \ref{tab:a_candiadtes_periods}) with the results of Table \ref{tab:correction_psds_results}. They differ from the $\geq$40\% in 15\% of cases (e.g., S5 1044+71), $\geq$30\% in 50\% of cases (e.g., PKS 0405$-$385), and $\geq$20\% in 35\% of cases. These results are consistent with the fact that TK tends to produce higher test statistics $\geq3\sigma$ \citep{jorge_2022}. 

Furthermore, we evaluate the impact on estimating the test statistics according to the residual of the most significant peak. These residuals are shown in Table \ref{tab:correction_psds_results}. In 65\% of the cases, the highest test statistics are obtained when the residual is the largest, and in 20\%, the highest test statistics result from the smallest test statistics. Therefore, there is no clear relation between the residual of the most significant peak and the test statistics.  

Finally, for the PL model, 13\% of our sample has a test statistics $\geq$3.0$\sigma$, 13\% $\geq$2.5$\sigma$, 50\% $\geq$2.0$\sigma$, and the remaining 24\% has a test statistics of $<$2.0$\sigma$. In the case of the BPL model, 26\% of the results, the test statistics is 2.5$\sigma$, 34\% has a test statistics of $\approx$2.0$\sigma$, and the remaining 40\% a test statistics is $<$2.0$\sigma$.

To sort the blazars, we quantify the periodicity test statistic by computing the median test statistic across all methods of Table \ref{tab:correction_psds_results}. We select the results of applying the BPL to set the test statistics to the inferred periods, due to the BPL being the model with a better fit for the PSDs. Consequently, we discard the blazars with test statistics $<$2.0$\sigma$, which are MG2 J083121+2629, S4 0110+49, PKS 1903$-$80, PKS 1824$-$582, PKS 0736+017, and PKS 0451$-$28. Therefore, the remaining sample includes 10 blazars (0.7\% of the 1492 sources), which are listed in Table \ref{tab:candidates_list}.

\section{Corrections of the Test Statistics} \label{sec:correction}
In statistical analysis, improving the robustness and reliability of test statistics is paramount. Therefore, we refine and adjust these statistical measures to account for different aspects of our methodology.  

\subsection{Uncertainty in the Power-Spectral Density }\label{sec:significance_correction}
We test the effects in the test statistics due to the uncertainty in the parameters obtained from fitting each PSD model, according to the approach of \citet{benkhali_power_spectrum}. The artificial LCs are generated based on PSD models outlined in $\S$\ref{sec:power_indices_significance}. We consider three different models for each PSD. These models encompass both PL and BPL formulations. Each model constitutes a set of parameters for each PSD model, encompassing values within the range while considering the uncertainties for each parameter outlined in Table \ref{tab:slopes}. Specifically for the power law, we consider three different models, combining the PSD indices and normalization values: the index and the normalization minus the corresponding uncertainties, the index and the normalization, and the index and the normalization plus the corresponding uncertainties. Regarding the bending power law, we also considered three models: index, bending frequency, and normalization minus their corresponding uncertainties; index, bending frequency, and normalization; and index, bending frequency, and normalization plus their corresponding uncertainties.

We assess the test statistics while considering the impact of errors in the parameters of the PSD model. Specifically, the results in Table \ref{tab:correction_power_law_new} and Table \ref{tab:correction_bpl1_new} suggest that an accurate fit of the PSD will be essential for assessing the real test statistics of the periods \citep[][]{benkhali_power_spectrum}. We also observe the effects of changing the assumed spectrum of the PSD on the test statistics, finding (as expected) that a ``redder'' spectrum (with more power at low frequencies) results in an increased chance probability (and hence lower test statistics) of finding long-frequency periods in the PL model \citep[][]{benkhali_power_spectrum}. Regarding BPL, these variations in the test statistics are not so relevant, denoting more robustness against the uncertainty in the PSD estimation. 

\subsection{Global significance} \label{sec:global_significance}
We implement the look-elsewhere effect to adjust the ``local significance'' representing the test statistic acquired for each method within the pipeline according to the techniques described in $\S$\ref{sec:methodology}. This correction helps determine the ``global significance'', which is calculated by comparing the likelihood of observing the excess at a specific fixed value to the likelihood of observing it anywhere within the entire value range considered in the analysis \citep{Gross_Vitells_Trial}. The ``global significance'' is obtained by 
\begin{equation}\label{eq:trial}
p_{\mathrm{global}}=1-(1-p_{\mathrm{local}})^{N},
\end{equation}
The trial factor, denoted as $N$, involves two parameters in its determination. The first parameter relates to the number of sources in our sample, 1492. Since we lack prior knowledge about which sources might exhibit periodic behavior, this count becomes significant. The second parameter is associated with the uncertainty of not having prior knowledge about the period for each source. This lack of prior information contributes to the estimation of the trial factor. Consequently, the trial factor is 
\begin{equation}
\label{eq:p}
 N=P \times B,    
\end{equation}

The trial factor, symbolized as $N$, is a function of $P$, representing the independent periods (frequencies) in each periodogram, and $B$, which signifies the number of sources in our sample. The count of methods employed in the analysis is not factored into the trial factor, as all method results are equally considered in the periodicity analysis of each blazar.

In our periodograms, we incorporate 100 periods to strike a balance between computational efficiency and resolution in the periodograms. However, for a 12-year LC with roughly 12 samples per year (totaling approximately $M=144$ data points in the light curve), our period range spanning [1-6] years corresponds to 11 independent frequencies that are sampled.

\begin{figure}
	\includegraphics[width=\columnwidth]{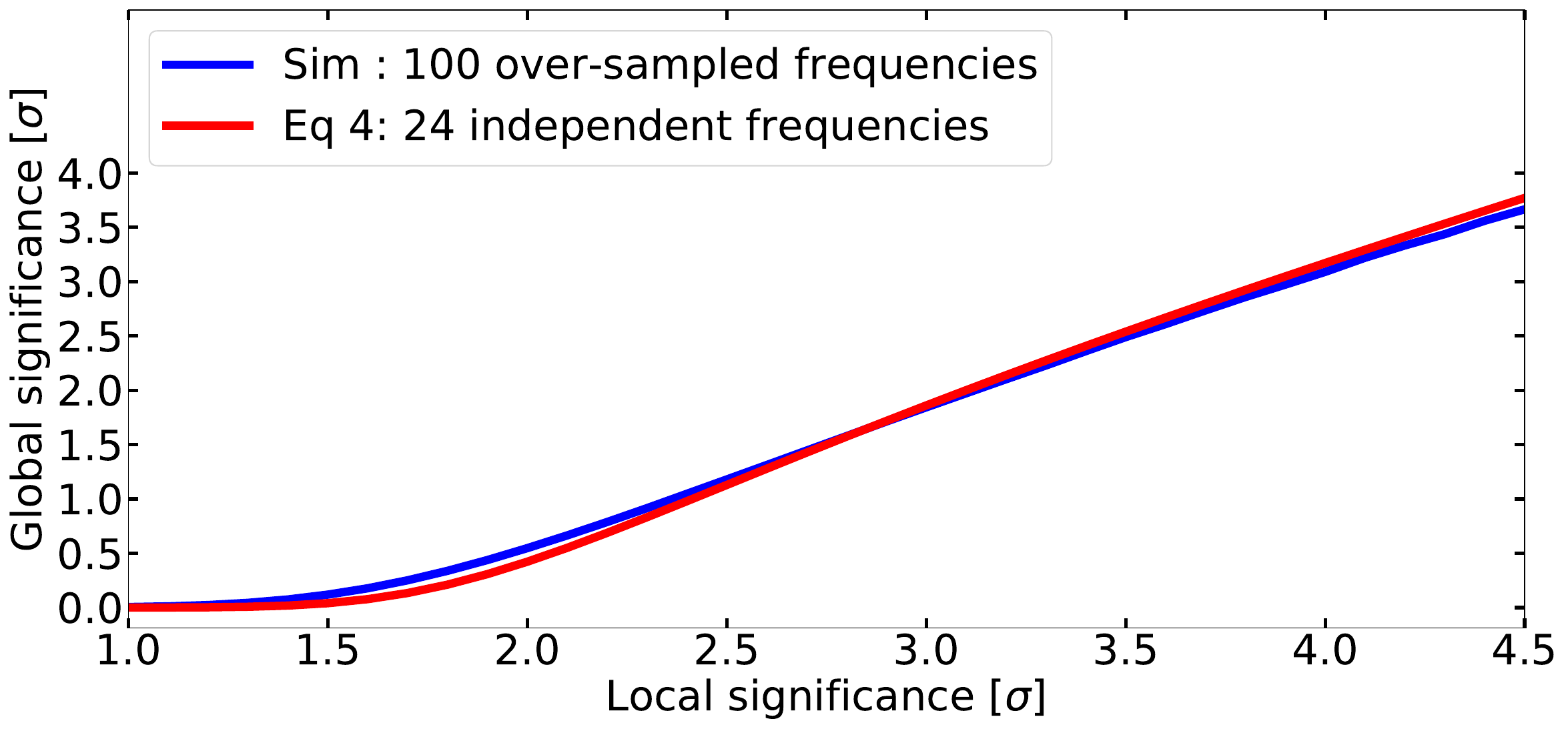}
		\caption{Trial-factor correction applied to the experimental relationship between local-global significance to estimate {\it P} in the periodicity analysis. {\it Eq. 1} represents the results obtained by applying Equation \ref{eq:trial} for a particular number of independent frequencies, reporting 24. The $P$ is chosen to adjust the test statistics of $\approx2.5\sigma$, thereby improving the correction of the most significant blazars of Table \ref{tab:candidates_list}.} \label{fig:trials}
\end{figure}

The estimation of $P$ is conducted through Monte Carlo simulations, specifically employing the algorithm described in \citep[][]{penil_2022}. By utilizing $10^{8}$ simulated LCs using the TK technique, we derive the empirical relationship between local-global significance (as depicted by the blue line in Figure \ref{fig:trials}). To determine the most suitable $P$, we explore various values by applying Equation \ref{eq:trial} to best align with the experimental relationship of local-global significance. The selection of $P$ aims to correct the ``local significance'' to be $\approx2.5\sigma$ (see $\S$\ref{sec:power_indices_significance}), enhancing the identification of highly significant jetted AGN of our sample. In this exploration of $P$, we consider lower and upper limits of 11 and 100, respectively.

As depicted in Figure \ref{fig:trials}, selecting $P=24$ provides the best alignment with the empirical relationship of local-global significance. Consequently, this choice leads to a trials factor of 35,808 ($1492$$\times$$24$). Subsequently, according to this correction, the ``global significance'' of the periods presented in Table \ref{tab:candidates_list} is 0\,$\sigma$.

\section{Results} \label{sec:results}
By employing the systematic periodicity-search pipeline and the statistical corrections, we do not find any significant detection.  

As a secondary result, we identify 10 blazars that exhibit a hint of periodic $\gamma$-ray emissions with $\geq$2$\sigma$, which are presented in Table \ref{tab:candidates_list}\footnote{Information obtained from the 4FGL-DR2 catalog \url{https://fermi.gsfc.nasa.gov/ssc/data/access/lat/10yr_catalog/}}.
The 10 blazars are categorized as follows: seven are flat-spectrum radio quasars (FSRQ), two are BL Lacertae (BL Lac), and one is classified as a blazar candidate of uncertain type (BCU) as per source identifications from 4FGL \citep{abdollahi_4fgl}. The jetted AGN with the highest test statistics, 2.5$\sigma$, are PKS 0405$-$385, S5 1044+71, and MH 2136$-$428. The LCs of the blazars are presented in Figure \ref{fig:lc_candidates_low}.

\begin{figure}
	\includegraphics[width=\columnwidth]{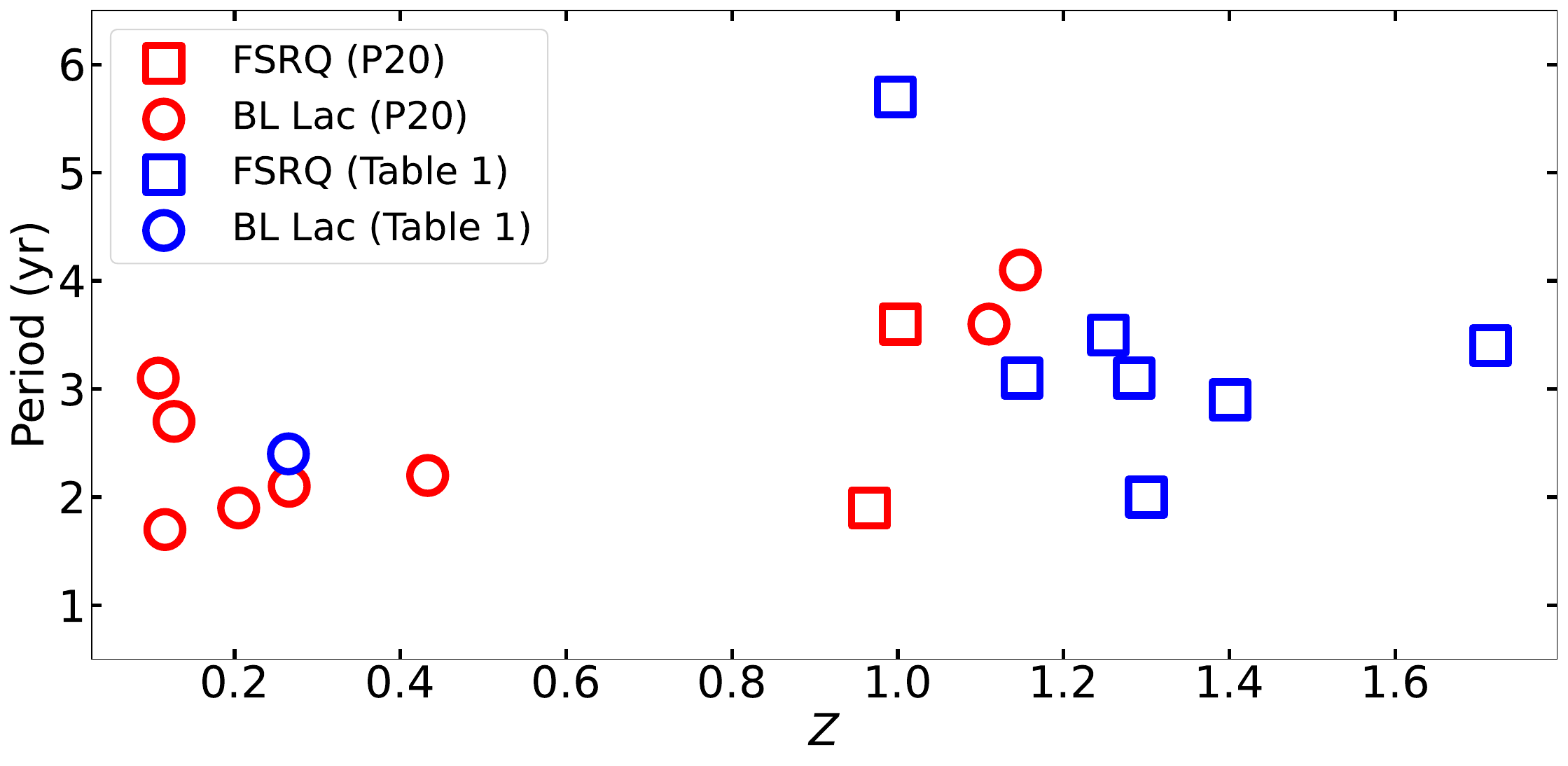}
	\caption{Period versus redshift for the blazars listed in Table \ref{tab:candidates_list}. Upon examination, no apparent correlation between period and redshift is observed. This lack of correlation suggests that the periodic behavior of jetted AGN, as detected in this study, does not exhibit a systematic dependence on redshift.}
	\label{fig:period_redshift}
\end{figure}

Figure \ref{fig:period_redshift} shows the relationship between the periods of the blazars inferred in this analysis and those presented in P20 against their redshifts. The majority of the candidates are observed at moderate to high redshifts ({\it z}$\geq$1), a trend that might support the theory linking their periodic behavior to SMBHB systems, as highlighted in studies such as \citet{rieger_2007}. Nevertheless, it remains challenging to draw robust conclusions from our reduced sample size. As depicted in Figure \ref{fig:period_redshift}, none of the sources are situated within the \textit{z}=0.4 to \textit{z}=1 range. This absence could be attributed to an observational bias, presumably of instrumental origin.

Additionally, no clear correlation between period and redshift is discernible in Figure \ref{fig:period_redshift}. This figure reveals that approximately 80\% of the FSRQs have {\it z}$\geq$1, while roughly 75\% of the BL Lacs have {\it z}$\leq$0.5. This distribution aligns with expectations, as BL Lacs objects tend to be discovered at lower redshifts, whereas FSRQs are more frequently observed at higher redshifts.

\begin{table*}
\centering
\caption{List of the ten $\sim$2$\sigma$ sources presented in $\S$\ref{sec:results}, including their Fermi-LAT name, coordinates, AGN type, redshift, and association name. We obtain this information from the 4FGL-DR2 catalog. We include the percentage of upper limits (UL) in the LC. Additionally, we include the period (expressed in years) determined through our periodicity-search pipeline. The candidates are sorted according to the median of their test statistics for the bending-power law described by Equation \ref{eqn:bpl1}, as detailed in the results presented in $\S$\ref{sec:significance_correction}. Note that this median significance does not have an actual statistical meaning; it is used as an arbitrary way that combines all test statistics to sort the candidates. We also included the global significance resulting from the method described in $\S$\ref{sec:global_significance}. The sources are categorized into two groups: those presented in this work and the ones included in P20 with test statistics $\geq$2$\sigma$. It is worth noting that some sources exhibit two significant periods, organized by their peak amplitude and denoted by an $\star$.  \label{tab:candidates_list}}
{%
\begin{tabular}{ccccccccccc}
\hline
\hline
4FGL Source Name & RAJ2000 & DecJ2000 & Type & Redshift & Association Name & UL (\%) & Period (yr) &
Local (S/N) & Global (S/N) \\
J0407.0$-$3826 & 61.7627 & $-$38.4394 & fsrq & 1.285 & PKS 0405$-$385 & 7.5\% & 3.1$\pm$0.5 & 2.5$\sigma$ & $\approx$0$\sigma$ \\
J1048.4+7143 & 162.1067 & 71.7297 & fsrq & 1.15 & S5 1044+71 & 0.6\% & 3.1$\pm$0.5 & 2.5$\sigma$ & $\approx$0$\sigma$ \\
J2139.4$-$4235 & 324.8546 & -42.5895 & bll & -- & MH 2136$-$428 & 0.6\% & 1.8$\pm$0.1 & 2.5$\sigma$ & $\approx$0$\sigma$ \\
J0526.2$-$4830$\star$ & 81.5714 & -48.5151 & fsrq & 1.3 & PKS 0524$-$485 & 3.1\% & \makecell{2.0$\pm$0.2 \\ 4.0$\pm$0.5} & \makecell{2.4$\sigma$ \\ 2.2$\sigma$} & $\approx$0$\sigma$ \\
J0217.8+0144 & 34.4621 & 1.7346 & fsrq & 1.715 & PKS 0215+015 & 1.8\% & 3.4$\pm$0.4 & 2.2$\sigma$ & $\approx$0$\sigma$ \\
J0112.1+2245 & 18.0294 & 22.7515 & bll & 0.265 & S2 0109+22 & 0.6\% & 2.4$\pm$0.3 & 2.1$\sigma$ & $\approx$0$\sigma$ \\        
J0533.3+4823$\star$ & 83.3313 & -55.8247 & bcu & -- & PMN J0533$-$5549 & 2.5\% & \makecell{2.8$\pm$0.3 \\ 2.0$\pm$0.3} & \makecell{2.0$\sigma$ \\ 2.2$\sigma$} & $\approx$0$\sigma$ \\
J1310.5+3221 & 197.6324 & 32.3547 & fsrq & 0.997 & OP 313 & 6.2\% & 5.7$\pm$0.8 & 2.0$\sigma$  & $\approx$0$\sigma$ \\
J1033.9+6050 & 158.4849 & 60.8493 & fsrq & 1.401 & S4 1030+61 & 3.1\% & 2.9$\pm$0.4 & 2.0$\sigma$ & $\approx$0$\sigma$ \\
J1044.6+8053 & 161.1638 & 80.8941 & fsrq & 1.254 & S5 1039+81 & 15.1\% & 3.5$\pm$0.4 & 2.0$\sigma$ & $\approx$0$\sigma$ \\ 
        \hline
        \hline
        J1555.7+1111 & 238.93169 & 11.18768 & bll & 0.433 & PG 1553+113 & 0\% & 2.2$\pm$0.2 & 4.5$\sigma$ & $\approx$1.8$\sigma$\\
        J2158.8$-$3013 & 329.71409 & -30.22556 & bll & 0.116 & PKS 2155$-$304 & 0\% & 1.7$\pm$0.1 & 3.3$\sigma$ & $\approx$0$\sigma$\\
	J0811.3+0146 & 122.86418 & 1.77344 & bll & 1.148 & OJ 014 & 0.6\% & 4.1$\pm$0.5 & 2.9$\sigma$ & $\approx$0$\sigma$\\
        J0457.0$-$2324 & 74.26096 & -23.41384 & fsrq & 1.003 & PKS 0454$-$234 & 0\% & 3.6$\pm$0.4 & 2.8$\sigma$ & $\approx$0$\sigma$\\      
        J0721.9+7120$\star$ & 110.48882 & 71.34127 & bll & 0.127 & S5 0716+714 & 0\% & \makecell{2.7$\pm$0.4 \\ 0.9$\pm$0.1 } & \makecell{2.8$\sigma$ \\ 1.9$\sigma$} & $\approx$0$\sigma$\\
        J0043.8+3425 & 10.96782 & 34.42687 & fsrq & 0.966 & GB6 J0043+3426 & 4.9\% & 1.9$\pm$0.2 & 2.7$\sigma$ & $\approx$0$\sigma$\\
        J0521.7+2113 & 80.44379 & 21.21369 & bll & 0.108 & TXS 0518+211 & 0\% & 3.1$\pm$0.4 & 2.6$\sigma$ & $\approx$0$\sigma$ \\
        J1649.4+5238 & 252.35208 & 52.58336 & bll & -- & 87GB 164812.2+524023 & 12\% & 2.8$\pm0.6$ & 2.2$\sigma$ & $\approx$0$\sigma$\\        
        J0449.4$-$4350 & 72.36042 & -43.83719 & bll & 0.205 & PKS 0447$-$439 & 0\% & 1.9$\pm$0.2 & 2.1$\sigma$ & $\approx$0$\sigma$ \\
        J0428.6$-$3756 & 67.17261 & -37.94081 & bll & 1.11 & PKS 0426$-$380 & 0\% & 3.6$\pm$0.5 & 2.1$\sigma$ & $\approx$0$\sigma$ \\ 
        J0303.4$-$2407 & 45.86259 & -24.12074 & bll & 0.266 & PKS 0301$-$243 & 0\% & 2.1$\pm$0.2 & 2.0$\sigma$ & $\approx$0$\sigma$ \\
\hline
\hline
\end{tabular}%
}
\end{table*}

We note, however, that reporting periodicities in blazar LCs is inherently challenging due to the pronounced variability and stochastic nature of these sources, which can easily mimic or mask quasi-periodic signatures \citep[e.g.,][]{power_law, vaughan_criticism, covino_negation}. A cautious interpretation is therefore essential. This stance is further motivated by recent studies demonstrating that even genuine periodic signals can be significantly altered by transient flaring activity. In particular, \citet{penil_flare_2025} showed that the presence of intense flares may reduce, distort, or even obscure an underlying periodic component, complicating its reliable detection. Such distortions can result in an underestimation of the true significance of candidate signals, especially when their apparent strength lies close to the 2$\sigma$ threshold. For these reasons, we regard the candidate periods reported here as tentative and stress that they should be interpreted as preliminary indicators rather than firm detections. As an initial step toward confirming or discarding possible periodic behavior in these blazars, we extend our analysis by providing forward predictions for their future $\gamma$-ray flux variability (see $\S$\ref{sec:predictions}). These forecasts offer an independent and testable means of assessing whether the observed modulations persist in subsequent data, thereby enhancing the robustness of our conclusions. Overall, this complementary strategy allows us to balance the reporting of intriguing candidate signals with the necessary caution dictated by the data. By explicitly accounting for the limitations imposed by variability, noise properties, and flaring events in blazar LCs, we aim to provide a framework that is both rigorous and transparent, ensuring that the tentative periodicities presented here can be critically evaluated and tested with future observations.

\subsection{Complementary Analysis}
The results from the complementary methods presented in $\S$\ref{sec:pipeline} are summarized in Table \ref{tab:mcmc_bayesian_arfima_results}. 

Three different scenarios resulting from the application of ARFIMA/ARIMA methods:
\begin{enumerate}
	\item A significant period is obtained as those listed in Table \ref{tab:a_candiadtes_periods} (e.g., S2 0109+22 and PKS 1824$-$582).	
	\item The modeling provides a compatible period but not significant ($<$2$\sigma$). For instance, OP 313 and PKS 0405$-$385 (in this case, in a secondary peak).
	\item No compatible period is reported as, for instance,  S5 1044+71 or PKS 0736+017. 
\end{enumerate}

The periods reported by  Z-DFC, most of the periods inferred are compatible with the results of Table \ref{tab:a_candiadtes_periods}, except two of them as S2 0109+22 or PKS 0215+015. The MCMC Sine Fitting method yields a period consistent with the results of Table \ref{tab:a_candiadtes_periods}. 

\subsection{Periodic-Emitter Blazars in the literature} \label{sec:sub_candidates_a}
PKS 0405$-$385 and S5 1044+71 have been previously identified in the literature as showcasing periodic behavior. For PKS 0405$-$385, \citet{gong_pks_0405_385} reported a period of roughly 2.8 yr, aligning with the period noted in Table \ref{tab:candidates_list} but with a higher test statistic of 4.3$\sigma$ (compared to our 2.5$\sigma$). \citet{alba_ssa} identified a period of 3.0 yr with a test statistics of 4.8$\sigma$\footnote{This represents the maximum test statistics achievable based on the number of artificial LCs used for estimation.}. This study utilizes Singular Spectrum Analysis (SSA) as an innovative approach for periodicity analysis \citep[][]{ssa_greco, SSA_algorithm}. SSA separates a given LC into distinct components, including oscillations, trends, and noise, thereby facilitating an accurate analysis of periodicities by minimizing distortions. This decomposition could account for variations in inferred test statistics, as it effectively diminishes factors that typically affect period detection \citep[e.g.,][]{penil_2025_trends, penil_flare_2025}.

Regarding S5 1044+71, \citet{wang_s5_1044+71} and \citet{ren_s5_1044+71} inferred periods of 3.0 yr (at 3.6$\sigma$) and about 3.1 yr (at $\sim$2.0$\sigma$, see their Figure 3), respectively. In our case, we inferred a similar period with a test statistic of 2.5$\sigma$. The differences in test statistics could be due to differing methodologies used to estimate them. Specifically, in \citet{ren_s5_1044+71}, the PSD is fitted using a smooth bending power-law with added white noise, as detailed in \citet{vaughan_bayesian}. However, this paper does not include the parameters and plots for each PSD fit, hindering a direct comparison with our estimations. Regarding \citet{gong_pks_0405_385}, a power-law is utilized to fit the PSD, and the TK method is employed to generate artificial LCs \citep[in][, EM is used]{ren_s5_1044+71}. As detailed in $\S$ $\ref{sec:power_indices_significance}$, the TK method tends to result in higher test statistics. This discrepancy in methodologies could contribute to the differing test statistic values between these studies. \citet{alba_ssa} reported a period of 3.0 yr and a test statistics of 4.8$\sigma$. 

Recent publications have reported blazars exhibiting periodic behavior, yet these are not included in Table \ref{tab:candidates_list}. For instance, \citet{zhang_pks0521_36} claimed periodic $\gamma$-ray emissions from PKS 0521$-$36 with a period of 1.1 years ($\approx$5$\sigma$). In our analysis, we derived a period of 2.2 years ($\approx$1.1$\sigma$). Our analysis covers 12 years of {\it Fermi}-LAT observations, whereas \citet{zhang_pks0521_36} focus on a shorter, 7-year segment of the LC (2012–2019), excluding the full observational period. The absence of oscillatory behavior in the remaining 5 years leads to a reduced periodicity detection in our results, as the full data set dilutes the periodic signal observed in the shorter segment. Additionally, the test statistics in \citet{zhang_pks0521_36} was determined using a bootstrap method, which, as discussed in $\S$\ref{sec:boostrap}, is not ideally suited for the properties of blazar LC. Additionally, \citet{zhang_pks0521_36} also apply the EM method to assess test statistics, modeling the PSD with a power-law fit. As shown in $\S$\ref{sec:power_indices_significance}, this approach can impact the resulting test statistics estimates, highlighting the methodological differences that contribute to the variation in test statistics between the two studies. 

In the study by \citet{ren_s5_1044+71}, they examined a sample of the brightest blazars using CWT, utilizing a similar 12-year span of {\it Fermi}-LAT observations as in our analysis. Seven blazars from their sample overlap with the 115 blazars subject to detailed analysis in our work (refer to $\S$\ref{sec:pipeline}). The observed discrepancies in test statistics are likely due to variations in the PSD fitting used in each case, which can influence the estimated test statistics.

Similarly, \citet{alba_ssa} reported 46 objects with test statistics of $\geq$2.0$\sigma$. Significant discrepancies persist between the test statistics obtained in this study and those reported by \citet{alba_ssa}, as observed in cases such as PKS 0405$-$385 and S5 1044+71. For instance, in the case of MH 2136$-$428, the same period is found with a test statistic of 4.8$\sigma$. These differences can be attributed to the use of SSA, which enables periodicity analysis while minimizing the influence of potential distortion factors.    

\begin{figure}[ht]
	\centering
        \includegraphics[scale=0.21]{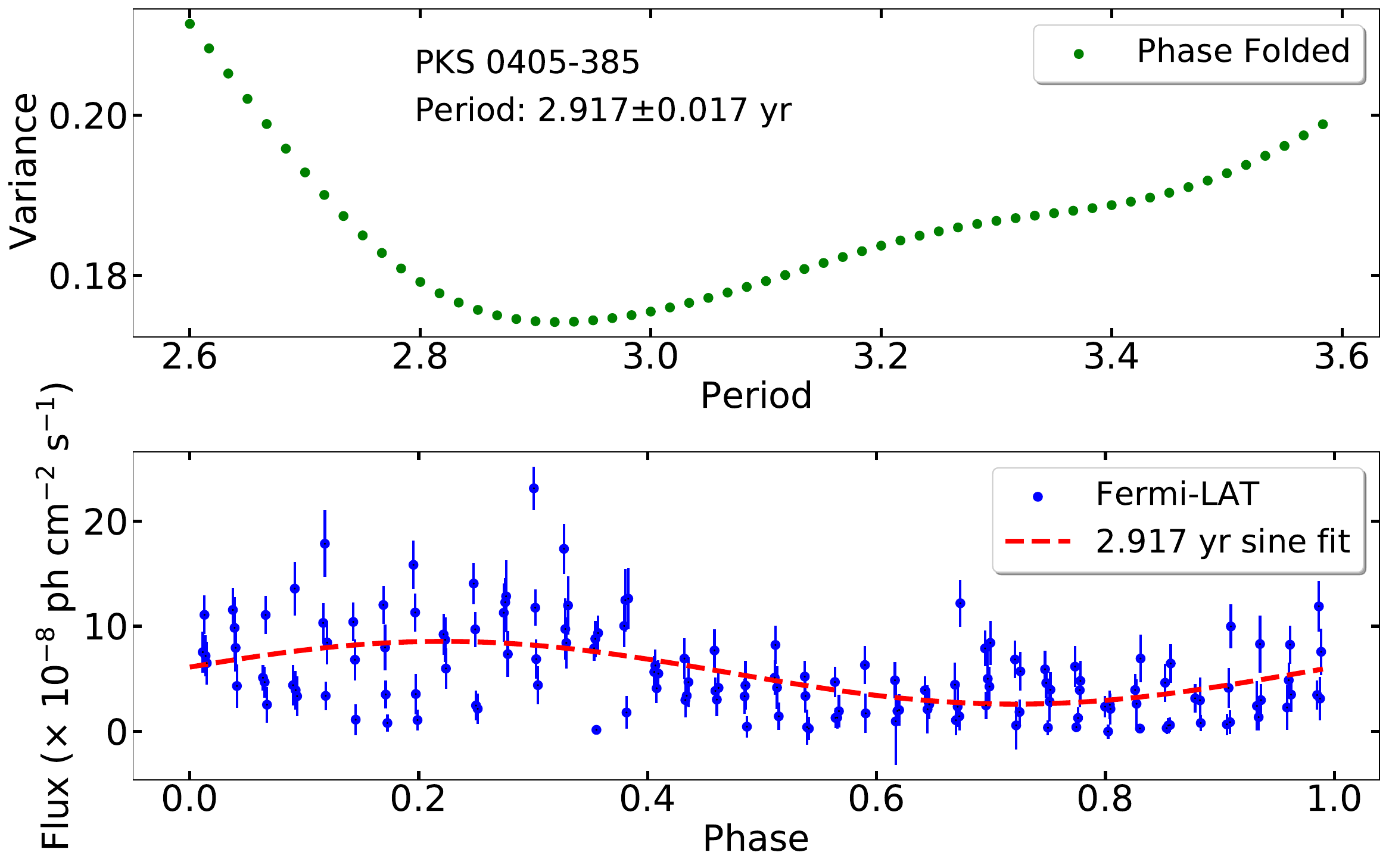}
	\caption{Period estimation for predicting the future behavior of PKS 0405$-$385. \textit{Top}: The period was estimated by analyzing variance through a sine fit with phase folding. The minimum point on the graph indicates the best-fit period at 2.917 yr. \textit{Bottom}: The LC of PKS 0405$-$385 after phase folding with the oscillating signal using the estimated period}\label{fig:qpo}
\end{figure}

\subsection{Quasi-periodic and periodic oscillations}\label{sec:predictions}
In the literature, the periodic behavior detected in blazar emissions can be categorized as either quasi-periodic oscillations (QPOs) or genuine periodicity. These categories exhibit distinct mathematical properties that are crucial to consider. A QPO refers to any behavior displaying irregular periodicity with a stochastic (unpredictable) element, rendering estimation of future signal oscillations challenging. In contrast, genuine periodicity enables the estimation of future oscillations in a signal.

Consequently, we aim to predict the upcoming low and high flux states of the blazars listed in Table \ref{tab:candidates_list}. To forecast the future oscillations of these blazars, we apply a sine fit to the LCs, considering variables such as amplitude, offset, phase, and period.

To refine the precision of the inferred periods presented in the LCs and optimize our predictions, we implement a method using phase folding analysis. Initially, we define a range of periods based on the period determined by the pipeline and its associated uncertainty. Subsequently, we apply phase folding to each value within the period range, assessing the variance between the data and the resulting phase-folded curve (refer to Figure \ref{fig:qpo}). The selected period is identified as the one yielding the minimum variance (as shown in Figure \ref{fig:qpo}). This methodology presents the predicted emissions for the selected blazars in Table \ref{tab:qpo}. As an example, the next high-flux emission state of PG 1553+113, the most promising candidate to have a periodic emission \citep[e.g.][]{ackermann_pg1553, penil_mwl_pg1553}, will take place in May-June of 2025. 

This approach offers a framework for testing the presence of periodic behaviors in blazar emissions, providing a more concrete basis to confirm or refute the existence of such patterns. Additionally, the ability to optimize predictions has practical implications for future observational strategies, optimizing the use of telescopic time and resources to focus on periods of heightened interest.

\section{Summary \& Conclusions} \label{sec:summary}
In this research, we conducted a systematic investigation into the periodic behavior of jetted AGN listed in the 4FGL catalog, which encompasses a total of 1492 sources, selected based on their variability according to the variability index. This represents the most comprehensive analysis of $\gamma$-ray jetted AGN undertaken to date in the time domain.

Our systematic analysis was conducted through an enhanced pipeline, ensuring more robust results by integrating new methods of analysis and addressing various limitations in the estimation of test statistics. Using this thorough approach, we do not find any evidence for periodic signals in the 1492 jetted AGN $\gamma$-ray LCs analyzed here. However, we selected ten blazars with hints of periodic oscillations with $\geq$2$\sigma$ ($\approx$0$\sigma$ post-trials), which could be candidates to be monitored in the next years. We also investigated autoregressive models as a framework for fitting the AGN LCs. Our analysis shows that the analyzed LCs are best described by ARFIMA ($p$,$d$,$q$) models, which naturally capture both short-memory autocorrelations and long-memory, 1/$f$-type red-noise variability.

Additionally, we complemented the periodicity investigation with supplementary analyses to ascertain the hints of these periodic signals. For example, we predicted future high-low flux states to evaluate the accuracy of such forecasts, which could aid in planning observations, such as those conducted by Cherenkov telescopes. The success of these predictions could contribute to more precise and effective observations in the future. 

Multi-wavelength monitoring is essential to validate the nature of the detected QPOs. Unfortunately, continuous multi-wavelength data are currently scarce for most AGN, mainly due to technical and observational limitations. Furthermore, theoretical studies focused on how these $\gamma$-ray QPOs relate to binary black hole dynamics could provide key estimates of physical parameters such as mass ratios, orbital separations, and possible jet precession periods. Such models could also help differentiate between various mechanisms, such as jet precession or disk instabilities, that may contribute to the observed periodic behavior. For these reasons, we encourage (1) future efforts in multi-wavelength monitoring, as they could reveal whether the periodic patterns observed in this paper are a broadband phenomenon or confined to specific energy ranges due to different emission mechanisms within the jet or the accretion disk, and (2) further theoretical modeling in this area to enhance our interpretation of the observational data. Including such models in future studies would bridge the gap between observations and the physical processes occurring near the central engines of AGN.

\section{Acknowledgments}
The \textit{Fermi}-LAT Collaboration acknowledges generous ongoing support from a number of agencies and institutes that have supported both the development and the operation of the LAT as well as scientific data analysis. These include the National Aeronautics and Space Administration and the
Department of Energy in the United States, the Commissariat \`a l'Energie Atomique and the Centre National de la Recherche Scientifique / Institut National de Physique Nucl\'eaire et de Physique des Particules in France, the Agenzia Spaziale Italiana and the Istituto Nazionale di Fisica Nucleare in Italy, the Ministry of Education, Culture, Sports, Science and Technology (MEXT), High Energy Accelerator Research Organization (KEK) and Japan Aerospace Exploration Agency (JAXA) in Japan, and the K.~A.~Wallenberg Foundation, the Swedish Research Council and the Swedish National Space Board in Sweden.

Additional support for science analysis during the operations phase is gratefully acknowledged from the Istituto Nazionale di Astrofisica in Italy and the Centre National d'\'Etudes Spatiales in France. This work was performed in part under DOE Contract DE-AC02-76SF00515.

P.P. and M.A. acknowledge funding under NASA contract 80NSSC20K1562. S.B. acknowledges financial support by the European Research Council for the ERC Starting grant MessMapp, under contract no. 949555, and by the German Science Foundation DFG, research grant “Relativistic Jets in Active Galaxies” (FOR 5195, grant No. 443220636). A.D. is thankful for the support of the Ram{\'o}n y Cajal program from the Spanish MINECO, Proyecto PID2021-126536OA-I00 funded by MCIN / AEI / 10.13039/501100011033, and Proyecto PR44/21‐29915 funded by the Santander Bank and Universidad Complutense de Madrid.

This work was supported by the European Research Council, ERC Starting grant \emph{MessMapp}, S.B. Principal Investigator, under contract no. 949555, and by the German Science Foundation DFG, research grant “Relativistic Jets in Active Galaxies” (FOR 5195, grant No. 443220636).

\section*{Data Availability}

All the data used in this work are publicly available or available on request to the responsible for the corresponding observatory/facility.

\section{Software}
\begin{enumerate}
        \item arfima R-software \citep{arima_hyndman_2008, arima_hyndman_2024},
	\item astroML \citep{astroml},
	\item astropy \citep{astropy_2013, astropy_2018, astropy_2022}, 
	\item colorednoise \url{https://github.com/felixpatzelt/colorednoise},
	\item emcee \citep {emcee}, 
	\item fermipy software package \citep{Wood:2017yyb},
	\item PyAstronomy \citep{PyAstronomy},	  
        \item PyCWT\footnote{We have modified the sources of the code to estimate the test statistics with  \citet{connolly_code}}  \url{https://pypi.org/project/pycwt/},
		\item rpy2 \url{https://rpy2.github.io/doc/latest/html/index.html},
	\item stats R-software \citep{r_manual},  
	\item statsmodels \citep[][]{stats_scipy_2010},, 
	\item SciPy \citep {SciPy},
	\item Simulating light curves \citep{connolly_code},
        \item Z-DFC \citep{zdfc_alexander}
\end{enumerate}

\bibliographystyle{mnras}
\bibliography{literature} 

\clearpage

\appendix
\setcounter{table}{0}
\setcounter{figure}{0}
\renewcommand{\thetable}{A\arabic{table}}
\renewcommand{\thefigure}{A\arabic{figure}}
\section*{Appendix}\label{sec:appendix}

\subsection{Light Curves}
\begin{figure*}
	\centering
        \includegraphics[scale=0.2195]{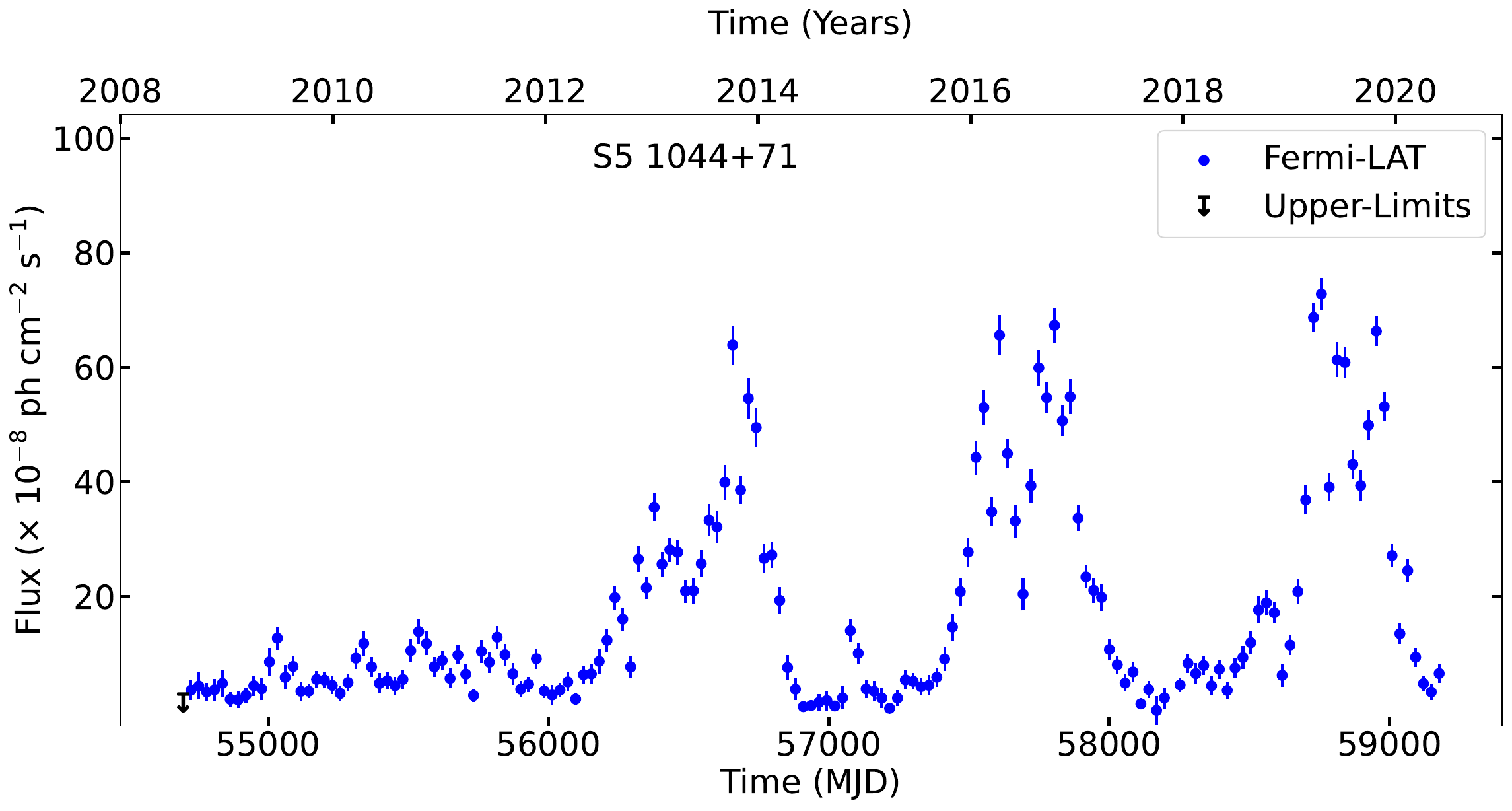}
         \includegraphics[scale=0.2195]{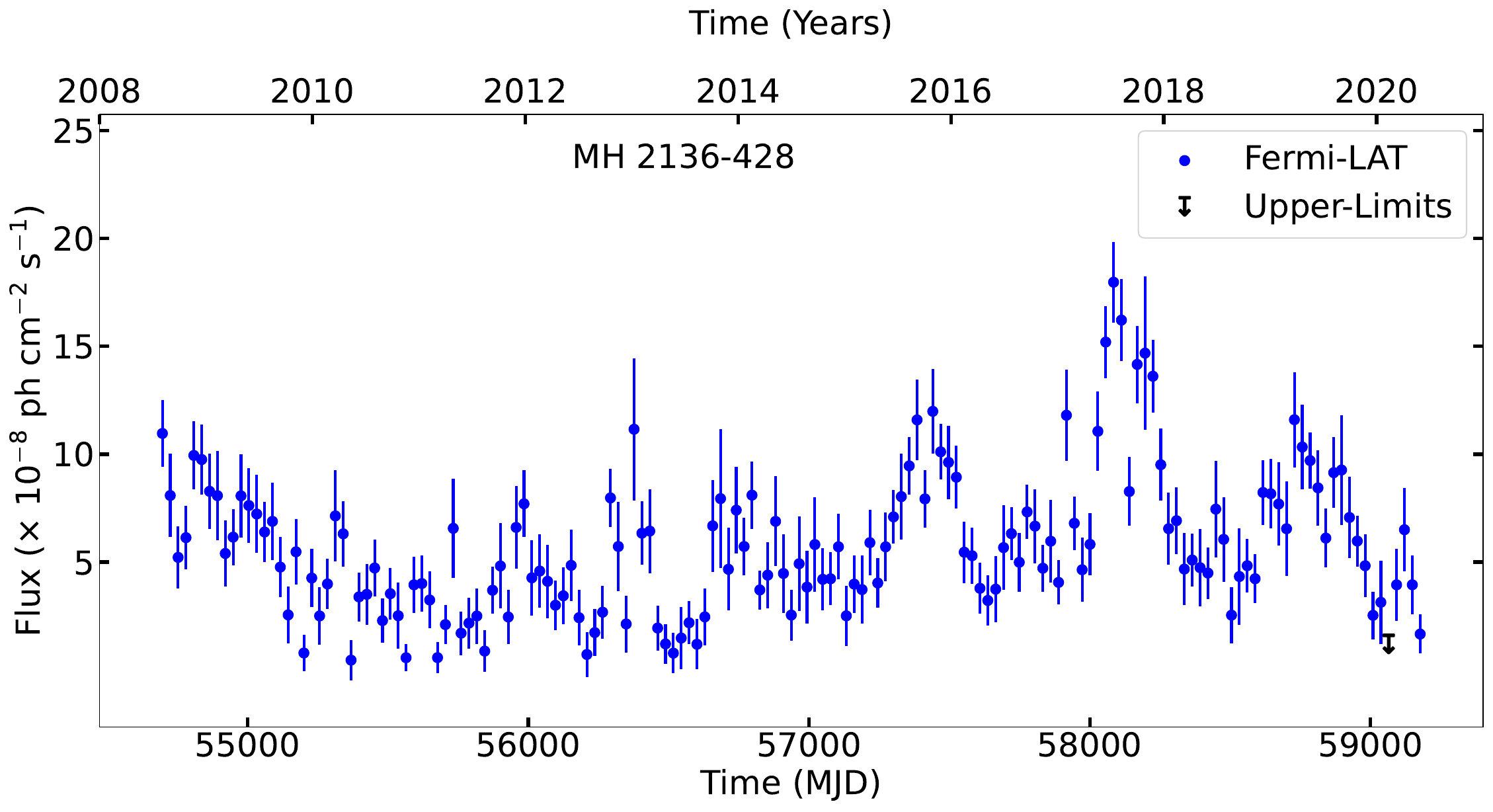}         
         \includegraphics[scale=0.2195]{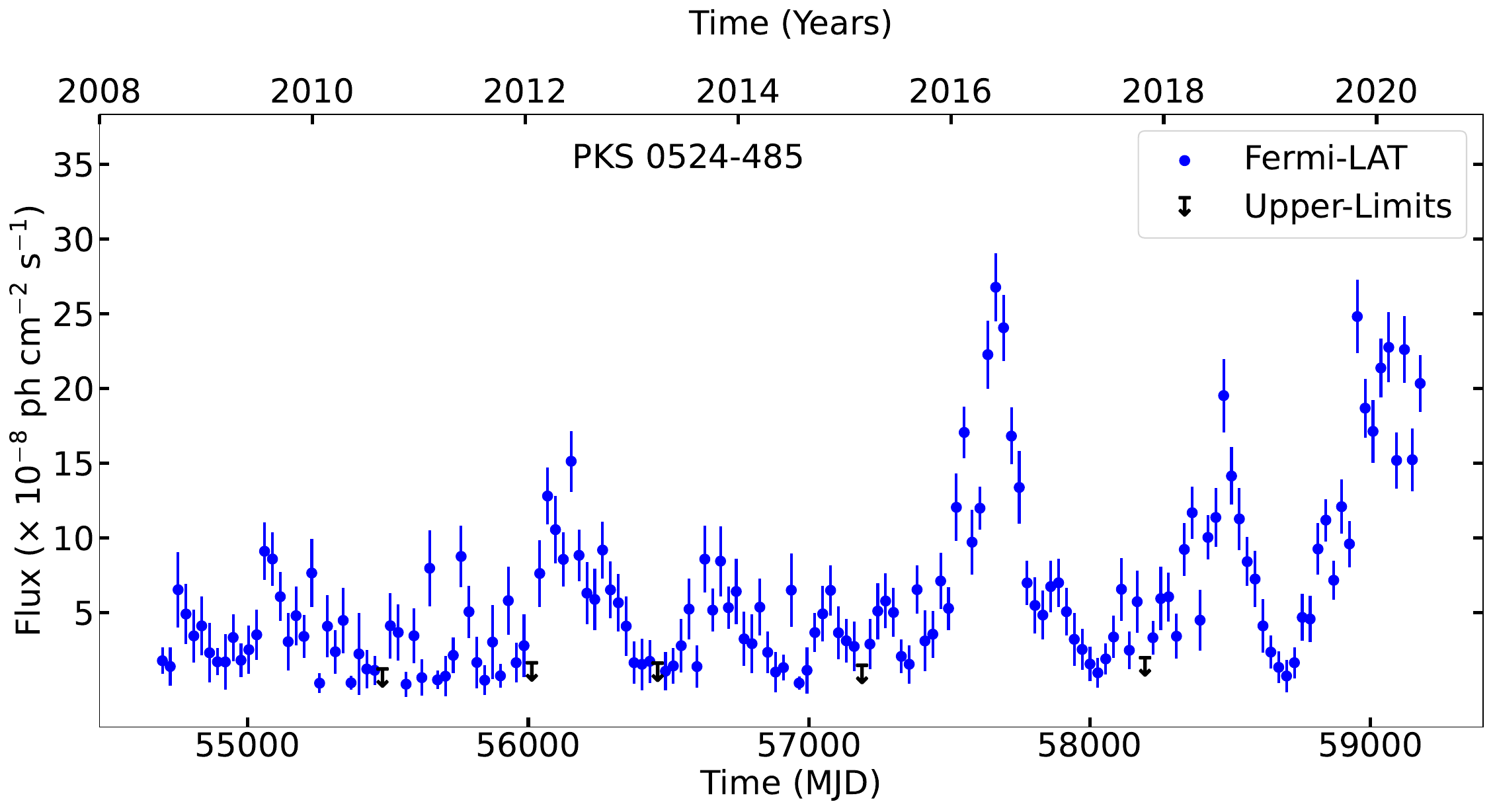}
         \includegraphics[scale=0.2195]{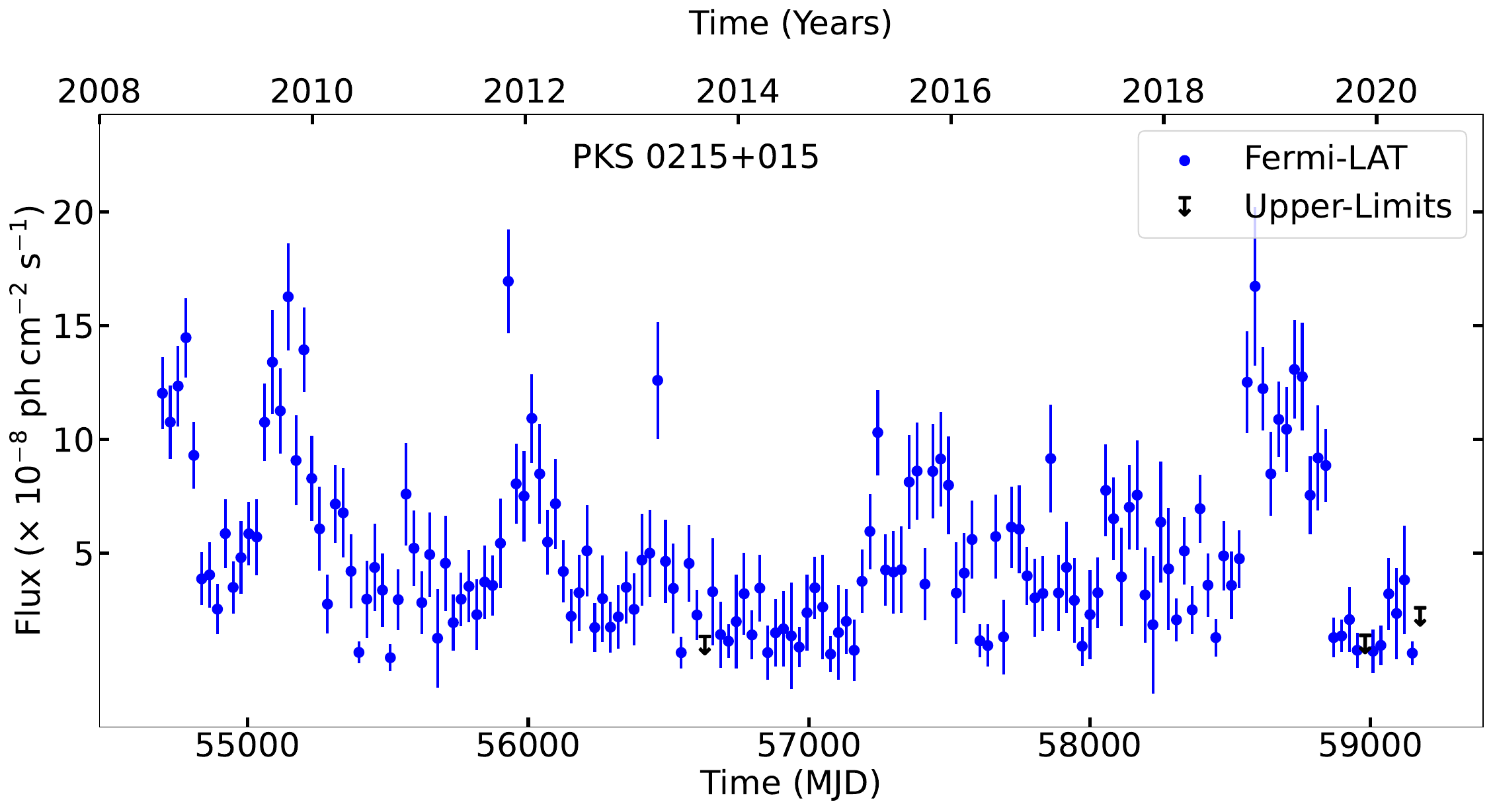}
         \includegraphics[scale=0.2195]{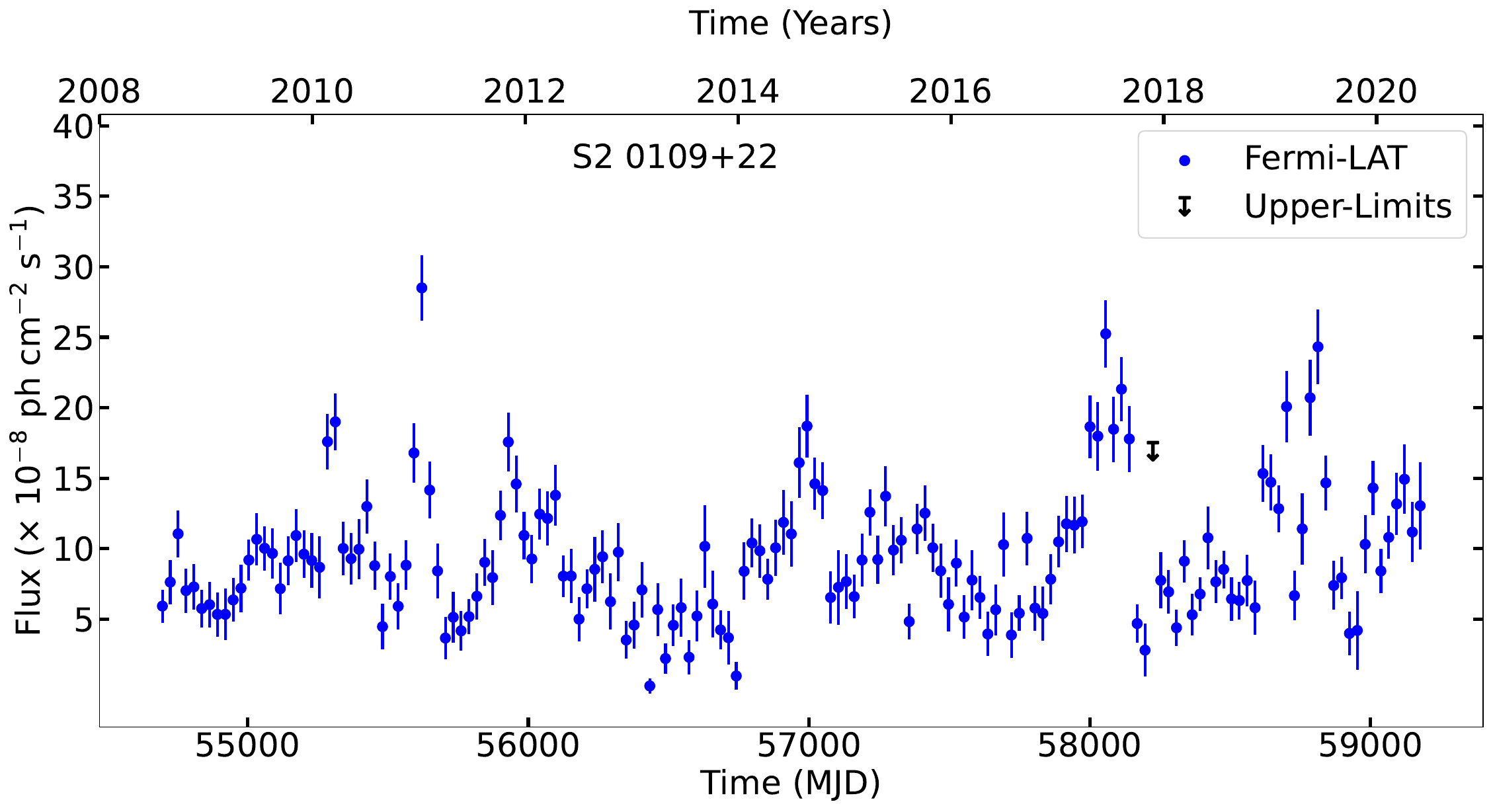} 
         \includegraphics[scale=0.2195]{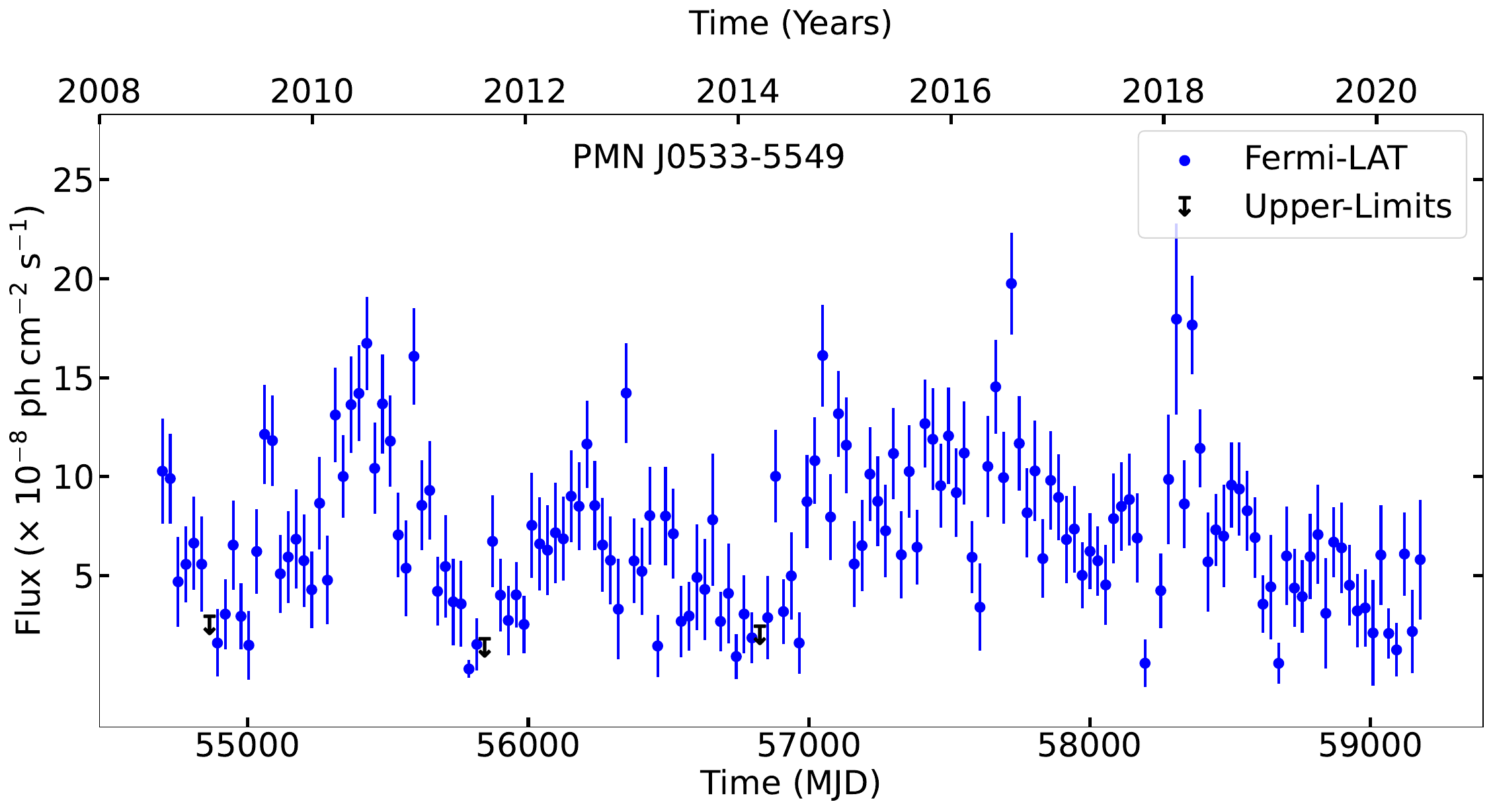} 
         \includegraphics[scale=0.2195]{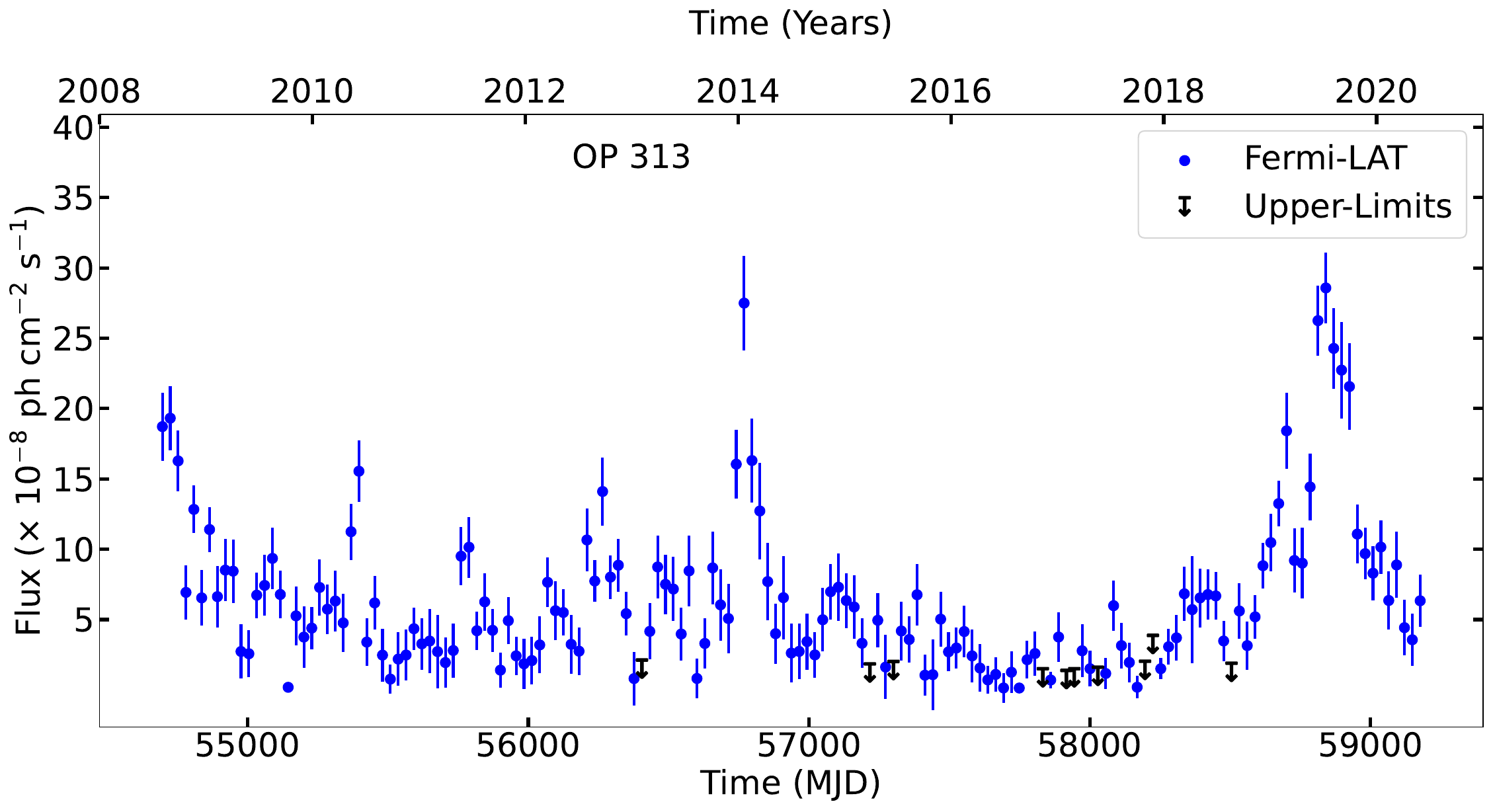}
         \includegraphics[scale=0.2195]{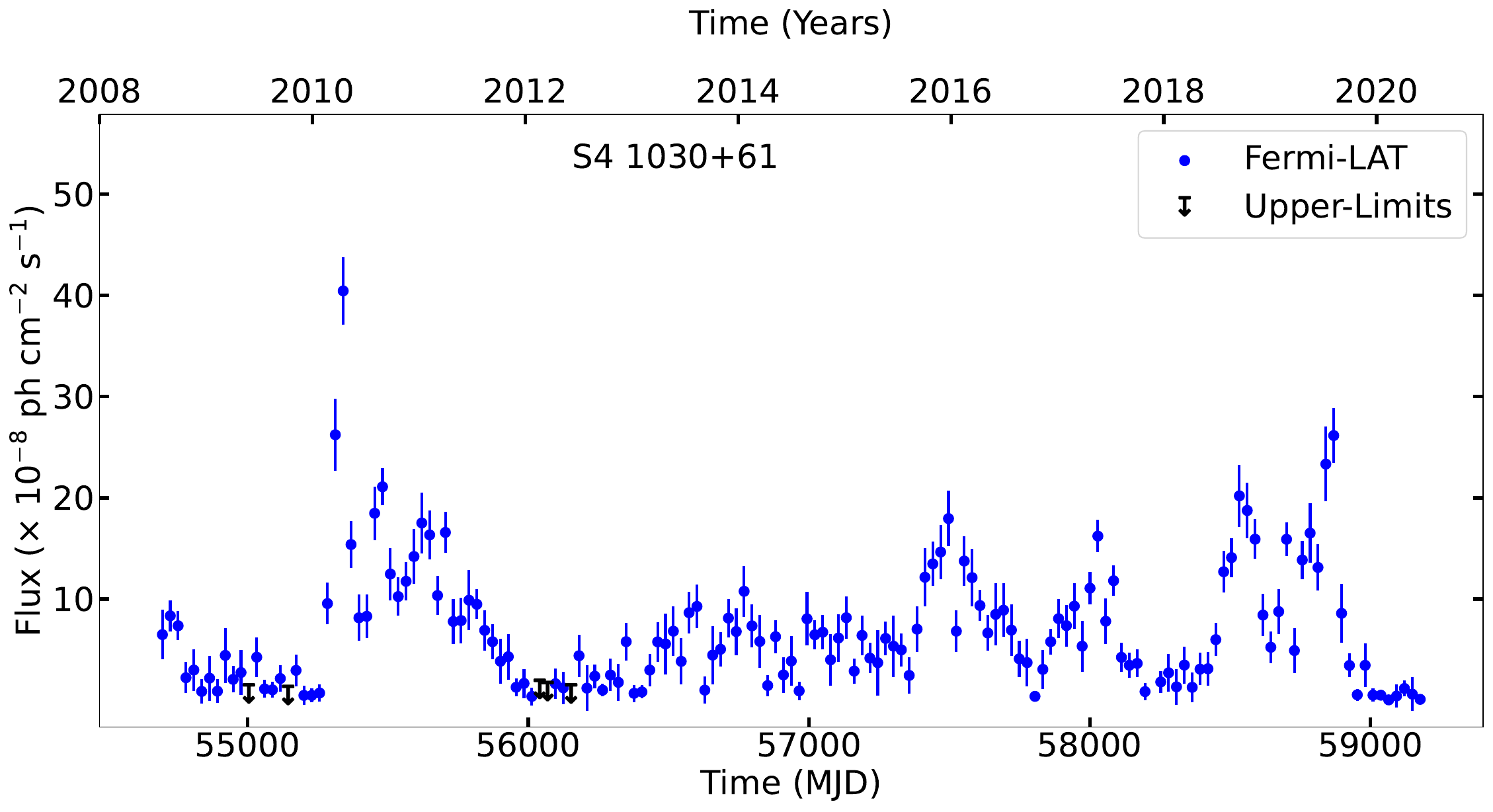}
         \includegraphics[scale=0.2195]{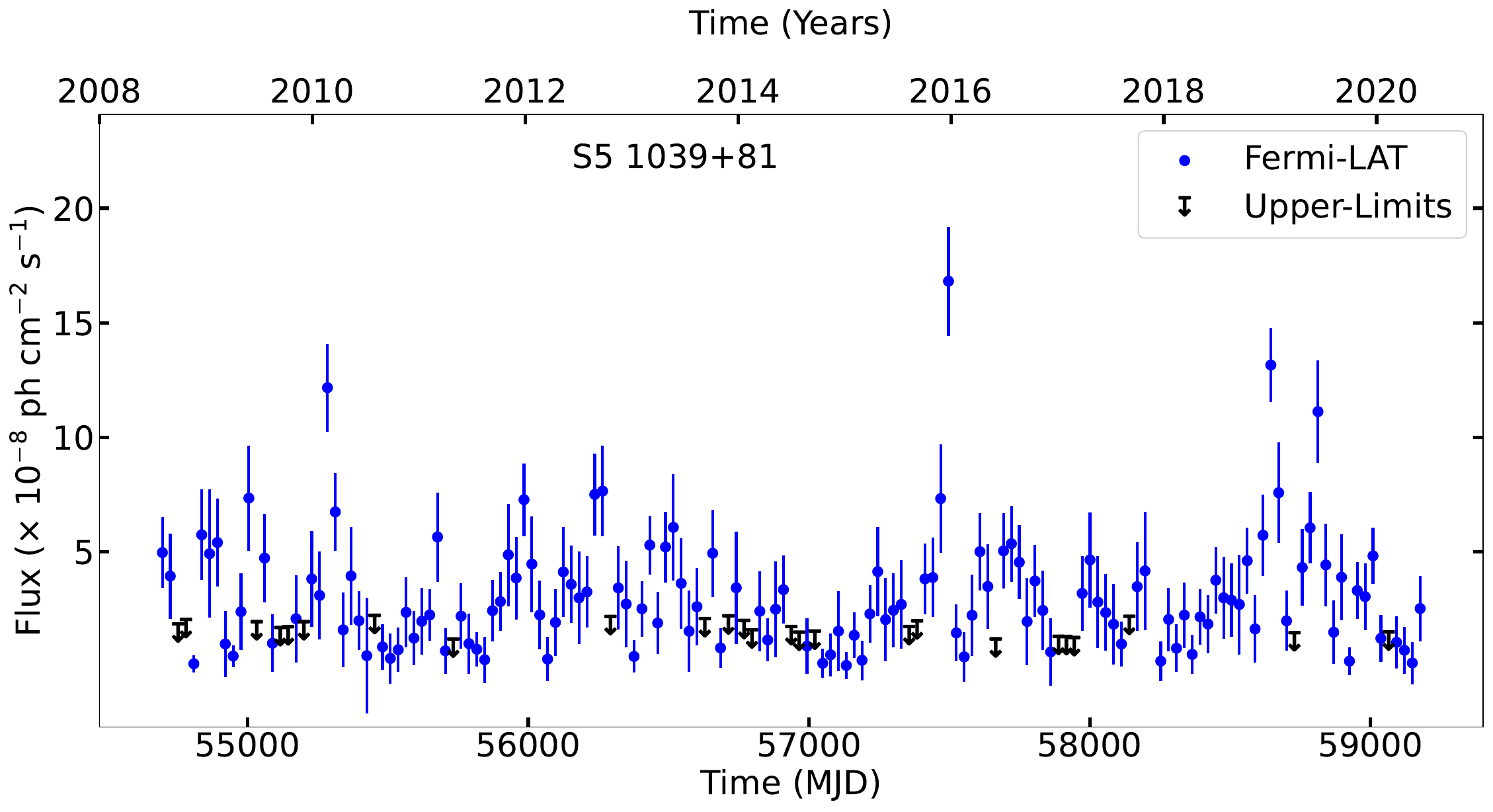}
        \caption{Light curves of the ten blazars with test statistics $<$2.5$\sigma$, as listed in Table \ref{tab:candidates_list}.}
	\label{fig:lc_candidates_low}
\end{figure*}

Figure \ref{fig:lc_candidates_low} reports the LCs of blazars.

\clearpage
\subsection{Power Spectral Densities}
Figure \ref{fig:psd_blazars} reports the PSD fit of the blazars. 
\begin{figure*}
	\centering
        \includegraphics[scale=0.185]{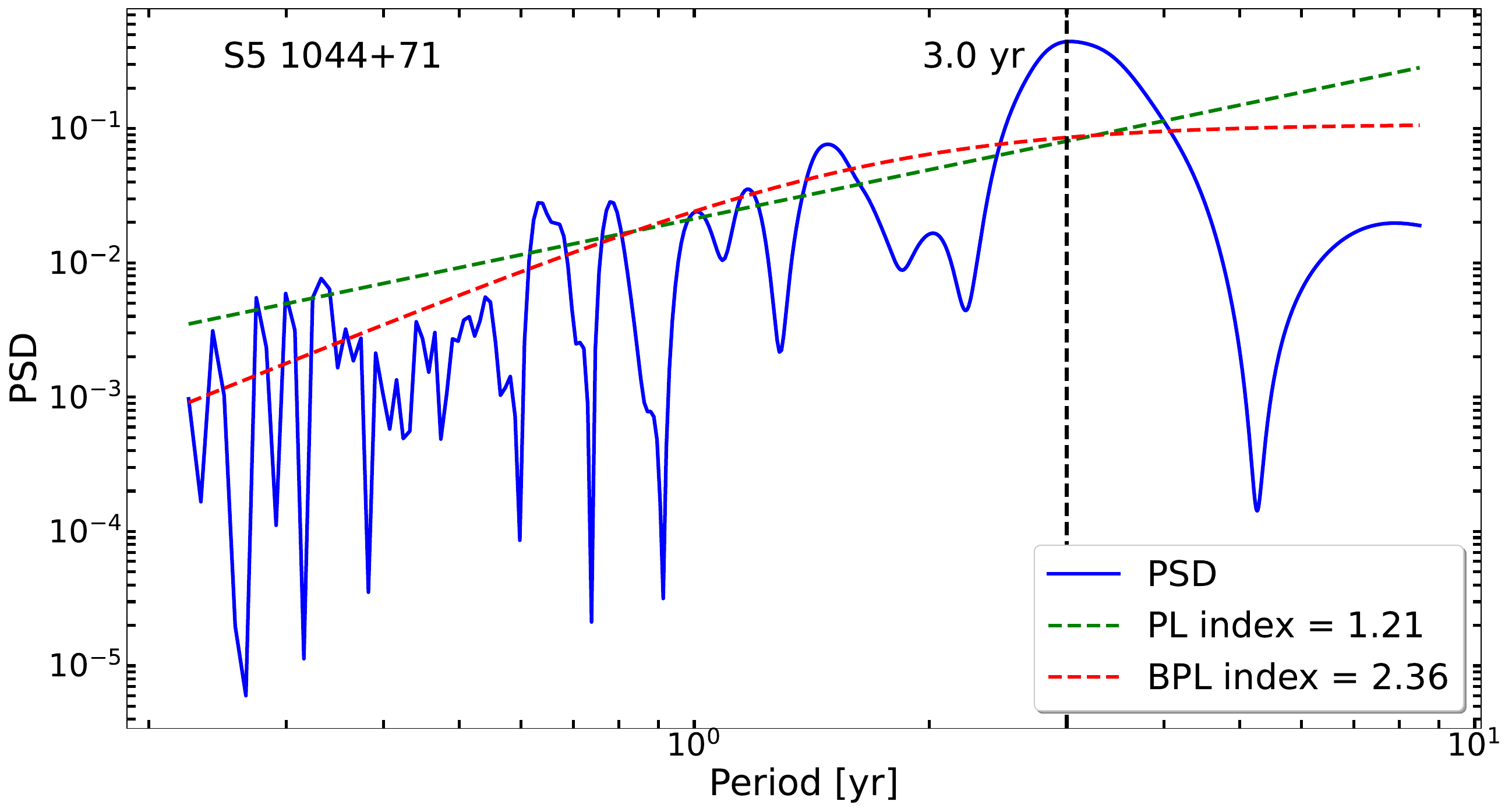}
        \includegraphics[scale=0.185]{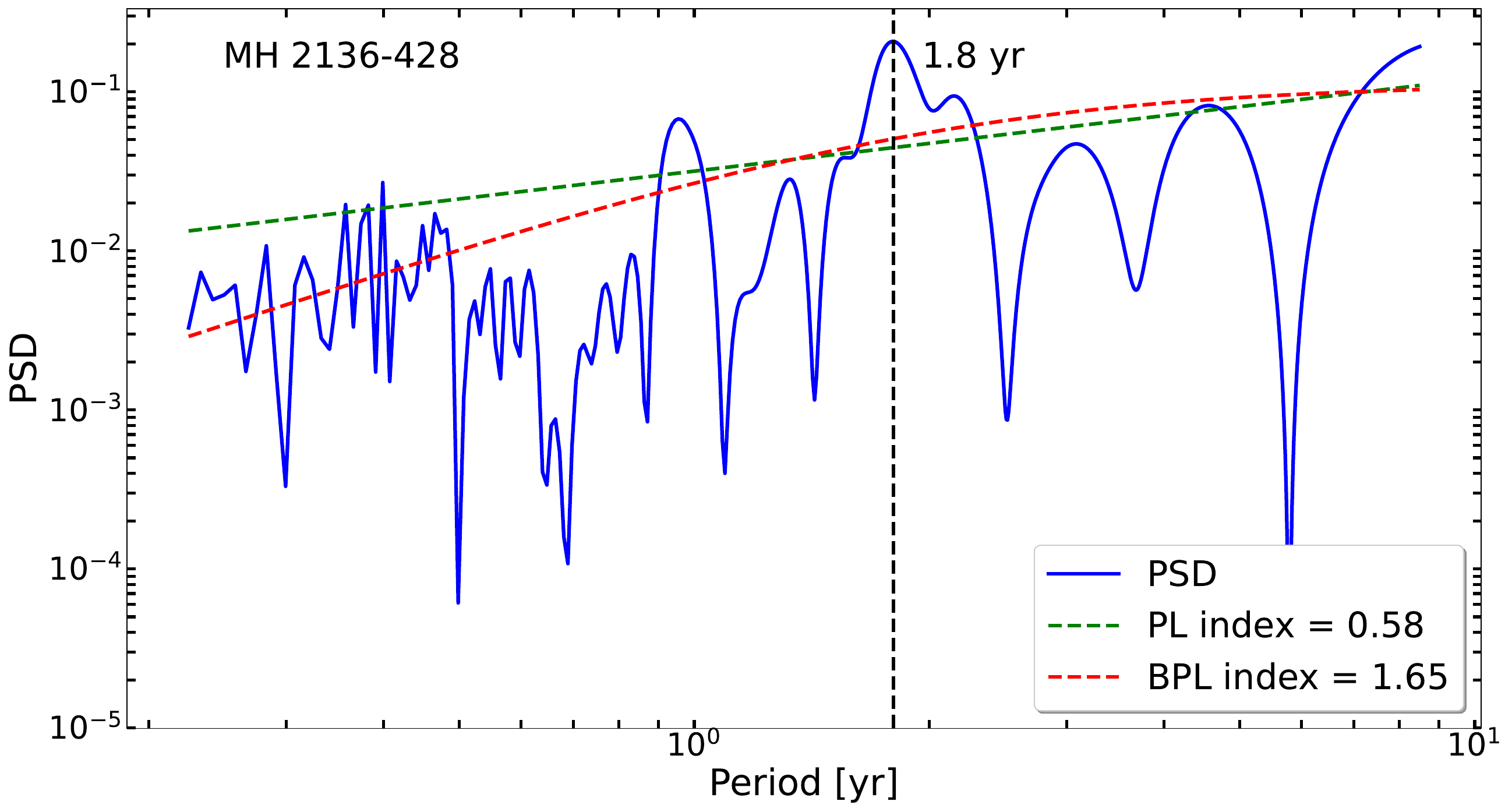} 
        \includegraphics[scale=0.185]{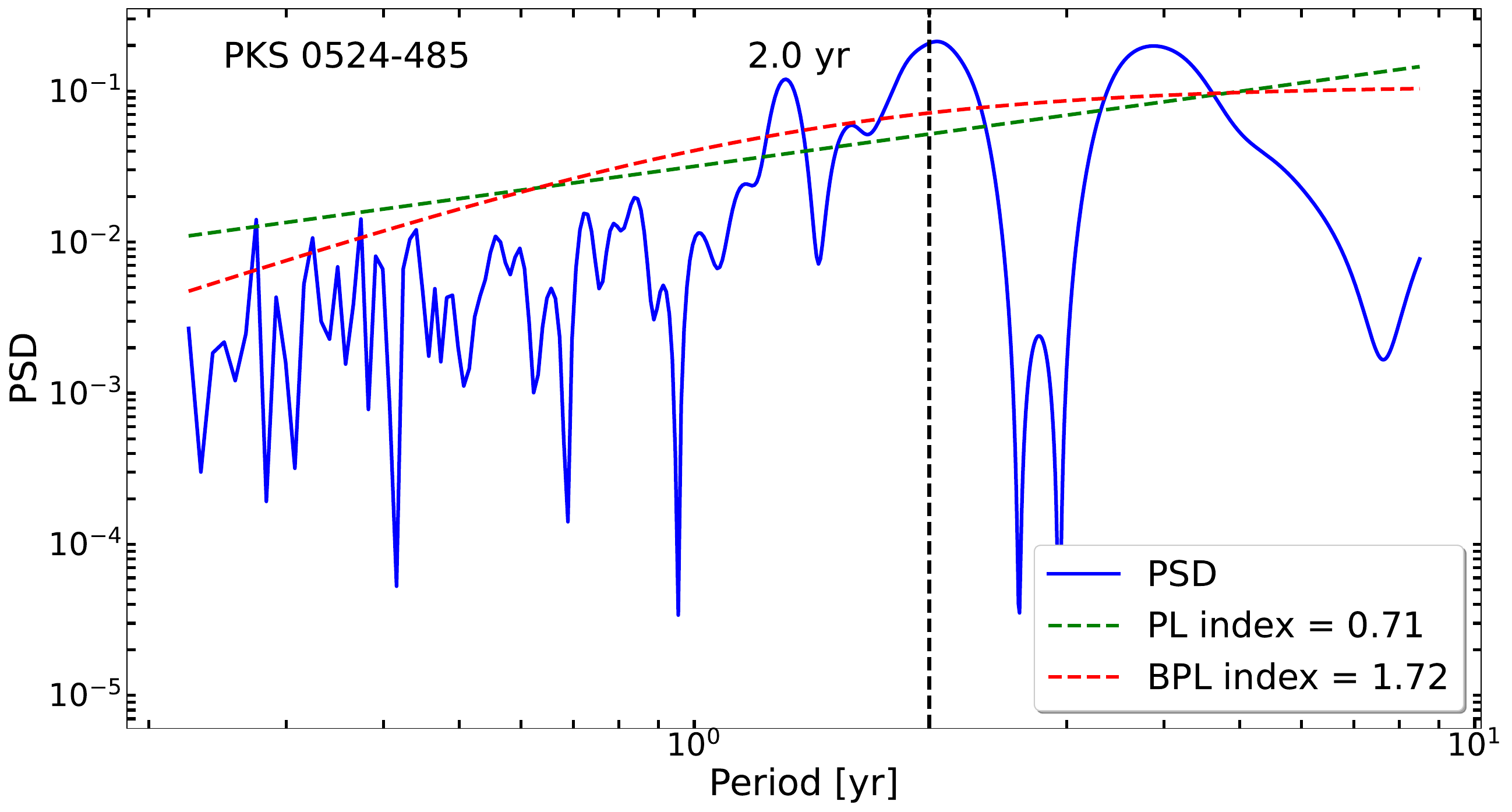}
        \includegraphics[scale=0.185]{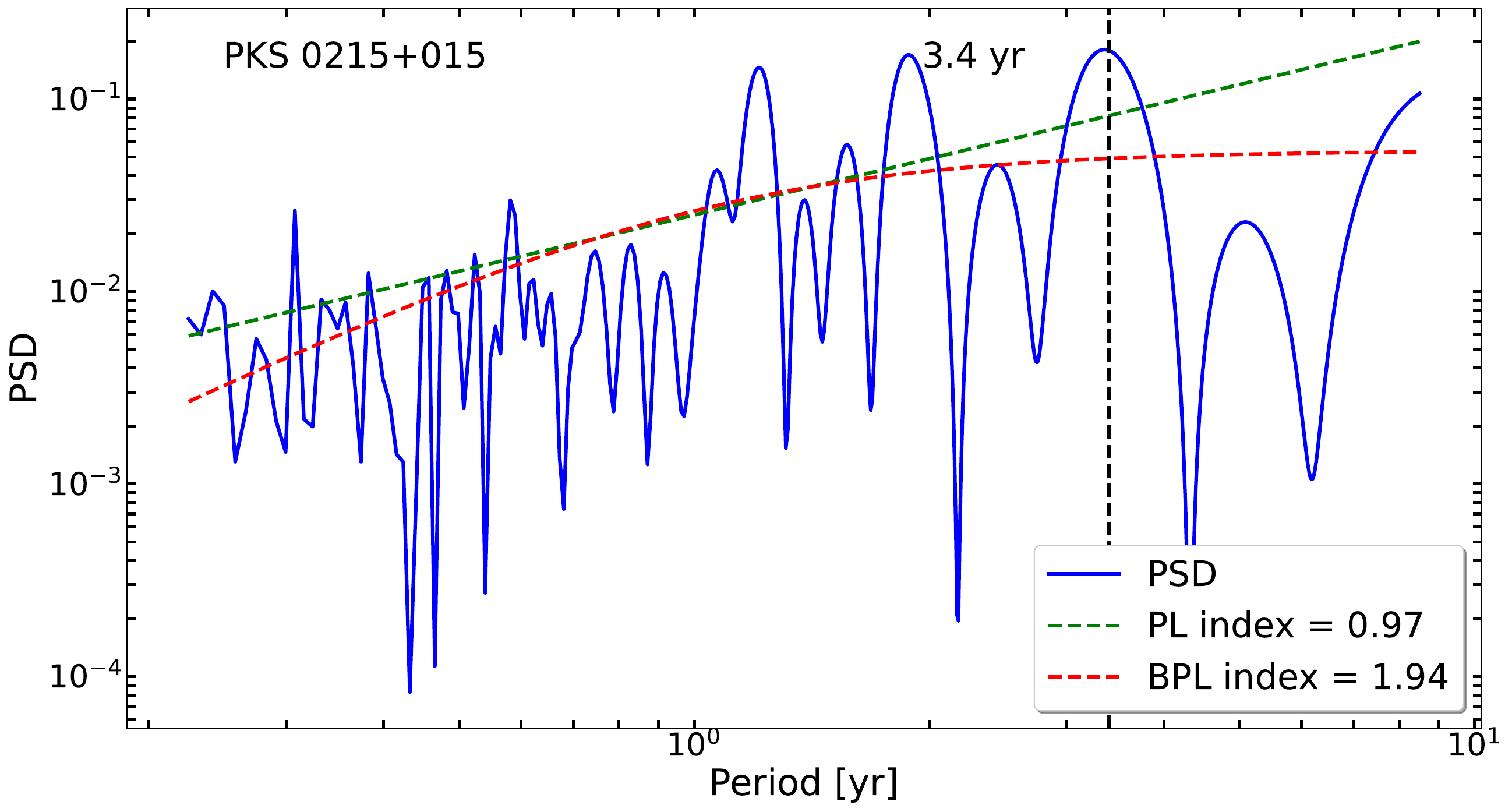}
        \includegraphics[scale=0.185]{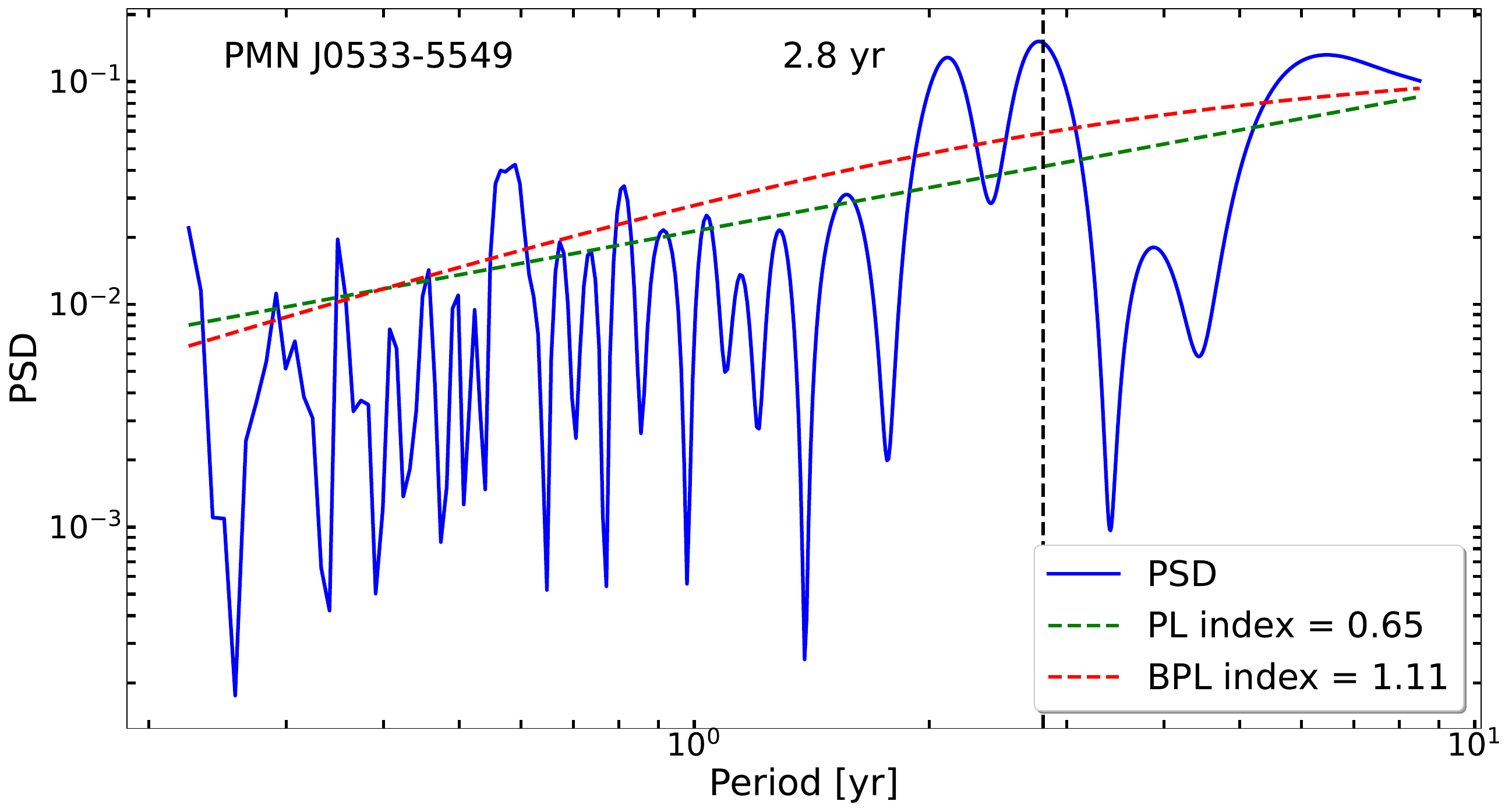}
        \includegraphics[scale=0.185]{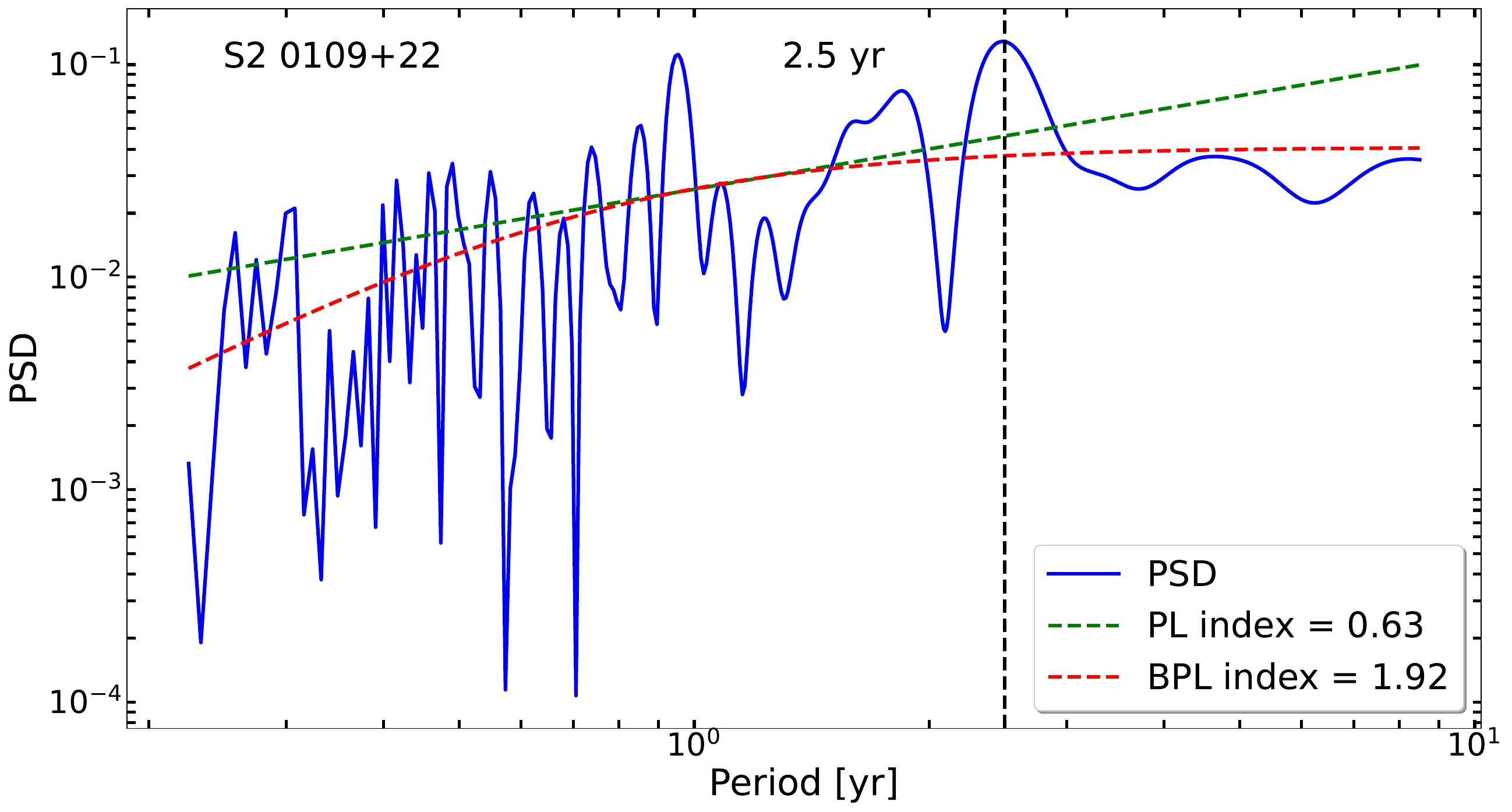}  
        \includegraphics[scale=0.185]{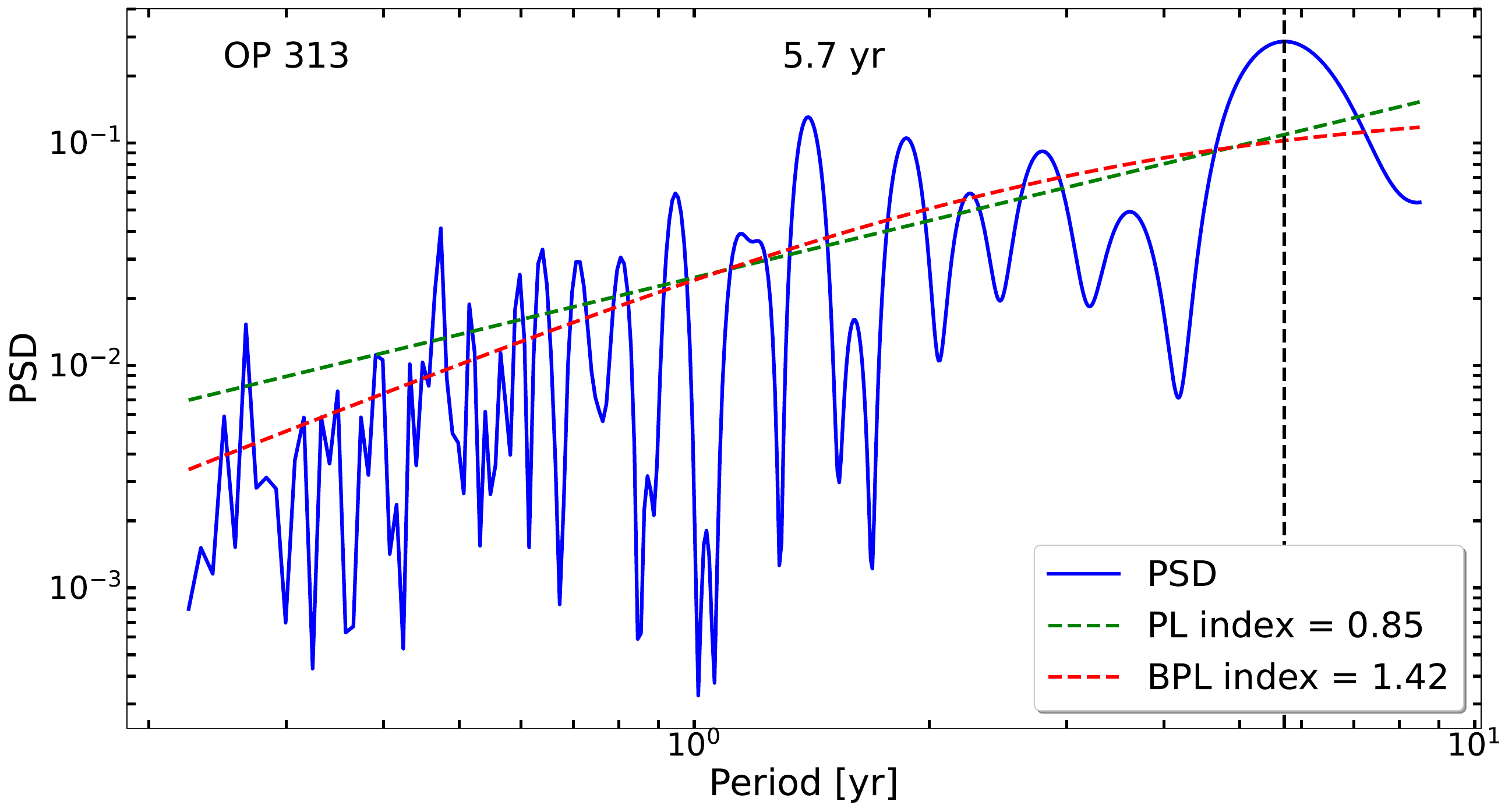}
        \includegraphics[scale=0.185]{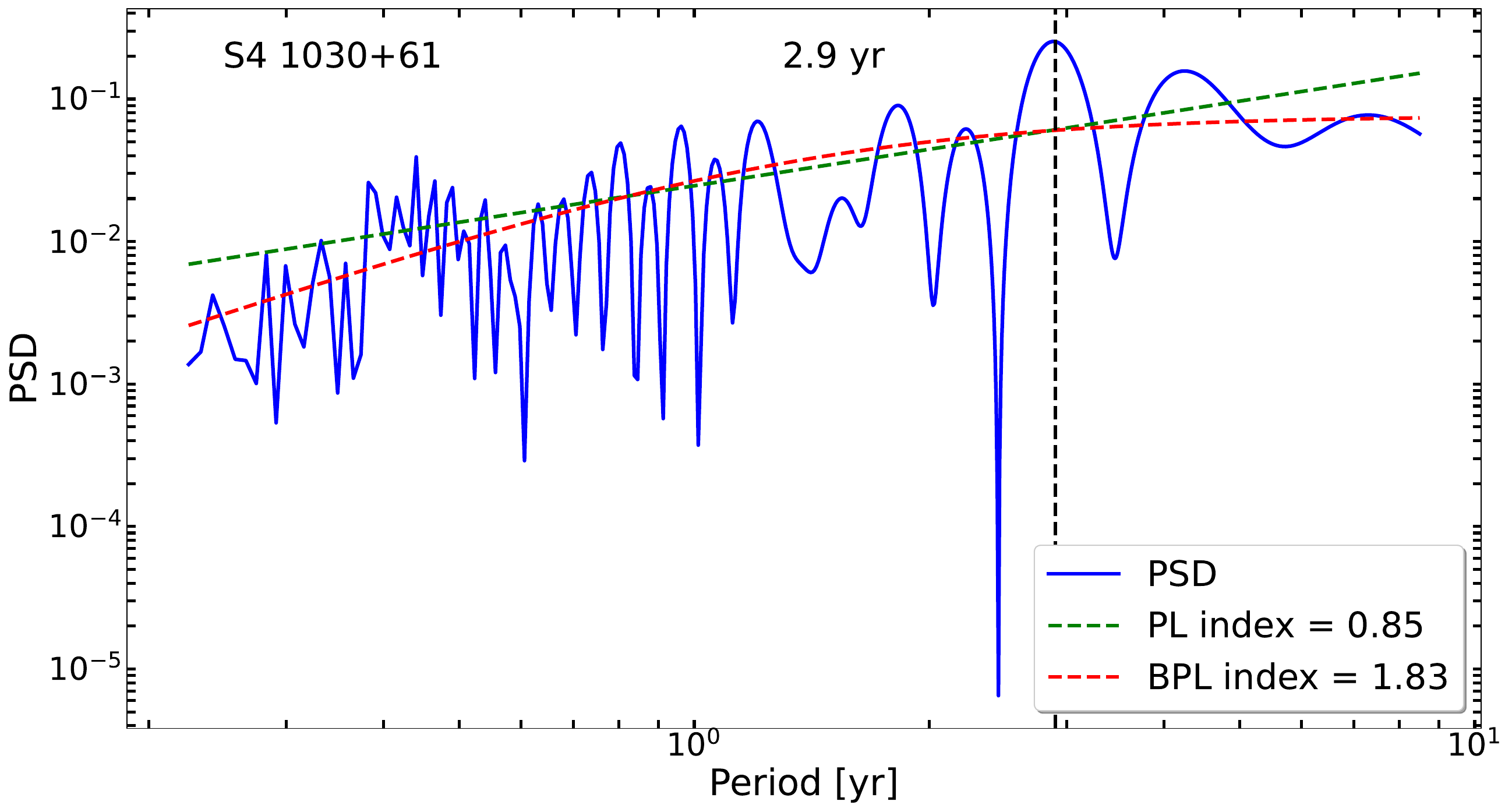}
	\caption{Power spectral densities using both the PL and BPL approaches, utilizing the power-spectral indices in Table \ref{tab:slopes}. The vertical line denotes the peak associated with the period observed in the LC.}
	\label{fig:psd_blazars} 
\end{figure*}
\begin{figure*}
	\centering
        \includegraphics[scale=0.185]{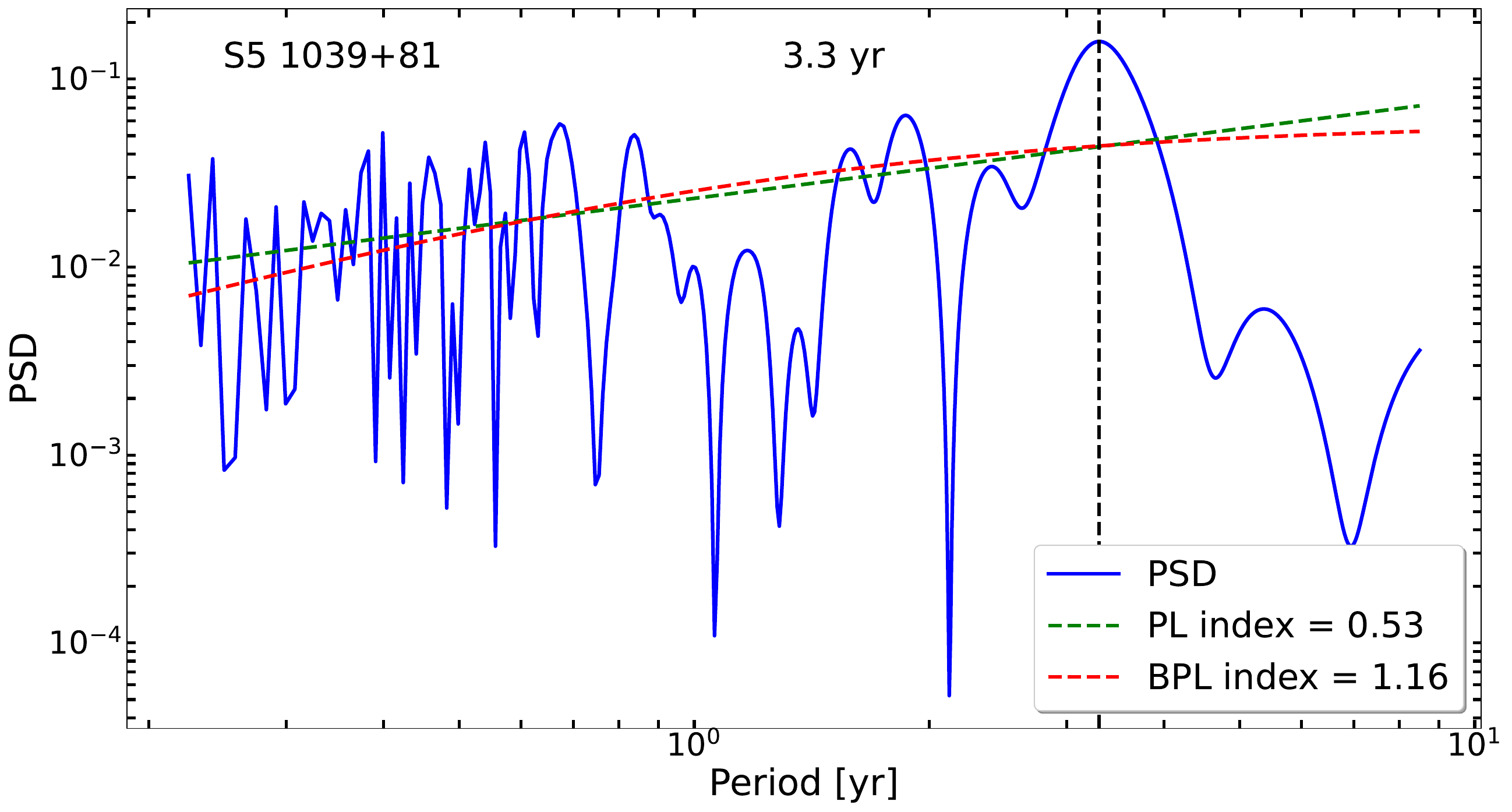}
	\caption{(Continued).}
\end{figure*}

\clearpage
\subsection{Tables}
This section presents Table \ref{tab:a_candiadtes_periods}, Table \ref{tab:slopes}, Table \ref{tab:correction_psds_results}, Table \ref{tab:correction_power_law_new}, Table \ref{tab:correction_bpl1_new}, Table \ref{tab:mcmc_bayesian_arfima_results}, and Table \ref{tab:qpo}, which summarize the estimations used to determine the periods of the selected blazars.

\begin{table*}
\centering
\caption{List of periods and their uncertainties (top), along with their associated test statistics obtained by applying TK (bottom), for the periodic-emission candidates inferred from the pipeline of Figure \ref{fig:study_flow}. Note that some sources exhibit two periods with high test statistics, organized by the amplitude of the peak and denoted by an asterisk $\star$. The symbol $\dagger$ denotes PDM results that exhibit the harmonic effect described in P20, while the symbol $\ddagger$ represents periods derived from the CWT analysis of 12 years of LC data. All periods are expressed in years. The high-significance blazars, as defined in the ``High-Significance Candidate Constraint'' outlined in Section \ref{sec:pipeline}, are indicated by a $\diamond$ symbol.\label{tab:a_candiadtes_periods}}
{%
\begin{tabular}{l|cccccccccc}
\hline
\hline
Association Name & LSP1 & GLSP & LSP2 & PDM & CWT & DFT-Welch & Period (yr) \\	
\hline
\hline
PKS 0405$-$385 $\diamond$ & $3.1^{\pm0.5}_{4.0\sigma}$ & $3.2^{\pm0.5}_{2.9\sigma}$ & $3.1^{\pm0.6}_{3.4\sigma}$ & $3.0^{\pm0.6}_{4.0\sigma}$ & $\ddagger$$3.3^{\pm0.5}_{4.2\sigma}$ & $3.1^{\pm0.7}_{4.3\sigma}$ & 3.1$\pm$0.5 (4.0$\sigma$)\\
 	S5 1044+71 $\diamond$ & $3.1^{\pm0.3}_{5.0\sigma}$ & $3.0^{\pm0.5}_{4.4\sigma}$ & $3.0^{\pm0.5}_{3.2\sigma}$ & $3.1^{\pm0.4}_{4.4\sigma}$ & $3.2^{\pm0.5}_{3.3\sigma}$ & $3.1^{\pm0.6}_{4.4\sigma}$ & 3.1$\pm$0.5 (4.4$\sigma$) \\
        MH 2136$-$428 & $1.8^{\pm0.1}_{3.9\sigma}$ & $1.8^{\pm0.1}_{3.8\sigma}$ & $1.8^{\pm0.1}_{3.6\sigma}$ & $1.8^{\pm0.1}_{3.0\sigma}$ & $1.9^{\pm0.4}_{2.5\sigma}$ & $1.9^{\pm0.2}_{4.4\sigma}$ & 1.8$\pm$0.1 (3.7$\sigma$)\\
        PKS 0524$-$485$\star$ & $2.0^{\pm0.2}_{3.4\sigma}$ & $3.9^{\pm0.5}_{2.9\sigma}$ & \makecell{$2.0^{\pm0.2}_{2.9\sigma}$ \\ $3.9^{\pm0.5}_{2.0\sigma}$} &  $4.0^{\pm0.4}_{3.8\sigma}$  & \makecell {$\ddagger$$4.2^{\pm0.6}_{4.5\sigma}$ \\ $2.1^{\pm0.3}_{4.4\sigma}$} & \makecell {$2.1^{\pm0.3}_{4.0\sigma}$ \\ $4.0^{\pm0.7}_{3.2\sigma}$} & \makecell{2.0$\pm$0.2 (3.7$\sigma$) \\ 4.0$\pm$0.5 (3.1$\sigma$)} \\        
        PKS 0215+015$\star$ & $3.3^{\pm0.4}_{3.8\sigma}$ & $3.4^{\pm0.4}_{3.0\sigma}$ & \makecell{$3.3^{\pm0.4}_{2.7\sigma}$ \\ $1.9^{\pm0.2}_{3.1\sigma}$} & $3.6^{\pm0.4}_{3.5\sigma}$ & $3.3^{\pm0.7}_{2.0\sigma}$ & \makecell{$2.0^{\pm0.3}_{4.2\sigma}$ \\ $3.6^{\pm0.6}_{4.0\sigma}$} & 3.4$\pm$0.4 (3.2$\sigma$) \\        
        S2 0109+22 & $2.5^{\pm0.3}_{3.0\sigma}$ & $2.6^{\pm0.4}_{3.8\sigma}$ & $2.5^{\pm0.3}_{2.4\sigma}$ & $\dagger$$5.0^{\pm0.5}_{2.8\sigma}$ & $\ddagger$$2.7^{\pm0.3}_{3.7\sigma}$ & $2.2^{\pm0.3}_{3.6\sigma}$ & 2.4$\pm$0.3 (3.5$\sigma$) \\
        PMN J0533$-$5549$\star$ & \makecell {$2.8^{\pm0.2}_{3.0\sigma}$ \\ $2.0^{\pm0.2}_{3.0\sigma}$} & \makecell {$2.8^{\pm0.3}_{3.3\sigma}$ \\ $2.0^{\pm0.1}_{3.1\sigma}$} & \makecell {$2.8^{\pm0.3}_{2.3\sigma}$ \\ $2.0^{\pm0.1}_{2.1\sigma}$} & $2.8^{\pm0.3}_{2.8\sigma}$ & $2.1^{\pm0.4}_{2.3\sigma}$ & $2.9^{\pm0.3}_{3.0\sigma}$ & \makecell{2.8$\pm$0.3 (3.0$\sigma$) \\ 2.0$\pm$0.3(2.7$\sigma$)} \\
        OP 313 & $5.6^{\pm1.2}_{1.9\sigma}$ & $5.9^{\pm0.5}_{2.5\sigma}$ & $5.7^{\pm0.8}_{1.9\sigma}$ & $5.7^{\pm0.4}_{4.2\sigma}$ & $\ddagger$$6.0^{\pm0.7}_{3.4\sigma}$ & $5.7^{\pm0.8}_{4.1\sigma}$ & 5.7$\pm$0.8 (3.0$\sigma$) \\
        S4 1030+61 & $2.9^{\pm0.2}_{3.0\sigma}$ & $2.7^{\pm0.3}_{1.8\sigma}$ & $2.9^{\pm0.2}_{2.6\sigma}$ & $2.8^{\pm0.3}_{3.1\sigma}$ & $\ddagger$$3.2^{\pm0.7}_{4.4\sigma}$ & $2.9^{\pm0.6}_{2.6\sigma}$ & 2.9$\pm$0.4 (2.8$\sigma$)\\
        S5 1039+81 & $3.3^{\pm0.4}_{4.0\sigma}$ & $3.1^{\pm0.3}_{2.3\sigma}$ & $3.3^{\pm0.4}_{2.4\sigma}$ & $3.3^{\pm0.3}_{3.0\sigma}$ & $\ddagger$$3.3^{\pm0.5}_{1.8\sigma}$ & $2.9^{\pm0.5}_{4.1\sigma}$ & 3.5$\pm$0.4 (2.7$\sigma$)\\ 
        PKS 0451$-$28 & $4.9^{\pm1.2}_{2.0\sigma}$ & $5.0^{\pm0.3}_{3.1\sigma}$ & $4.9^{\pm0.8}_{1.6\sigma}$ & $5.0^{\pm0.3}_{3.4\sigma}$ & $\ddagger$$5.2^{\pm0.8}_{3.6\sigma}$ & $4.8^{\pm0.8}_{3.1\sigma}$ & 5.0$\pm$0.7 (3.1$\sigma$)\\
        PKS 0736+017$\star$ & $4.6^{\pm0.9}_{3.1\sigma}$ & \makecell{$2.2^{\pm0.2}_{2.7\sigma}$ \\ $4.5^{\pm0.8}_{2.0\sigma}$} & $4.7^{\pm0.9}_{1.6\sigma}$ & $4.4^{\pm0.5}_{3.5\sigma}$ & $\ddagger$$4.5^{\pm0.9}_{2.7\sigma}$ & $4.3^{\pm1.1}_{4.0\sigma}$ & 4.5$\pm$0.8 (2.9$\sigma$)\\
        PKS 1824$-$582 & $2.1^{\pm0.2}_{3.1\sigma}$ & $2.1^{\pm0.1}_{1.2\sigma}$ & $2.1^{\pm0.2}_{2.4\sigma}$ & $2.0^{\pm0.2}_{3.0\sigma}$ & $2.3^{\pm0.4}_{2.1\sigma}$ & $2.1^{\pm0.4}_{3.3\sigma}$ & 2.4$\pm$0.1 (2.7$\sigma$)\\    
        PKS 1903$-$80$\star$ & $2.4^{\pm0.2}_{3.1\sigma}$ & $2.4^{\pm0.2}_{2.1\sigma}$ & $2.4^{\pm0.2}_{2.5\sigma}$ &  $\dagger$$4.8^{\pm0.4}_{3.3\sigma}$ & \makecell{$\ddagger$$4.0^{\pm0.6}_{3.1\sigma}$ \\ $\ddagger$$2.5^{\pm0.5}_{2.6\sigma}$} & $2.5^{\pm0.5}_{3.8\sigma}$ & 2.4$\pm$0.2 (2.9$\sigma$)\\       	
        S4 0110+49 & $3.7^{\pm0.8}_{2.0\sigma}$ & $3.8^{\pm0.8}_{1.8\sigma}$ & $3.7^{\pm0.9}_{1.9\sigma}$ & $3.7^{\pm0.9}_{3.2\sigma}$ & $\ddagger$$3.9^{\pm0.8}_{3.0\sigma}$ & $4.2^{\pm0.9}_{4.2\sigma}$ & 3.7$\pm$0.8 (2.5$\sigma$) \\
        MG2 J083121+2629 & $3.3^{\pm0.4}_{3.0\sigma}$ & $3.5^{\pm0.6}_{2.2\sigma}$ & $3.3^{\pm0.4}_{3.1\sigma}$ & $3.6^{\pm0.5}_{2.6\sigma}$ & $3.3^{\pm0.7}_{3.1\sigma}$ & $1.3^{\pm0.1}_{2.2\sigma}$ &3.3$\pm$0.5 (2.6$\sigma$) \\
\hline
\hline
\end{tabular}%
}
\end{table*}

\begin{table*}
\centering
\caption{List of PSD fits for the blazars in Table \ref{tab:candidates_list}. The fitting includes the ML-MCMC method applied to both PL and BPL approaches. The parameter ``A'' represents the normalization in units of $rms^2/year^{-1}$, and $\nu_{Bending}$ denotes the bending frequency.\label{tab:slopes}}
{%
\begin{tabular}{l|cccccccccc}
\hline
\hline
Association Name & A / PL index & A / BPL index / $\nu_{Bending}$ \\
\hline
\hline
PKS 0405$-$385 & \makecell{1.79$\mathrm{x}10^{-2}$$\pm$1.39$\mathrm{x}10^{-3}$ \\ 1.11$\pm$0.03} & \makecell{1.31$\mathrm{x}10^{-1}$$\pm$3.22$\mathrm{x}10^{-2}$ \\ 1.62$\pm$0.08 \\ 0.45$\pm$0.07} \\
       \hline
S5 1044+71 & \makecell{2.13$\mathrm{x}10^{-2}$$\pm$1.46$\mathrm{x}10^{-3}$ \\ 1.21$\pm$0.10} & \makecell{1.08$\mathrm{x}10^{-1}$ $\pm$5.83$\mathrm{x}10^{-2}$ \\ 2.36$\pm$0.07 \\ 0.59$\pm$0.05} \\
        \hline
MH 2136$-$428 & \makecell{3.06$\mathrm{x}10^{-2}$$\pm$2.73$\mathrm{x}10^{-3}$ \\ 0.58$\pm$0.04} & \makecell{1.13$\mathrm{x}10^{-1}$$\pm$8.21$\mathrm{x}10^{-2}$ \\ 1.65$\pm$0.12 \\ 0.49$\pm$0.05} \\
        \hline
PKS 0524$-$485 & \makecell{3.17$\mathrm{x}10^{-2}$$\pm$2.24$\mathrm{x}10^{-3}$ \\ 0.71$\pm$0.07} & \makecell{1.04$\mathrm{x}10^{-1}$$\pm$6.84$\mathrm{x}10^{-2}$ \\ 1.72$\pm$0.07 \\ 0.74$\pm$0.07} \\
        \hline
PKS 0215+015 & \makecell{2.50$\mathrm{x}10^{-2}$$\pm$1.18$\mathrm{x}10^{-3}$ \\ 0.97$\pm$0.05} & \makecell{6.33$\mathrm{x}10^{-2}$$\pm$3.75$\mathrm{x}10^{-3}$ \\ 1.94$\pm$0.14 \\ 0.97$\pm$0.09} \\
        \hline
S2 0109+22 & \makecell{2.59$\mathrm{x}10^{-2}$$\pm$5.15$\mathrm{x}10^{-3}$ \\ 0.63$\pm$0.05} & \makecell{4.59$\mathrm{x}10^{-2}$$\pm$2.85$\mathrm{x}10^{-3}$ \\ 1.92$\pm$0.53 \\ 1.34$\pm$0.17} \\
        \hline
PMN J0533$-$5549 & \makecell{2.13$\mathrm{x}10^{-2}$$\pm$1.28$\mathrm{x}10^{-3}$ \\ 0.65$\pm$0.04} & \makecell{1.15$\mathrm{x}10^{-1}$$\pm$9.10$\mathrm{x}10^{-2}$\\ 1.11$\pm$0.08 \\ 0.33$\pm$0.08} \\
	\hline
OP 313 & \makecell{2.48$\mathrm{x}10^{-2}$$\pm$1.04$\mathrm{x}10^{-3}$ \\ 0.85$\pm$0.04} & \makecell{1.46$\mathrm{x}10^{-1}$$\pm$9.26$\mathrm{x}10^{-2}$ \\ 1.42$\pm$0.08 \\ 0.32$\pm$0.05} \\ 
        \hline
S4 1030+61 & \makecell{2.46$\mathrm{x}10^{-2}$$\pm$1.67$\mathrm{x}10^{-3}$ \\ 0.85$\pm$0.05} & \makecell{8.84$\mathrm{x}10^{-2}$$\pm$3.42$\mathrm{x}10^{-3}$ \\ 1.83$\pm$0.08 \\ 0.71$\pm$0.06} \\
        \hline
S5 1039+81 & \makecell{2.32$\mathrm{x}10^{-2}$$\pm$3.45$\mathrm{x}10^{-3}$ \\ 0.53$\pm$0.04} & \makecell{5.24$\mathrm{x}10^{-2}$$\pm$3.62$\mathrm{x}10^{-3}$ \\ 1.16$\pm$0.15 \\ 0.80$\pm$0.20} \\    
        \hline
PKS 0451$-$28 & \makecell{1.89$\mathrm{x}10^{-2}$$\pm$8.75$\mathrm{x}10^{-3}$ \\ 0.83$\pm$0.03} & \makecell{1.37$\mathrm{x}10^{-1}$$\pm$8.17$\mathrm{x}10^{-2}$ \\ 1.14$\pm$0.06 \\  0.23$\pm$0.06} \\
        \hline
PKS 0736+017 & \makecell{2.38$\mathrm{x}10^{-2}$$\pm$1.01$\mathrm{x}10^{-3}$ \\ 0.89$\pm$0.04} & \makecell{1.34$\mathrm{x}10^{-1}$$\pm$7.54$\mathrm{x}10^{-2}$ \\ 1.37$\pm$0.07 \\ 0.38$\pm$0.06} \\ 
        \hline
PKS 1824$-$582 & \makecell{2.96$\mathrm{x}10^{-2}$$\pm$1.16$\mathrm{x}10^{-3}$ \\ 0.64$\pm$0.05} & \makecell{3.79$\mathrm{x}10^{-2}$$\pm$6.24$\mathrm{x}10^{-3}$ \\ 2.13$\pm$0.18 \\ 1.90$\pm$0.20} \\
        \hline
PKS 1903$-$80 & \makecell{2.98$\mathrm{x}10^{-2}$$\pm$3.29$\mathrm{x}10^{-3}$ \\ 0.68$\pm$0.04} & \makecell{1.42$\mathrm{x}10^{-1}$$\pm$8.29$\mathrm{x}10^{-2}$\\ 1.21$\pm$0.07 \\ 0.40$\pm$0.08} \\ 
        \hline  
S4 0110+49 & \makecell{2.61$\mathrm{x}10^{-2}$$\pm$9.55$\mathrm{x}10^{-3}$ \\ 1.21$\pm$0.05} & \makecell{1.93$\mathrm{x}10^{-1}$$\pm$1.74$\mathrm{x}10^{-2}$ \\ 1.81$\pm$0.05 \\ 0.36$\pm$0.03} \\  
        \hline
MG2 J083121+2629 & \makecell{2.28$\mathrm{x}10^{-2}$$\pm$8.74$\mathrm{x}10^{-3}$ \\ 0.54$\pm$0.03} & \makecell{5.78$\mathrm{x}10^{-2}$$\pm$7.24$\mathrm{x}10^{-3}$\\ 1.32$\pm$0.34 \\ 1.00$\pm$0.20} \\     
\hline
\hline
\end{tabular}%
}
\end{table*}

\begin{table*}
\centering
\caption{List of test statistics obtained with both PL and BPL using the parameters from Table \ref{tab:slopes} for the sources listed in Table \ref{tab:a_candiadtes_periods}. We calculate the new test statistics using the method described in \citet{emma_lc}. Additionally, we present the residuals of the various fits at the peak of the PSD, indicating the primary period in the LC. The AIC values for both PSD models, as well as the RML for model comparison, are provided. Objects with test statistics $<$2$\sigma$ are indicated by a double-line separation, signifying their exclusion. \label{tab:correction_psds_results}}
{%
\begin{tabular}{l|cccccccccc}
\hline
\hline
Association Name & PSD Fit & GLSP & LSP2 & PDM & CWT & DFT-Welch & Period (yr) & Residual$_{peak}$ & AIC & RML \\
\hline
\hline
\multirow{3}{*}{PKS 0405$-$385} 
                                        & PL & 2.4$\sigma$ & 2.8$\sigma$ & 2.5$\sigma$ & 2.7$\sigma$ & 3.0$\sigma$ & 3.1$\pm$0.5 (2.7$\sigma$) & 0.249 & -1573.9 \\ 
                                        & BPL & 2.3$\sigma$ & 2.7$\sigma$ & 2.2$\sigma$ & 2.5$\sigma$ & 2.9$\sigma$ & 3.1$\pm$0.5 (2.5$\sigma$) & 0.233 & -1613.8 & $2.2\mathrm{x}10^{-9}$\\
        \hline
 	\multirow{3}{*}{S5 1044+71}  
                                            & PL & 3.1$\sigma$ & 3.4$\sigma$ & 2.8$\sigma$ & 2.9$\sigma$ & 3.7$\sigma$ & 3.1$\pm$0.5 (3.1$\sigma$) & 0.393 & -945.3\\ 
                                            & BPL & 2.5$\sigma$ & 2.7$\sigma$ & 2.3$\sigma$ & 2.5$\sigma$ & 3.2$\sigma$ & 3.1$\pm$0.5 (2.5$\sigma$) & 0.373 & -1243.4 & $1.8\mathrm{x}10^{-65}$ \\ 
        \hline
        \multirow{3}{*}{MH 2136$-$428} 
                                    & PL & 3.0$\sigma$ & 3.6$\sigma$ & 2.5$\sigma$ & 2.2$\sigma$ & 4.1$\sigma$ & 1.8$\pm$0.1 (3.0$\sigma$) & 0.178 & -1810.2 \\ 
                                    & BPL & 2.5$\sigma$ & 2.8$\sigma$ & 2.1$\sigma$ & 2.0$\sigma$ & 3.1$\sigma$ & 1.8$\pm$0.1 (2.5$\sigma$) & 0.182 & -1813.5 & $3.0\mathrm{x}10^{-3}$ \\     
        \hline
        \multirow{3}{*}{PKS 0524$-$485$\star$} 
                                    & PL & 2.3$\sigma$ & \makecell{3.4$\sigma$ \\ 2.6$\sigma$} & 2.5$\sigma$ & \makecell{2.1$\sigma$ \\ 2.4$\sigma$} & \makecell{4.5$\sigma$\\ 3.0$\sigma$} & \makecell{2.0$\pm$0.2 (2.4$\sigma$) \\ 4.0$\pm$0.5 (2.4$\sigma$)} & \makecell{0.178 \\ 0.168} & -1722.6 \\ 
                                    & BPL & 2.1$\sigma$ & \makecell{2.7$\sigma$ \\ 2.3$\sigma$} & 2.1$\sigma$ & \makecell{2.1$\sigma$ \\ 1.9$\sigma$} & \makecell{3.9$\sigma$ \\ 2.8$\sigma$} & \makecell{2.0$\pm$0.2 (2.4$\sigma$) \\ 4.0$\pm$0.5 (2.2$\sigma$)} & \makecell{0.168 \\ 0.160} & -1846.4 & $1.3\mathrm{x}10^{-27}$\\                                      
        \hline
        \multirow{3}{*}{PKS 0215+015$\star$} 
                                    & PL & 2.4$\sigma$ & \makecell{2.1$\sigma$\\ 2.6$\sigma$} & 2.0$\sigma$ & 1.6$\sigma$ & 2.7$\sigma$ & 3.4$\pm$0.4 (2.4$\sigma$) & 0.108 & -2059.3\\ 
                                    & BPL & 2.2$\sigma$ & \makecell{2.2$\sigma$\\ 2.4$\sigma$} & 1.5$\sigma$ & 1.8$\sigma$ & 2.3$\sigma$ & 3.4$\pm$0.4 (2.2$\sigma$) & 0.130 & -2073.9 & $6.7\mathrm{x}10^{-4}$\\         
        \hline 
        \multirow{3}{*}{S2 0109+22} 
                                    & PL & 2.6$\sigma$ & 2.4$\sigma$ & 1.5$\sigma$ & 1.7$\sigma$ & 2.4$\sigma$ & 2.4$\pm$0.3 (2.4$\sigma$) & 0.087 & -2448.9 \\ 
                                    & BPL & 2.3$\sigma$ & 2.1$\sigma$ & 1.4$\sigma$ & 1.5$\sigma$ & 2.1$\sigma$ & 2.4$\pm$0.3 (2.1$\sigma$) & 0.088 & -2519.7 & $3.8\mathrm{x}10^{-3}$\\
        \hline 
        \multirow{3}{*}{PMN J0533$-$5549$\star$} 
                                    & PL & \makecell{2.7$\sigma$ \\ 2.6$\sigma$} & \makecell{2.4$\sigma$ \\ 2.2$\sigma$} & 1.8$\sigma$ & 1.7$\sigma$ & 1.4$\sigma$ & \makecell{2.8$\pm$0.3 (2.1$\sigma$) \\ 2.0$\pm$0.3(2.2$\sigma$)} & \makecell{0.121 \\ 0.105} & -2104.8\\ 
                                    & BPL & \makecell{2.3$\sigma$ \\ 2.2$\sigma$} & \makecell{2.1$\sigma$ \\ 2.0$\sigma$} & 1.4$\sigma$ & 1.8$\sigma$ & 2.0$\sigma$ & \makecell{2.8$\pm$0.3 (2.0$\sigma$) \\ 2.0$\pm$0.3(2.2$\sigma$)} & \makecell{0.106 \\ 0.094} & -2123.6 & $8.2\mathrm{x}10^{-5}$ \\
        \hline
        \multirow{3}{*}{OP 313} 
                                    & PL & 2.1$\sigma$ & 2.2$\sigma$ & 1.8$\sigma$ & 2.2$\sigma$ & 2.6$\sigma$ & 5.7$\pm$0.8 (2.2$\sigma$) & 0.209 & -1663.1\\ 
                                    & BPL & 2.0$\sigma$ & 2.1$\sigma$ & 1.8$\sigma$ & 2.0$\sigma$ & 2.6$\sigma$ & 5.7$\pm$0.8 (2.0$\sigma$) & 0.208 & -1710.1 & $6.2\mathrm{x}10^{-11}$\\
        \hline 
        \multirow{3}{*}{S4 1030+61} 
                                    & PL & 2.0$\sigma$ & 2.7$\sigma$ & 2.2$\sigma$ & 2.0$\sigma$ & 2.3$\sigma$ & 2.9$\pm$0.4 (2.2$\sigma$) & 0.198 & -1909.8\\ 
                                    & BPL & 2.0$\sigma$ & 2.5$\sigma$ & 2.0$\sigma$ & 1.9$\sigma$ & 2.0$\sigma$ & 2.9$\pm$0.4 (2.0$\sigma$) & 0.194 & -2027.3 & $2.9\mathrm{x}10^{-26}$\\
        \hline
        \multirow{3}{*}{S5 1039+81} 
                                    & PL & 2.1$\sigma$ & 2.6$\sigma$ & 1.8$\sigma$ & 2.1$\sigma$ & 2.6$\sigma$ & 3.5$\pm$0.4 (2.1$\sigma$) & 0.126 & -2201.2\\ 
                                    & BPL & 2.0$\sigma$ & 2.4$\sigma$ & 1.7$\sigma$ & 2.0$\sigma$ & 2.5$\sigma$ & 3.5$\pm$0.4 (2.0$\sigma$) & 0.125 & -2215.2 & $9.1\mathrm{x}10^{-4}$\\
        \hline 
        \hline 
        \multirow{3}{*}{PKS 0451$-$28} 
                                    & PL & 2.1$\sigma$ & 2.1$\sigma$ & 1.6$\sigma$ & 1.9$\sigma$ & 1.3$\sigma$ & 5.0$\pm$0.7 (1.9$\sigma$) & 0.155 & -1858.1\\ 
                                    & BPL & 2.1$\sigma$ & 2.1$\sigma$ & 1.6$\sigma$ & 1.9$\sigma$ & 1.2$\sigma$ & 5.0$\pm$0.7 (1.9$\sigma$) & 0.152 & -1881.5 & $8.3\mathrm{x}10^{-6}$ \\
        \hline 
        \multirow{3}{*}{PKS 0736+017$\star$} 
                                    & PL & \makecell{2.0$\sigma$ \\ 1.3$\sigma$} & 2.2$\sigma$ & 1.9$\sigma$ & 1.9$\sigma$ & 2.0$\sigma$ & 4.5$\pm$0.8 (1.9$\sigma$) & 0.142 & -1917.4 \\ 
                                    & BPL & \makecell{1.5$\sigma$ \\ 1.4$\sigma$} & 2.1$\sigma$ & 1.8$\sigma$ & 1.9$\sigma$ & 2.0$\sigma$ & 4.5$\pm$0.8 (1.9$\sigma$) & 0.126 & -1966.1 & $2.6\mathrm{x}10^{-11}$ \\
        \hline 
        \multirow{3}{*}{PKS 1824$-$582} 
                                    & PL & 1.2$\sigma$ & 2.4$\sigma$ & 2.1$\sigma$ & 1.8$\sigma$ & 2.7$\sigma$ & 2.4$\pm$0.1 (2.1$\sigma$) & 0.087 & -2454.3 \\ 
                                    & BPL & 1.2$\sigma$ & 2.3$\sigma$ & 1.9$\sigma$ & 1.7$\sigma$ & 2.7$\sigma$ & 2.4$\pm$0.1 (1.9$\sigma$) & 0.086 & -2569.2 & $1.2\mathrm{x}10^{-25}$ \\
        \hline 
        \multirow{3}{*}{PKS 1903$-$80$\star$} 
                                    & PL & 1.9$\sigma$ & 2.6$\sigma$ & 1.8$\sigma$ & 1.8$\sigma$ & 2.5$\sigma$ & 2.4$\pm$0.2 (1.9$\sigma$) & 0.145 & -1949.2 \\ 
                                    & BPL & 1.7$\sigma$ & 2.2$\sigma$ & 1.6$\sigma$ & 1.5$\sigma$ & 1.9$\sigma$ & 2.4$\pm$0.2 (1.7$\sigma$) & 0.135 & -1982.4 & $6.1\mathrm{x}10^{-8}$ \\        
        \hline        
        \multirow{3}{*}{S4 0110+49} 
                                    & PL & 1.4$\sigma$ & 2.1$\sigma$ & 1.4$\sigma$ & 1.9$\sigma$ & 1.9$\sigma$ & 3.7$\pm$0.8 (1.9$\sigma$) & 0.179 & -1770.5\\ 
                                    & BPL & 1.3$\sigma$ & 1.8$\sigma$ & 1.4$\sigma$ & 1.7$\sigma$ & 2.0$\sigma$ & 3.7$\pm$0.8 (1.7$\sigma$) & 0.150 & -1897.5 & $2.6\mathrm{x}10^{-28}$\\
        \hline 
        \multirow{3}{*}{MG2 J083121+2629} 
                                    & PL & 2.0$\sigma$ & 2.1$\sigma$ & 1.3$\sigma$ & 1.6$\sigma$ & 1.3$\sigma$ & 3.3$\pm$0.5 (1.6$\sigma$) & 0.064 & -2609.1\\ 
                                    & BPL & 1.8$\sigma$ & 1.8$\sigma$ & 1.3$\sigma$ & 1.5$\sigma$ & 1.3$\sigma$ & 3.3$\pm$0.5 (1.5$\sigma$) & 0.067 & -2625.7 & $2.4\mathrm{x}10^{-4}$\\
\hline
\hline
\end{tabular}%
}
\end{table*}

\begin{table*}
\centering
\caption{Results of the correction to the test statistics using the PL models from Table \ref{tab:slopes}. Three different models are examined based on the fit parameters and their uncertainties. \label{tab:correction_power_law_new}}
{%
\begin{tabular}{l|cccccccccc}
\hline
\hline
Association Name & Model & GLSP & LSP & PDM & CWT & DFT-Welch \\
\hline
\hline
\multirow{2}{*}{PKS 0405$-$385} 
                & \makecell{\makecell{$\beta_{min}$=1.08 \\ $A_{min}$=1.68$\mathrm{x}10^{-2}$ \\ \\ }\\ \makecell{$\beta$=1.11 \\ $A$=1.79$\mathrm{x}10^{-2}$ \\ \\ }\\ \makecell{$\beta_{max}$=1.14 \\ $A_{max}$=1.92$\mathrm{x}10^{-2}$ \\ \\ } } 
                & \makecell{2.6$\sigma$ \\ \\ \\ 2.4$\sigma$ \\ \\ \\ 2.2$\sigma$ \\ \\ \\ } 
                & \makecell{3.0$\sigma$ \\ \\ \\ 2.8$\sigma$ \\ \\ \\ 2.5$\sigma$ \\ \\ \\ } 
                & \makecell{2.8$\sigma$ \\ \\ \\ 2.5$\sigma$ \\ \\ \\ 2.3$\sigma$ \\ \\ \\ } 
                & \makecell{2.9$\sigma$ \\ \\ \\ 2.7$\sigma$ \\ \\ \\ 2.4$\sigma$ \\ \\ \\ } 
                & \makecell{3.2$\sigma$ \\ \\ \\ 3.0$\sigma$ \\ \\ \\ 2.7$\sigma$ \\ \\ \\ } 
            \\
        \hline
         \multirow{2}{*}{S5 1044+71} 
                & \makecell{\makecell{$\beta_{min}$=1.11 \\ $A_{min}$=1.98$\mathrm{x}10^{-2}$ \\ \\ }\\ \makecell{$\beta$=1.21 \\ $A$=2.13$\mathrm{x}10^{-2}$ \\ \\ }\\ \makecell{$\beta_{max}$=1.31 \\ $A_{max}$=2.27$\mathrm{x}10^{-2}$ \\ \\ } } 
                & \makecell{3.2$\sigma$ \\ \\ \\ 3.1$\sigma$ \\ \\ \\ 2.9$\sigma$ \\ \\ \\ } 
                & \makecell{3.5$\sigma$ \\ \\ \\ 3.4$\sigma$ \\ \\ \\ 3.1$\sigma$ \\ \\ \\ } 
                & \makecell{2.9$\sigma$ \\ \\ \\ 2.8$\sigma$ \\ \\ \\ 2.7$\sigma$ \\ \\ \\ } 
                & \makecell{3.1$\sigma$ \\ \\ \\ 2.9$\sigma$ \\ \\ \\ 2.7$\sigma$ \\ \\ \\ } 
                & \makecell{3.9$\sigma$ \\ \\ \\ 3.7$\sigma$ \\ \\ \\ 3.5$\sigma$ \\ \\ \\ } 
            \\            
        \hline
        \multirow{2}{*}{MH 2136$-$428} 
                & \makecell{\makecell{$\beta_{min}$=0.54 \\ $A_{min}$=2.78$\mathrm{x}10^{-2}$ \\ \\ }\\ \makecell{$\beta$=0.58 \\ $A$=3.06$\mathrm{x}10^{-2}$ \\ \\ }\\ \makecell{$\beta_{max}$=0.64 \\ $A_{max}$=3.33$\mathrm{x}10^{-2}$ \\ \\ } } 
                & \makecell{3.2$\sigma$ \\ \\ \\ 3.0$\sigma$ \\ \\ \\ 2.8$\sigma$ \\ \\ \\ } 
                & \makecell{3.8$\sigma$ \\ \\ \\ 3.6$\sigma$ \\ \\ \\ 3.4$\sigma$ \\ \\ \\ } 
                & \makecell{2.6$\sigma$ \\ \\ \\ 2.5$\sigma$ \\ \\ \\ 2.4$\sigma$ \\ \\ \\ } 
                & \makecell{2.3$\sigma$ \\ \\ \\ 2.2$\sigma$ \\ \\ \\ 2.0$\sigma$ \\ \\ \\ } 
                & \makecell{4.3$\sigma$ \\ \\ \\ 4.1$\sigma$ \\ \\ \\ 4.0$\sigma$ \\ \\ \\ } 
            \\            
        \hline
        \multirow{2}{*}{PKS 0524$-$485} 
                & \makecell{\makecell{$\beta_{min}$=0.64 \\ $A_{min}$=2.94$\mathrm{x}10^{-2}$ \\ \\ }\\ \makecell{$\beta$=0.71 \\ $A$=3.17$\mathrm{x}10^{-2}$ \\ \\ }\\ \makecell{$\beta_{max}$=0.78 \\ $A_{max}$=3.39$\mathrm{x}10^{-2}$ \\ \\ } } 
                & \makecell{2.5$\sigma$ \\ \\ \\ 2.3$\sigma$ \\ \\ \\ 2.2$\sigma$ \\ \\ \\ } 
                & \makecell{2.8$\sigma$ \\ \\ \\ 2.6$\sigma$ \\ \\ \\ 2.3$\sigma$ \\ \\ \\ } 
                & \makecell{2.6$\sigma$ \\ \\ \\ 2.5$\sigma$ \\ \\ \\ 2.4$\sigma$ \\ \\ \\ } 
                & \makecell{2.2$\sigma$ \\ \\ \\ 2.1$\sigma$ \\ \\ \\ 2.0$\sigma$ \\ \\ \\ } 
                & \makecell{3.2$\sigma$ \\ \\ \\ 3.0$\sigma$ \\ \\ \\ 2.8$\sigma$ \\ \\ \\ } 
            \\            
\hline
\hline
\end{tabular}%
}
\end{table*}

\begin{table*}
\centering
\setcounter{table}{3}
\caption{\textit{(continued)} \label{tab:correction_power_law_b_new}}
{%
\begin{tabular}{l|cccccccccc}
\hline
\hline
Association Name & Model & GLSP & LSP & PDM & CWT & DFT-Welch \\
\hline
\hline
\multirow{2}{*}{PKS 0215+015} 
                & \makecell{\makecell{$\beta_{min}$=0.92 \\ $A_{min}$=2.38$\mathrm{x}10^{-2}$ \\ \\ }\\ \makecell{$\beta$=0.97 \\ $A$=2.50$\mathrm{x}10^{-2}$ \\ \\ }\\ \makecell{$\beta_{max}$=1.02 \\ $A_{max}$=2.61$\mathrm{x}10^{-2}$ \\ \\ } } 
                & \makecell{2.6$\sigma$ \\ \\ \\ 2.4$\sigma$ \\ \\ \\ 2.3$\sigma$ \\ \\ \\ } 
                & \makecell{2.8$\sigma$ \\ \\ \\ 2.6$\sigma$ \\ \\ \\ 2.3$\sigma$ \\ \\ \\ } 
                & \makecell{2.0$\sigma$ \\ \\ \\ 2.0$\sigma$ \\ \\ \\ 1.9$\sigma$ \\ \\ \\ } 
                & \makecell{1.7$\sigma$ \\ \\ \\ 1.6$\sigma$ \\ \\ \\ 1.5$\sigma$ \\ \\ \\ } 
                & \makecell{2.8$\sigma$ \\ \\ \\ 2.7$\sigma$ \\ \\ \\ 2.5$\sigma$ \\ \\ \\ } 
            \\            
        \hline
        \multirow{2}{*}{S2 0109+22} 
                & \makecell{\makecell{$\beta_{min}$=0.58 \\ $A_{min}$=2.07$\mathrm{x}10^{-2}$ \\ \\ }\\ \makecell{$\beta$=0.63 \\ $A$=2.59$\mathrm{x}10^{-2}$ \\ \\ }\\ \makecell{$\beta_{max}$=0.68 \\ $A_{max}$=3.10$\mathrm{x}10^{-2}$ \\ \\ } } 
                & \makecell{2.6$\sigma$ \\ \\ \\ 2.6$\sigma$ \\ \\ \\ 2.5$\sigma$ \\ \\ \\ } 
                & \makecell{2.5$\sigma$ \\ \\ \\ 2.4$\sigma$ \\ \\ \\ 2.3$\sigma$ \\ \\ \\ } 
                & \makecell{1.5$\sigma$ \\ \\ \\ 1.5$\sigma$ \\ \\ \\ 1.4$\sigma$ \\ \\ \\ } 
                & \makecell{1.8$\sigma$ \\ \\ \\ 1.7$\sigma$ \\ \\ \\ 1.6$\sigma$ \\ \\ \\ } 
                & \makecell{2.4$\sigma$ \\ \\ \\ 2.4$\sigma$ \\ \\ \\ 2.3$\sigma$ \\ \\ \\ } 
            \\            
        \hline
        \multirow{2}{*}{PMN J0533$-$5549} 
                & \makecell{\makecell{$\beta_{min}$=0.61 \\ $A_{min}$=2.02$\mathrm{x}10^{-2}$ \\ \\ }\\ \makecell{$\beta$=0.65 \\ $A$=2.13$\mathrm{x}10^{-2}$ \\ \\ }\\ \makecell{$\beta_{max}$=0.69 \\ $A_{max}$=2.25$\mathrm{x}10^{-2}$ \\ \\ } } 
                & \makecell{2.8$\sigma$ \\ \\ \\ 2.7$\sigma$ \\ \\ \\ 2.6$\sigma$ \\ \\ \\ } 
                & \makecell{2.5$\sigma$ \\ \\ \\ 2.4$\sigma$ \\ \\ \\ 2.3$\sigma$ \\ \\ \\ } 
                & \makecell{2.0$\sigma$ \\ \\ \\ 1.8$\sigma$ \\ \\ \\ 1.7$\sigma$ \\ \\ \\ } 
                & \makecell{1.8$\sigma$ \\ \\ \\ 1.7$\sigma$ \\ \\ \\ 1.6$\sigma$ \\ \\ \\ } 
                & \makecell{1.6$\sigma$ \\ \\ \\ 1.4$\sigma$ \\ \\ \\ 1.3$\sigma$ \\ \\ \\ } 
            \\            
        \hline
        \multirow{2}{*}{OP 313} 
                & \makecell{\makecell{$\beta_{min}$=0.81 \\ $A_{min}$=2.37$\mathrm{x}10^{-2}$ \\ \\ }\\ \makecell{$\beta$=0.85 \\ $A$=2.48$\mathrm{x}10^{-2}$ \\ \\ }\\ \makecell{$\beta_{max}$=0.89 \\ $A_{max}$=2.58$\mathrm{x}10^{-2}$ \\ \\ } } 
                & \makecell{2.2$\sigma$ \\ \\ \\ 2.1$\sigma$ \\ \\ \\ 2.1$\sigma$ \\ \\ \\ } 
                & \makecell{2.3$\sigma$ \\ \\ \\ 2.2$\sigma$ \\ \\ \\ 2.0$\sigma$ \\ \\ \\ } 
                & \makecell{1.8$\sigma$ \\ \\ \\ 1.8$\sigma$ \\ \\ \\ 1.7$\sigma$ \\ \\ \\ } 
                & \makecell{2.3$\sigma$ \\ \\ \\ 2.2$\sigma$ \\ \\ \\ 2.0$\sigma$ \\ \\ \\ } 
                & \makecell{2.7$\sigma$ \\ \\ \\ 2.6$\sigma$ \\ \\ \\ 2.4$\sigma$ \\ \\ \\ } 
            \\            
        \hline
        \multirow{2}{*}{S4 1030+61} 
                & \makecell{\makecell{$\beta_{min}$=0.80 \\ $A_{min}$=2.29$\mathrm{x}10^{-2}$ \\ \\ }\\ \makecell{$\beta$=0.85 \\ $A$=2.46$\mathrm{x}10^{-2}$ \\ \\ }\\ \makecell{$\beta_{max}$=0.90 \\ $A_{max}$=2.62$\mathrm{x}10^{-2}$ \\ \\ } } 
                & \makecell{2.1$\sigma$ \\ \\ \\ 2.0$\sigma$ \\ \\ \\ 2.0$\sigma$ \\ \\ \\ } 
                & \makecell{2.8$\sigma$ \\ \\ \\ 2.7$\sigma$ \\ \\ \\ 2.7$\sigma$ \\ \\ \\ } 
                & \makecell{2.3$\sigma$ \\ \\ \\ 2.2$\sigma$ \\ \\ \\ 2.1$\sigma$ \\ \\ \\ } 
                & \makecell{2.0$\sigma$ \\ \\ \\ 2.0$\sigma$ \\ \\ \\ 1.9$\sigma$ \\ \\ \\ } 
                & \makecell{2.4$\sigma$ \\ \\ \\ 2.3$\sigma$ \\ \\ \\ 2.2$\sigma$ \\ \\ \\ } 
            \\            
        \hline
        \multirow{2}{*}{S5 1039+81} 
                & \makecell{\makecell{$\beta_{min}$=0.49 \\ $A_{min}$=1.97$\mathrm{x}10^{-2}$ \\ \\ }\\ \makecell{$\beta$=0.53 \\ $A$=2.32$\mathrm{x}10^{-2}$ \\ \\ }\\ \makecell{$\beta_{max}$=0.57 \\ $A_{max}$=2.65$\mathrm{x}10^{-2}$ \\ \\ } } 
                & \makecell{2.2$\sigma$ \\ \\ \\ 2.1$\sigma$ \\ \\ \\ 2.1$\sigma$ \\ \\ \\ } 
                & \makecell{2.8$\sigma$ \\ \\ \\ 2.6$\sigma$ \\ \\ \\ 2.5$\sigma$ \\ \\ \\ } 
                & \makecell{1.8$\sigma$ \\ \\ \\ 1.8$\sigma$ \\ \\ \\ 1.7$\sigma$ \\ \\ \\ } 
                & \makecell{2.1$\sigma$ \\ \\ \\ 2.0$\sigma$ \\ \\ \\ 1.8$\sigma$ \\ \\ \\ } 
                & \makecell{2.6$\sigma$ \\ \\ \\ 2.6$\sigma$ \\ \\ \\ 2.5$\sigma$ \\ \\ \\ } 
            \\
\hline
\hline
\end{tabular}%
}
\end{table*}

\begin{table*}
\centering
\caption{Results of the correction to the significance using the BPL models from Table \ref{tab:slopes}. Three different models are analyzed based on the fit parameters and their uncertainties.\label{tab:correction_bpl1_new}} 
{%
\begin{tabular}{l|cccccccccc}
\hline
\hline
Association Name & Model & GLSP & LSP & PDM & CWT & DFT-Welch \\
\hline
\hline
\multirow{2}{*}{PKS 0405$-$385} 
                & \makecell{\makecell{$\alpha_{min}$=1.54 $\nu_{Bending\_min}$=0.38 \\ $A_{min}$=9.88$\mathrm{x}10^{-2}$ \\ \\ } \\ \makecell{$\alpha$=1.62 $\nu_{Bending}$=0.45 \\ $A$=1.31$\mathrm{x}10^{-1}$ \\ \\ } \\ \makecell{$\alpha_{max}$=1.70 $\nu_{Bending\_max}$=0.52 \\ $A_{max}$=1.63$\mathrm{x}10^{-1}$ \\ \\ } \\} 
                & \makecell{2.3$\sigma$ \\ \\ \\ 2.3$\sigma$ \\ \\ \\ 2.3$\sigma$ \\ \\ \\ } 
                & \makecell{2.7$\sigma$ \\ \\ \\ 2.7$\sigma$ \\ \\ \\ 2.7$\sigma$ \\ \\ \\ } 
                & \makecell{2.2$\sigma$ \\ \\ \\ 2.2$\sigma$ \\ \\ \\ 2.1$\sigma$ \\ \\ \\ } 
                & \makecell{2.5$\sigma$ \\ \\ \\ 2.5$\sigma$ \\ \\ \\ 2.5$\sigma$ \\ \\ \\ } 
                & \makecell{3.9$\sigma$ \\ \\ \\ 3.9$\sigma$ \\ \\ \\ 3.8$\sigma$ \\ \\ \\ } 
            \\
        \hline
         \multirow{2}{*}{S5 1044+71} 
                & \makecell{\makecell{$\alpha_{min}$=2.29 $\nu_{Bending\_min}$=0.54 \\ $A_{min}$=4.97$\mathrm{x}10^{-2}$ \\ \\ } \\ \makecell{$\alpha$=2.36 $\nu_{Bending}$=0.59 \\ $A$=1.08$\mathrm{x}10^{-1}$ \\ \\ } \\ \makecell{$\alpha_{max}$=2.43 $\nu_{Bending\_max}$=0.63 \\ $A_{max}$=1.66$\mathrm{x}10^{-1}$ \\ \\ } \\} 
                & \makecell{2.5$\sigma$ \\ \\ \\ 2.5$\sigma$ \\ \\ \\ 2.6$\sigma$ \\ \\ \\ } 
                & \makecell{2.7$\sigma$ \\ \\ \\ 2.7$\sigma$ \\ \\ \\ 2.7$\sigma$ \\ \\ \\ } 
                & \makecell{2.3$\sigma$ \\ \\ \\ 2.3$\sigma$ \\ \\ \\ 2.3$\sigma$ \\ \\ \\ } 
                & \makecell{2.5$\sigma$ \\ \\ \\ 2.5$\sigma$ \\ \\ \\ 2.6$\sigma$ \\ \\ \\ } 
                & \makecell{3.1$\sigma$ \\ \\ \\ 3.2$\sigma$ \\ \\ \\ 3.2$\sigma$ \\ \\ \\ } 
            \\            
            \hline
            \multirow{2}{*}{MH 2136$-$428} 
                & \makecell{\makecell{$\alpha_{min}$=1.53 $\nu_{Bending\_min}$=0.44 \\ $A_{min}$=3.09$\mathrm{x}10^{-2}$ \\ \\ } \\ \makecell{$\alpha$=1.65 $\nu_{Bending}$=0.49 \\ $A$=1.13$\mathrm{x}10^{-1}$ \\ \\ } \\ \makecell{$\alpha_{max}$=1.77 $\nu_{Bending\_max}$=0.54 \\ $A_{max}$=1.95$\mathrm{x}10^{-1}$ \\ \\ } \\} 
                & \makecell{2.5$\sigma$ \\ \\ \\ 2.5$\sigma$ \\ \\ \\ 2.4$\sigma$ \\ \\ \\ } 
                & \makecell{2.8$\sigma$ \\ \\ \\ 2.8$\sigma$ \\ \\ \\ 2.7$\sigma$ \\ \\ \\ } 
                & \makecell{2.1$\sigma$ \\ \\ \\ 2.1$\sigma$ \\ \\ \\ 2.1$\sigma$ \\ \\ \\ } 
                & \makecell{2.0$\sigma$ \\ \\ \\ 2.0$\sigma$ \\ \\ \\ 2.0$\sigma$ \\ \\ \\ } 
                & \makecell{3.2$\sigma$ \\ \\ \\ 3.1$\sigma$ \\ \\ \\ 3.0$\sigma$ \\ \\ \\ } 
            \\            
            \hline  
            \multirow{2}{*}{PKS 0524$-$485} 
                & \makecell{\makecell{$\alpha_{min}$=1.63 $\nu_{Bending\_min}$=0.67 \\ $A_{min}$=0.12 \\ \\ } \\ \makecell{$\alpha$=1.72 $\nu_{Bending}$=0.74 \\ $A$=1.04$\mathrm{x}10^{-1}$ \\ \\ } \\ \makecell{$\alpha_{max}$=1.79 $\nu_{Bending\_max}$=0.81 \\ $A_{max}$=1.72$\mathrm{x}10^{-1}$ \\ \\ } \\} 
                & \makecell{2.1$\sigma$ \\ \\ \\ 2.1$\sigma$ \\ \\ \\ 2.1$\sigma$ \\ \\ \\ } 
                & \makecell{2.7$\sigma$ \\ \\ \\ 2.7$\sigma$ \\ \\ \\ 2.7$\sigma$ \\ \\ \\ } 
                & \makecell{2.1$\sigma$ \\ \\ \\ 2.1$\sigma$ \\ \\ \\ 2.1$\sigma$ \\ \\ \\ } 
                & \makecell{2.1$\sigma$ \\ \\ \\ 2.1$\sigma$ \\ \\ \\ 2.1$\sigma$ \\ \\ \\ } 
                & \makecell{3.9$\sigma$ \\ \\ \\ 3.9$\sigma$ \\ \\ \\ 3.8$\sigma$ \\ \\ \\ } 
            \\            
\hline
\hline
\end{tabular}%
}
\end{table*}

\begin{table*}
\centering
\setcounter{table}{4}
\caption{\textit{(continued)}\label{tab:correction_bpl1_b_new}}
{%
\begin{tabular}{l|cccccccccc}
\hline
\hline
Association Name & Model & GLSP & LSP & PDM & CWT & DFT-Welch \\
\hline
\hline
\multirow{2}{*}{PKS 0215+015} 
                & \makecell{\makecell{$\alpha_{min}$=1.80 $\nu_{Bending\_min}$=0.88 \\ $A_{min}$=5.95$\mathrm{x}10^{-2}$ \\ \\ } \\ \makecell{$\alpha$=1.94 $\nu_{Bending}$=0.97 \\ $A$=6.33$\mathrm{x}10^{-2}$ \\ \\ } \\ \makecell{$\alpha_{max}$=2.08 $\nu_{Bending\_max}$=1.06 \\ $A_{max}$=6.70$\mathrm{x}10^{-2}$ \\ \\ } \\} 
                & \makecell{2.1$\sigma$ \\ \\ \\ 2.2$\sigma$ \\ \\ \\ 2.2$\sigma$ \\ \\ \\ } 
                & \makecell{2.3$\sigma$ \\ \\ \\ 2.4$\sigma$ \\ \\ \\ 2.4$\sigma$ \\ \\ \\ } 
                & \makecell{1.5$\sigma$ \\ \\ \\ 1.5$\sigma$ \\ \\ \\ 1.6$\sigma$ \\ \\ \\ }
                & \makecell{1.8$\sigma$ \\ \\ \\ 1.8$\sigma$ \\ \\ \\ 1.8$\sigma$ \\ \\ \\ }
                & \makecell{2.3$\sigma$ \\ \\ \\ 2.3$\sigma$ \\ \\ \\ 1.3$\sigma$ \\ \\ \\ } 
            \\            
            \hline
            \multirow{2}{*}{S2 0109+22} 
                & \makecell{\makecell{$\alpha_{min}$=1.39 $\nu_{Bending\_min}$=1.17 \\ $A_{min}$=4.30$\mathrm{x}10^{-2}$ \\ \\ } \\ \makecell{$\alpha$=1.92 $\nu_{Bending}$=1.34 \\ $A$=4.59$\mathrm{x}10^{-2}$ \\ \\ } \\ \makecell{$\alpha_{max}$=2.45 $\nu_{Bending\_max}$=1.51 \\ $A_{max}$=4.87$\mathrm{x}10^{-2}$ \\ \\ } \\} 
                & \makecell{2.5$\sigma$ \\ \\ \\ 2.3$\sigma$ \\ \\ \\ 2.2$\sigma$ \\ \\ \\ } 
                & \makecell{2.2$\sigma$ \\ \\ \\ 2.1$\sigma$ \\ \\ \\ 1.9$\sigma$ \\ \\ \\ } 
                & \makecell{1.6$\sigma$ \\ \\ \\ 1.4$\sigma$ \\ \\ \\ 1.2$\sigma$ \\ \\ \\ } 
                & \makecell{1.7$\sigma$ \\ \\ \\ 1.5$\sigma$ \\ \\ \\ 1.5$\sigma$ \\ \\ \\ } 
                & \makecell{2.3$\sigma$ \\ \\ \\ 2.1$\sigma$ \\ \\ \\ 2.0$\sigma$ \\ \\ \\ } 
            \\            
            \hline
            \multirow{2}{*}{PMN J0533$-$5549} 
                & \makecell{\makecell{$\alpha_{min}$=1.03 $\nu_{Bending\_min}$=0.25 \\ $A_{min}$=2.40$\mathrm{x}10^{-2}$ \\ \\ } \\ \makecell{$\alpha$=1.11 $\nu_{Bending}$=0.33 \\ $A$=1.15$\mathrm{x}10^{-1}$ \\ \\ } \\ \makecell{$\alpha_{max}$=1.19 $\nu_{Bending\_max}$=0.41 \\ $A_{max}$=2.06$\mathrm{x}10^{-1}$ \\ \\ } \\} 
                & \makecell{2.3$\sigma$ \\ \\ \\ 2.3$\sigma$ \\ \\ \\ 2.3$\sigma$ \\ \\ \\ } 
                & \makecell{2.1$\sigma$ \\ \\ \\ 2.1$\sigma$ \\ \\ \\ 2.1$\sigma$ \\ \\ \\ } 
                & \makecell{1.4$\sigma$ \\ \\ \\ 1.4$\sigma$ \\ \\ \\ 1.4$\sigma$ \\ \\ \\ } 
                & \makecell{1.7$\sigma$ \\ \\ \\ 1.8$\sigma$ \\ \\ \\ 1.8$\sigma$ \\ \\ \\ } 
                & \makecell{1.9$\sigma$ \\ \\ \\ 2.0$\sigma$ \\ \\ \\ 2.0$\sigma$ \\ \\ \\ } 
            \\            
            \hline
            \multirow{2}{*}{OP 313} 
                & \makecell{\makecell{$\alpha_{min}$=1.34 $\nu_{Bending\_min}$=0.27 \\ $A_{min}$=5.33$\mathrm{x}10^{-2}$ \\ \\ } \\ \makecell{$\alpha$=1.42 $\nu_{Bending}$=0.32 \\ $A$=1.46$\mathrm{x}10^{-1}$ \\ \\ } \\ \makecell{$\alpha_{max}$=1.50 $\nu_{Bending\_max}$=0.37 \\ $A_{max}$=2.38$\mathrm{x}10^{-1}$ \\ \\ } \\} 
                & \makecell{2.0$\sigma$ \\ \\ \\ 2.0$\sigma$ \\ \\ \\ 2.0$\sigma$ \\ \\ \\ } 
                & \makecell{2.1$\sigma$ \\ \\ \\ 2.1$\sigma$ \\ \\ \\ 2.1$\sigma$ \\ \\ \\ } 
                & \makecell{1.8$\sigma$ \\ \\ \\ 1.8$\sigma$ \\ \\ \\ 1.8$\sigma$ \\ \\ \\ } 
                & \makecell{2.0$\sigma$ \\ \\ \\ 2.0$\sigma$ \\ \\ \\ 2.0$\sigma$ \\ \\ \\ } 
                & \makecell{2.6$\sigma$ \\ \\ \\ 2.6$\sigma$ \\ \\ \\ 2.6$\sigma$ \\ \\ \\ } 
            \\            
            \hline
            \multirow{2}{*}{S4 1030+61} 
                & \makecell{\makecell{$\alpha_{min}$=1.75 $\nu_{Bending\_min}$=0.65 \\ $A_{min}$=8.49$\mathrm{x}10^{-2}$ \\ \\ } \\ \makecell{$\alpha$=1.83 $\nu_{Bending}$=0.71 \\ $A$=8.84$\mathrm{x}10^{-2}$ \\ \\ } \\ \makecell{$\alpha_{max}$=1.91 $\nu_{Bending\_max}$=0.77 \\ $A_{max}$=9.18$\mathrm{x}10^{-2}$ \\ \\ } \\} 
                & \makecell{2.0$\sigma$ \\ \\ \\ 2.0$\sigma$ \\ \\ \\ 2.0$\sigma$ \\ \\ \\ } 
                & \makecell{2.5$\sigma$ \\ \\ \\ 2.5$\sigma$ \\ \\ \\ 2.5$\sigma$ \\ \\ \\ } 
                & \makecell{2.1$\sigma$ \\ \\ \\ 2.2$\sigma$ \\ \\ \\ 2.2$\sigma$ \\ \\ \\ } 
                & \makecell{1.8$\sigma$ \\ \\ \\ 1.9$\sigma$ \\ \\ \\ 1.9$\sigma$ \\ \\ \\ } 
                & \makecell{2.0$\sigma$ \\ \\ \\ 2.0$\sigma$ \\ \\ \\ 2.0$\sigma$ \\ \\ \\ } 
            \\            
            \hline
            \multirow{2}{*}{S5 1039+81} 
                & \makecell{\makecell{$\alpha_{min}$=1.01 $\nu_{Bending\_min}$=0.60 \\ $A_{min}$=4.87$\mathrm{x}10^{-2}$ \\ \\ } \\ \makecell{$\alpha$=1.16 $\nu_{Bending}$=0.80 \\ $A$=5.24$\mathrm{x}10^{-2}$ \\ \\ } \\ \makecell{$\alpha_{max}$=1.31 $\nu_{Bending\_max}$=1.00 \\ $A_{max}$=5.60$\mathrm{x}10^{-2}$ \\ \\ } \\} 
                & \makecell{2.0$\sigma$ \\ \\ \\ 2.0$\sigma$ \\ \\ \\ 2.0$\sigma$ \\ \\ \\ } 
                & \makecell{2.4$\sigma$ \\ \\ \\ 2.4$\sigma$ \\ \\ \\ 2.5$\sigma$ \\ \\ \\ } 
                & \makecell{1.7$\sigma$ \\ \\ \\ 1.7$\sigma$ \\ \\ \\ 1.8$\sigma$ \\ \\ \\ } 
                & \makecell{2.0$\sigma$ \\ \\ \\ 2.0$\sigma$ \\ \\ \\ 2.0$\sigma$ \\ \\ \\ } 
                & \makecell{2.5$\sigma$ \\ \\ \\ 2.5$\sigma$ \\ \\ \\ 2.6$\sigma$ \\ \\ \\ } 
            \\
\hline
\hline
\end{tabular}%
}
\end{table*}

\begin{table*}
\centering
\caption{List of periods provided by the MCMC Sine Fitting method (MCMC SF), \textit{z}-DCF, and the autoregressive models for the periodic-emission candidates in Table~\ref{tab:candidates_list}. Additionally, the maximum sensitivity of the Bayesian-quasi-periodic oscillation analysis is included \citep{penil_2020}. The \myhash~symbol in the ARFIMA/ARIMA column indicates that the model is ARIMA due to the non-stationarity of the LC, as determined by the augmented Dickey-Fuller test. The period obtained with the ACF from the residuals generated from the original LC and the ARFIMA/ARIMA model is also displayed. Note that some sources have two periods (arranged by significance), denoted by $\star$. Lastly, an {\it X} signifies that the null hypothesis is rejected in the Dickey-Fuller and Box-Ljung tests. All periods are measured in years. \label{tab:mcmc_bayesian_arfima_results}}
{%
\begin{tabular}{l|cccccccccc}
\hline
\hline
Association & MCMC SF & \textit{z}-DCF & ARFIMA/ & ACF & Dicker-Fuller & Box-Ljung \\
Name &  &  & ARIMA &  & Residuals &  & \\
\hline
\hline
PKS 0405$-$385$\star$ & $3.3^{+0.3}_{-0.4}$ & $3.5^{\pm0.2}_{4.0\sigma}$ & [14, 0.456,14] & \makecell{1.0 (2.0$\sigma$) \\ 2.9 (1.9$\sigma$)} & \redcheck & \redcheck \\	
S5 1044+71 & 3.0$\pm$0.1 & $3.2^{\pm0.1}_{4.0\sigma}$ & [15, 0.462, 15] & 1.6 (2.0$\sigma$) & \redcheck & \redcheck \\
MH 2136$-$428 & 1.8$\pm0.1$ & $1.8^{\pm0.1}_{3.0\sigma}$ & \myhash[1,1,1] & 3.4 (1.0$\sigma$) & X & \redcheck \\
PKS 0524$-$485 & 2.0$\pm0.2$ & $2.1^{\pm0.1}_{2.0\sigma}$ & [15, 0.447, 14] & 1.4 (1.2$\sigma$) & \redcheck & \redcheck \\
PKS 0215+015 & $3.3^{+0.1}_{-1.4}$ & $1.2^{\pm0.2}_{1.5\sigma}$ & [15, 0.465, 15] & 1.2 (1.9$\sigma$) & \redcheck & \redcheck \\
S2 0109+22 & $2.4^{+0.3}_{-0.1}$ & $3.2^{\pm0.2}_{2.0\sigma}$ & [12, 0.471, 15] & 2.2 (2.9$\sigma$) & \redcheck & \redcheck \\
PMN J0533$-$5549$\star$ & $2.9^{+0.1}_{-0.9}$ & $2.4^{\pm0.4}_{2\sigma}$ & [14, 0.461, 13] & \makecell{1.4(2.0$\sigma$) \\ 1.8 (1.5$\sigma$)} & \redcheck & \redcheck \\
OP 313 & 5.5$\pm$0.3 & $5.7^{\pm0.6}_{2.5\sigma}$ & [14, 0.450, 14] & 6.1 (1.8$\sigma$) & \redcheck & \redcheck \\
S4 1030+61 & $3.0^{+1.1}_{-0.2}$ & $2.6^{\pm0.2}_{3.0\sigma}$ & [9, 0.264, 14] & 2.7 (2.1$\sigma$) & \redcheck & \redcheck \\
S5 1039+81 & 3.3$\pm$0.1 & $2.6^{\pm0.3}_{2.5\sigma}$ & [9, 0.264, 14] & 2.6 (2.1$\sigma$) & \redcheck & \redcheck \\
\hline
\hline
\end{tabular}%
}
\end{table*}

\begin{table*}
\centering
\caption{Results of the estimations for the upcoming high and low flux states of the blazars listed in Table \ref{tab:candidates_list}. The estimations are based on the parameters of the sine fit, which are also provided: {\it T} represents the period (in years), {\it A} denotes the amplitude of the oscillations and {\it O} indicates the offset (in flux $\times$ 10$^{-8}$ ph cm$^{-2}$ s$^{-1}$)), while {\it$\phi$} represents the phase (in radians). \label{tab:qpo}} 
{%
\begin{tabular}{l|cccccccccc}
\hline
\hline
Association & Sine Model & Date & Peak & Valley & Peak & Valley &  \\
Name &  & Error &  &  &  &  \\
\hline
\hline
PKS 0405$-$385 & \makecell{T=2.917 \\ A=2.9 \\ O=5.6 \\ $\phi$=0.7} & $\pm$7 days & -- & April 13, 2026 & September 24, 2027 & March 8, 2029 \\
        \hline
        S5 1044+71 & \makecell{T=3.067 \\ A=-17.7 \\ O=17.2 \\ $\phi$=6.7} & $\pm$7 days & February 3,  2026 & August 12, 2027 & February 24,  2029 & September 10, 2030 \\
        \hline
        MH 2136$-$428 & \makecell{T=1.820 \\ A=1.9 \\ O=5.7 \\ $\phi$=2.9} & $\pm$3 days & April 9, 2025 & March 10, 2026 & February 1, 2027 & January 3, 2028  \\
        \hline
        PKS 0524$-$485 & \makecell{T=2.073 \\ A=3.6 \\ O=6.3 \\ $\phi$=0.8} & $\pm$4 days & -- & December 1, 2025 & December 12, 2026 & December 22, 2027 \\
        \hline
        PKS 0215+015 & \makecell{T=3.361 \\ A=2.3 \\ O=4.9 \\ $\phi$=1.5} & $\pm$11 days & February 16, 2026 & October 23, 2027 & June 30, 2029 & February 23, 2031 \\
        \hline
        S2 0109+22 & \makecell{T=2.492 \\ A=2.5 \\ O=9.5 \\ $\phi$=0.2} & $\pm$5 days & January 26, 2025 & April 24, 2026 & July 25, 2027 & November 22, 2028 \\
        \hline
        PMN J0533$-$5549 & \makecell{T=2.840 \\ A=2.0 \\ O=7.3 \\ $\phi$=2.3} & $\pm$4 days & -- & January 2, 2026 & June 2, 2027 & October 31, 2028 \\
        \hline
        OP 313 & \makecell{T=5.727 \\ A=4.2 \\ O=6.2 \\ $\phi$=3.3} & $\pm$10 days & September 2, 2025 & July 11, 2028 & May 17, 2031 & April 6, 2034 \\
        \hline
        S4 1030+61 & \makecell{T=2.887 \\ A=4.5 \\ O=6.9 \\ $\phi$=0.9} & $\pm$7 days & April 26, 2025 & November 8, 2026 & March 13, 2028 & August 27, 2029 \\
        \hline
        S5 1039+81 & \makecell{T=3.647 \\ A=1.2 \\ O=3.1 \\ $\phi$=1.8} & $\pm$8 days & -- & July 25, 2025 & May 23, 2027 & March 12, 2029 \\
\hline
\hline
\end{tabular}%
}
\end{table*}

\begin{table*}
\centering
\setcounter{table}{6}
\caption{\textit{(continued)}} 
{%
\begin{tabular}{l|cccccccccc}
\hline
\hline
Association & Sine Model & Date & Peak & Valley & Peak & Valley &  \\
Name &  & Error &  &  &  &  \\
\hline
\hline
PG 1553+113 & \makecell{T=2.100 \\ A=-1.2 \\ O=5.2 \\ $\phi$=1.9} & $\pm$3 days & May 5, 2025 & May 26, 2026 & June 8, 2027 & June 27, 2028 \\
        \hline
        PKS 2155$-$304 & \makecell{T=1.710 \\ A=3.2 \\ O=11.4 \\ $\phi$=1.8} & $\pm$2 days & -- & June 3,2025 & April 13, 2026 & February 22, 2027 \\
        \hline
        OJ 014 & \makecell{T=4.300 \\ A=2.1 \\ O=3.9 \\ $\phi$=2.9} & $\pm$6 days & -- & July 3, 2026 & August 28, 2028 & October 23, 2030 \\
        \hline
        PKS 0454$-$234 & \makecell{T=3.547 \\ A=-13.1 \\ O=24.9 \\ $\phi$=-10.6} & $\pm$5 days & September 28, 2026 & July 7, 2028 & April 16, 2030 & January 24, 2032 \\
        \hline
        S5 0716+714 & \makecell{T=2.630 \\ A=-7.3 \\ O=21.9 \\ $\phi$=-2.6} & $\pm$5 days & September 15, 2025 & January 2, 2027 & April 28, 2028 & August 16, 2029 \\
        \hline
        GB6 J0043+3426 & \makecell{T=1.827 \\ A=0.75 \\ O=2.1 \\ $\phi$=1.3} & $\pm$3 days & November 9, 2025 & October 6, 2026 & September 11, 2027 & August 7, 2028 \\
        \hline
        TXS 0518+211 & \makecell{T=3.127 \\ A=-4.5 \\ O=9.1 \\ $\phi$=4.8} & $\pm$5 days & January 5, 2026 & July 29, 2027 & February 20, 2029 & September 13, 2030 \\
        \hline
        87GB 164812.2+524023 & \makecell{T=3.040 \\ A=0.5 \\ O=1.27 \\ $\phi$=-0.5} & $\pm$8 days & August 18, 2025 & March 1, 2027 & September 2, 2028 & March 7, 2030 \\
        \hline
        PKS 0447$-$439 & \makecell{T=1.813 \\ A=1.8 \\ O=6.5 \\ $\phi$=0.2} & $\pm$3 days & November 26, 2025 & October 27, 2026 & September 20, 2027 & August 28, 2028 \\
        \hline
        PKS 0426$-$380 & \makecell{T=3.547 \\ A=9.7 \\ O=20.3 \\ $\phi$=-1.9} & $\pm$6 days & -- & January 10, 2026 & October 21, 2027 & August 2, 2029 \\
        \hline
        PKS 0301$-$243 & \makecell{T=2.107 \\ A=1.4 \\ O=3.3 \\ $\phi$=0.2} & $\pm$4 days & -- & January 1, 2026 & February 1, 2027 & February 21, 2028 \\
\hline
\hline
\end{tabular}%
}
\end{table*}

\bsp	
\label{lastpage}
\end{document}